\RequirePackage{lineno}
\RequirePackage{fix2col} 
\RequirePackage{fix-cm}

\documentclass[twocolumn,epjc3]{svjour3}

\RequirePackage[T1]{fontenc}
\smartqed  
\RequirePackage{mathptmx}      
\RequirePackage[numbers,sort&compress]{natbib} 

\journalname{Eur.\ Phys.\ J.\ C}
\usepackage{subfig} 

\usepackage{color}\definecolor{darkblue}{rgb}{0,0,0.5}
\usepackage{color}\definecolor{pink}{rgb}{1,0,0.2}      
\usepackage{color}\definecolor{darkgreen}{rgb}{0,0.6,0.0} 
\usepackage{hyperref}
\hypersetup{
    unicode=false,          
    pdftoolbar=true,        
    pdfmenubar=true,        
    pdffitwindow=true,      
    pdftitle={My title},    
    pdfauthor={Author},     
    pdfsubject={Subject},   
    pdfnewwindow=true,      
    pdfkeywords={keywords}, 
    colorlinks=true,       
    linkcolor=darkblue,          
    citecolor=darkblue,        
    filecolor=magenta,      
    urlcolor=darkblue,         
}

\newcommand{\pt}{\ensuremath{p_\mathrm{T}}\xspace}
\newcommand{\mt}{\ensuremath{m_\mathrm{T}}\xspace}

\usepackage{xspace}
\xspaceaddexceptions{[]+} 

\newcommand{\hm}{\ensuremath{\mathrm{h}^-}\xspace}
\newcommand{\pip}{\ensuremath{\pi^+}\xspace}
\newcommand{\pim}{\ensuremath{\pi^-}\xspace}
\newcommand{\km}{\ensuremath{\mathrm{K}^-}\xspace}
\newcommand{\kzeros}{\ensuremath{\mathrm{K}^0_\mathrm{S}}\xspace}
\newcommand{\Lam}{\ensuremath{\mathrm{\Lambda}}\xspace}
\newcommand{\pbar}{\ensuremath{\bar{\textup{p}}}\xspace}

\newcommand{\avg}[1]{\langle{#1}\rangle}
\newcommand{\n}{\ensuremath{n}\xspace}
\newcommand{\dd}{\mathrm{d}}
\newcommand{\dedx}{\ensuremath{\dd E/\dd x}\xspace}
\newcommand{\NASixtyOne}{NA61\slash SHINE\xspace} 

\newcommand{\eeV}{\text{\texorpdfstring{e\kern-0.1em V}{eV}}\xspace}

\newcommand{\GGeVc}{\text{G\eeV/\ensuremath{c}}\xspace}

\newcommand{\GeV}{\text{~G\eeV}\xspace}
\newcommand{\GeVc}{\text{~G\eeV/\ensuremath{c}}\xspace}
\newcommand{\GeVcc}{\text{~G\eeV/}\ensuremath{c^2}\xspace}

\newcommand{\AGeVc}{\ensuremath{A}\text{~G\eeV/\ensuremath{c}}\xspace}

\newcommand{\MeVcc}{\text{~M\eeV/\ensuremath{c^2}}\xspace}

\newcommand{\MMeVcc}{\text{M\eeV/\ensuremath{c^2}}\xspace}

\newcommand{\tof}{\text{\emph{tof}}\xspace}

\usepackage[latin2]{inputenc}
\usepackage{indentfirst}
\usepackage{enumerate}

\usepackage{amsmath}
\usepackage{amssymb}
\DeclareMathOperator\atanh{atanh} 
\usepackage[english]{babel}
\usepackage{natbib}

\usepackage{graphicx} 

\usepackage[normalem]{ulem}

\usepackage{hyphenat} 

\begin{document}
\newcommand{\todo}[1]{\textbf{\textcolor{red}{Note: #1}}}
\pagenumbering{roman}

\title{
Measurement of negatively charged pion spectra
in inelastic p+p interactions at 
\textbf{\emph{p}}$_\mathrm{\textbf{lab}} = $ 20, 31, 40, 80 
and 158\GeV/\emph{c}
}

\author{N.~Abgrall\thanksref{Geneva}
 \and A.~Aduszkiewicz\thanksref{UniversityWarsaw}$^{,*}$
 \and Y.~Ali\thanksref{Cracow}
 \and T.~Anticic\thanksref{Zagreb}
 \and N.~Antoniou\thanksref{Athens}
 \and B.~Baatar\thanksref{Dubna}
 \and F.~Bay\thanksref{Zurich}
 \and A.~Blondel\thanksref{Geneva}
 \and J.~Blumer\thanksref{Karlsruhe}
 \and M.~Bogomilov\thanksref{Sofia}
% ...
 \\(\NASixtyOne Collaboration) 
}

\institute{University of Geneva, Geneva, Switzerland \label{Geneva}
\and
Faculty of Physics, University of Warsaw, Warsaw, Poland
\label{UniversityWarsaw}
\and
Jagiellonian University, Cracow, Poland \label{Cracow}
\and
Rudjer Bo\v{s}kovi\'c Institute, Zagreb, Croatia \label{Zagreb}
\and
University of Athens, Athens, Greece \label{Athens}
\and
Joint Institute for Nuclear Research, Dubna, Russia \label{Dubna}
\and
ETH, Zurich, Switzerland \label{Zurich}
\and
Karlsruhe Institute of Technology, Karlsruhe, Germany \label{Karlsruhe}
\and
Faculty of Physics, University of Sofia, Sofia, Bulgaria \label{Sofia}
% ...
}

\date{\today}

\maketitle

\begin{abstract}
\begin{sloppypar}
We present experimental results on inclusive spectra and mean multiplicities of
negatively charged pions produced in inelastic p+p interactions
at incident projectile momenta of 20, 31, 40, 80 and 158\GeVc
($\sqrt{s} = $ 6.3, 7.7, 8.8, 12.3 and 17.3\GeV, respectively).
The measurements were performed using
the large acceptance \NASixtyOne hadron spectrometer at the
CERN Super Proton Synchrotron.
\end{sloppypar}

Two-dimensional spectra are determined in terms of rapidity and transverse 
momentum.
Their properties such as the width of rapidity distributions and
the inverse slope parameter of transverse mass spectra are extracted and
their collision energy dependences are presented.
The results on inelastic p+p interactions are compared with the corresponding
data on central Pb+Pb collisions measured by the NA49 experiment
at the CERN SPS.

The results presented in this paper are
part of the \NASixtyOne ion program devoted to the study of the
properties of the onset of deconfinement and search for the
critical point of strongly interacting matter.
They are required for interpretation of
results on nucleus\hyp{}nucleus and proton\hyp{}nucleus collisions.

\end{abstract}

\pagenumbering{arabic}

\section{Introduction}

\begin{sloppypar}
This paper presents experimental results on
inclusive spectra and mean multiplicities of
negatively charged pions produced in inelastic p+p interactions
at 20, 31, 40, 80 and 158\GeVc.
The measurements were performed by the multi-purpose
\NASixtyOne{} --
the SPS Heavy Ion and Neutrino Experiment~\cite{proposala} --
at the CERN Super Proton Synchrotron (SPS).
They are part of the \NASixtyOne ion program devoted to the study of the
properties of the onset of deconfinement and search for the
critical point of strongly interacting matter.
Within this program data on p+p, Be+Be and  p+Pb collisions were 
recorded and data on Ar+Ca and Xe+La collisions will be registered within 
the coming years.
The started two dimensional scan in collision energy
and size of colliding nuclei is mainly motivated by
the observation of the onset of deconfinement~\cite{onseta, onsetb}
in central Pb+Pb collisions at about 30\AGeVc by the
NA49 experiment at the CERN SPS.
 Recently the NA49 results were confirmed
by the RHIC beam energy scan programme and their interpretation
by the onset of deconfinement is supported by the LHC results (see 
Ref.~\cite{Rustamov:2012np} and references therein).
\end{sloppypar}

\begin{sloppypar}
In addition to the ion programme, \NASixtyOne is conducting
precise hadron production measurements for improving calculations
of the initial neutrino flux in long-baseline
neutrino oscillation experiments~\cite{T2K-experiment, T2K-experiment2, 
NA61-future, T2K-flux}
as well as for more reliable simulations of cosmic-ray
air showers~\cite{Auger, KASCADE}.
\end{sloppypar}

An interpretation of the rich experimental results on nucleus\hyp{}nucleus
collisions relies to a large extent on a comparison to the
corresponding data on p+p and p+A interactions.
However, the available data concern
mainly basic features of unidentified charged hadron production
and they are sparse.
Many needed results on hadron spectra, fluctuations and
correlations are missing.
Detailed measurements of hadron spectra in a large acceptance
in the beam momentum range covered by the data presented in this
paper exist only for inelastic p+p
interactions at 158\GeVc~\cite{na49_pp_pions,na49_pp_protons,na49_pp_kaons}.
Thus the new high precision measurements of hadron production properties
in p+p and p+A interactions are necessary and they are performed
in parallel with the corresponding measurements in nucleus\hyp{}nucleus
collisions.
Among the many different hadrons produced in high energy collisions
pions are the lightest and by far the most abundant ones.
Thus, data on pion production properties are crucial for
constraining basic properties of models of strong interactions.
In particular, the most significant signals of the onset
of deconfinement (the "kink" and "horn")~\cite{onsetc}
require precise measurements of the mean pion multiplicity
at the same beam momenta per nucleon as the corresponding A+A data.
Moreover, the \NASixtyOne data are taken with the same detector and the same 
acceptance.

\begin{sloppypar}
In the CERN SPS beam momentum range of 10--450\GeVc the mean multiplicity
of negatively charged pions in inelastic p+p interactions
increases from about 0.7 at 10\GeVc to about 3.5 at 450\GeVc~\cite{Sasha}.
Among three charged states of pions
the most reliable measurements in the largest phase-space
are usually possible for \pim mesons.
Neutral pions decay electromagnetically into two photons and thus
measurements of their production properties require
measurements of both photons and then extraction of the $\pi^0$ signal
from the two-photon mass spectra.
Charged pions are easy to detect by ionisation detectors as they 
decay weakly with a long lifetime ($c\tau= 7.8$~m).
A significant fraction of positively charged hadrons are protons 
(25\%) and kaons (5\%)~\cite{na49_pp_pions,na49_pp_protons,na49_pp_kaons}.
Therefore measurements of $\pip$ mesons require their
identification by measurements of the energy loss and/or time-of-flight.
This identification is not as important for
\pim mesons because the contamination of negatively charged particles
by $\km$ mesons and anti-protons is below 
10\%~\cite{na49_pp_pions,na49_pp_protons,na49_pp_kaons} and can be subtracted 
reliably.
The latter method is used in this paper and it  allows to derive 
\pim spectra in the broadest phase-space region in a uniform way.
Results obtained using explicit pion identification are planned in future 
\NASixtyOne publications.
\end{sloppypar}

\begin{sloppypar}
The paper is organised as follows.
In Sec.~\ref{sec:set-up} the \NASixtyOne experimental set-up is described.
Details on the beam, trigger and event selection are given
in Sec.~\ref{sec:beam}.
Data reconstruction, simulation and detector performance are described in
Sec.~\ref{sec:data}.
Analysis techniques and final results are presented
in Secs.~\ref{sec:analysis} and~\ref{sec:results},
respectively. These results are compared with the corresponding data
on central Pb+Pb collisions in Sec.~\ref{sec:PbPb}.
A summary in Sec.~\ref{sec:summary} closes the paper.
\end{sloppypar}

\begin{figure*}
\centering
  \includegraphics[width=0.89\textwidth]{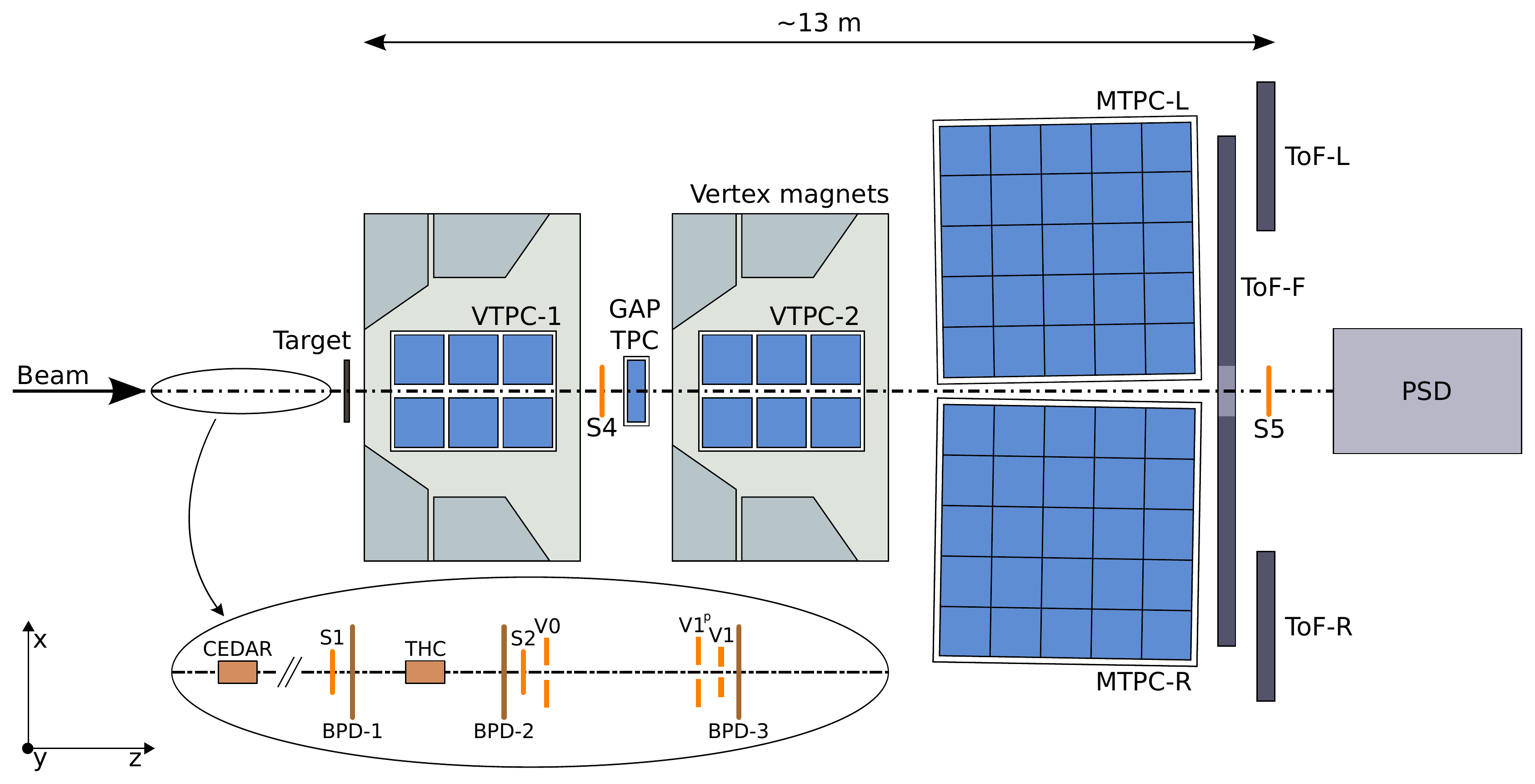}
  \caption{ 
  (Colour online)
  The schematic layout of the \NASixtyOne experiment at the CERN SPS 
  (horizontal cut, not to scale).
  The beam and trigger detector configuration used for data
  taking on p+p interactions in 2009 is shown.
  Alignment of the chosen coordinate system is shown on the plot;
  its origin lies in the middle of VTPC-2, on the beam axis.
  The nominal beam direction is along the $z$ axis.
  The magnetic field bends charged particle trajectories
  in the $x$--$z$ (horizontal) plane.
  The drift direction in the TPCs is along the $y$ (vertical) axis.
  }
\label{fig:detector}
\end{figure*}

The pion rapidity is calculated in the collision centre of
mass system: $y = \atanh(\beta_\mathrm{L})$,
where $\beta_\mathrm{L} = p_\mathrm{L}/E$ is the longitudinal component of the 
velocity,
$p_\mathrm{L}$ and $E$ are pion longitudinal momentum and energy given
in the collision centre of mass system. 
The transverse component of the momentum is denoted as \pt and
the transverse mass \mt is defined as $\mt = \sqrt{m_{\pi}^2 + \pt^2}$,
where $m_{\pi}$ is the charged pion mass.
The nucleon mass and collision energy per nucleon pair in the centre of mass 
system are denoted as $m_\mathrm{N}$ and $\sqrt{s_\mathrm{NN}}$, respectively.

\section{The \NASixtyOne facility}
\label{sec:set-up}

\begin{sloppypar}
The \NASixtyOne experimental facility consists of 
a large acceptance hadron spectrometer located in the 
CERN North Area Hall~887~(EHN1) and the H2 beam-line 
to which beams accelerated in the CERN accelerator
complex are delivered from the Super Proton Synchrotron.
\NASixtyOne profits from the long development of the
CERN proton and ion sources and the accelerator chain as well as
the H2 beam line of the CERN North Area. The latter
has recently been modified to also serve as a fragment separator
as needed to produce the Be beams for \NASixtyOne.
Numerous components of the \NASixtyOne set-up were inherited from
its predecessor, the NA49 experiment~\cite{NA49nim}.
\end{sloppypar}

The schematic layout of the \NASixtyOne detector 
is shown in Fig.~\ref{fig:detector}.

A set of scintillation and Cherenkov counters as well as beam position
detectors (BPDs) upstream of the spectrometer provide timing reference,
identification and position measurements of incoming beam particles.
The trigger scintillator counter S4 placed downstream of the target is used to
select events with collisions in the target area.
Details on this system are presented in
Sec.~\ref{sec:beam}.

The main tracking devices of the spectrometer
are large volume Time Projection Chambers (TPCs).
Two of them, the vertex TPCs (VTPC-1 and VTPC-2 in Fig.~\ref{fig:detector}),
are located in the magnetic fields of two super-conducting
dipole magnets with a maximum combined bending power of 9~Tm which
corresponds to about 1.5~T and 1.1~T fields in the upstream 
and downstream magnets, respectively.
This field configuration was used for data taking on p+p interactions at 
158\GeVc.
In order to optimise the acceptance of the detector at lower
collision momenta, the field in both magnets was lowered 
in proportion to the beam momentum. 

Two large TPCs (MTPC-L and \mbox{MTPC-R}) are positioned 
downstream of the magnets symmetrically to the beam line. 
The fifth small TPC (\mbox{GAP-TPC}) is placed
between \mbox{VTPC-1} and \mbox{VTPC-2} directly on the beam line. 
It closes the gap between the beam axis and the sensitive 
volumes of the other TPCs. 

The TPCs are filled with Ar:CO$_2$ gas mixtures in proportions 90:10 
for the VTPCs and the GAP-TPC, and 95:5 for the MTPCs.

\begin{sloppypar}
The particle identification capability of the TPCs
based on measurements of the specific energy loss, \dedx,
is augmented by time-of-flight (\tof) measurements using
Time-of-Flight (ToF) detectors.
The high resolution forward calorimeter,
the Projectile Spectator Detector (PSD),
measures energy flow around the beam
direction, which in nucleus-nucleus collisions is primarily given by
the projectile spectators.
\end{sloppypar}

\NASixtyOne uses various solid nuclear targets and a liquid
hydrogen target (see Sec.~\ref{sec:beam} for details).
The targets are positioned about 80~cm  upstream of the sensitive volume 
of VTPC-1.

The results presented in this paper were obtained using
information from the Time Projection Chambers, the Beam Position Detectors
as well as from the beam and trigger counters.

\section{Beams, target, triggers and data samples}
\label{sec:beam}

This section summarises basic information on the beams, 
target, triggers and recorded data samples which is
relevant for the results presented in this paper. 

Secondary beams of positively charged hadrons
at 20, 31, 40, 80 and 158\GeVc are produced from 400\GeV
protons extracted from the SPS in a slow extraction mode
with a flat-top of 10 seconds.
The secondary beam momentum and intensity is adjusted
by proper setting of the H2 beam-line magnet currents and
collimators.
The beam is transported along the H2 beam-line 
towards the experiment.
The precision of the setting of the beam magnet currents
is approximately 0.5\%. This was verified by a direct
measurement of the  beam momentum at 31\GeVc
by bending the incoming beam particles into the TPCs 
with the maximum magnetic field~\cite{pionpC}.
The selected beam properties are given in Table~\ref{tab:beams}. 

The set-up of beam detectors is illustrated in the inset
on Fig.~\ref{fig:detector}.
Protons from the secondary hadron beam are identified by two
Cherenkov counters, a CEDAR~\cite{CEDAR} (either CEDAR-W or CEDAR-N) 
and a threshold counter (THC).
The CEDAR counter, using a 
coincidence of six out of the eight photo-multipliers placed
radially along the Cherenkov ring,
provides positive identification of protons, while the
THC, operated at pressure lower than the proton
threshold, is used in anti-coincidence in the trigger logic.
Due to their limited range of operation two different CEDAR counters were used,
namely for beams at  20, 31, and 40\GeVc the CEDAR-W counter and
for beams at 80 and 158\GeVc the CEDAR-N counter. The threshold counter
was used for all beam energies.
A selection based on signals from the Cherenkov counters allowed
to identify beam protons with a purity of about 99\%.
A consistent value for the purity was found by bending the beam into
the TPCs with the full magnetic field and using the \dedx identification
method~\cite{Claudia}.
The fraction of protons in the beams is given in Table~\ref{tab:beams}.

\begin{table}
  \caption{
Basic beam properties and numbers of events recorded for
p+p interactions at  20, 31, 40, 80 and 158\GeVc.
The first column gives the beam momentum. The second and third 
columns list typical numbers of beam particles
at \NASixtyOne per spill (about 10~seconds) and 
the fraction of protons in the beam,
respectively. 
}
\label{tab:beams}
\centering
  \begin{tabular}{c | c | c }
  \hline\hline
    $p_\mathrm{beam}$ [\GGeVc] & particles per spill & proton fraction \\
    \hline 
    20  & 1000k &   12\%      \\
    31  & 1000k  &   14\%      \\
    40  & 1200k &   14\%      \\
    80  & 460k  &   28\%      \\
    158 & 250k  &   58\%      \\\hline\hline
  \end{tabular}
\end{table}

\begin{figure*}
\centering
\includegraphics[width=0.79\linewidth]{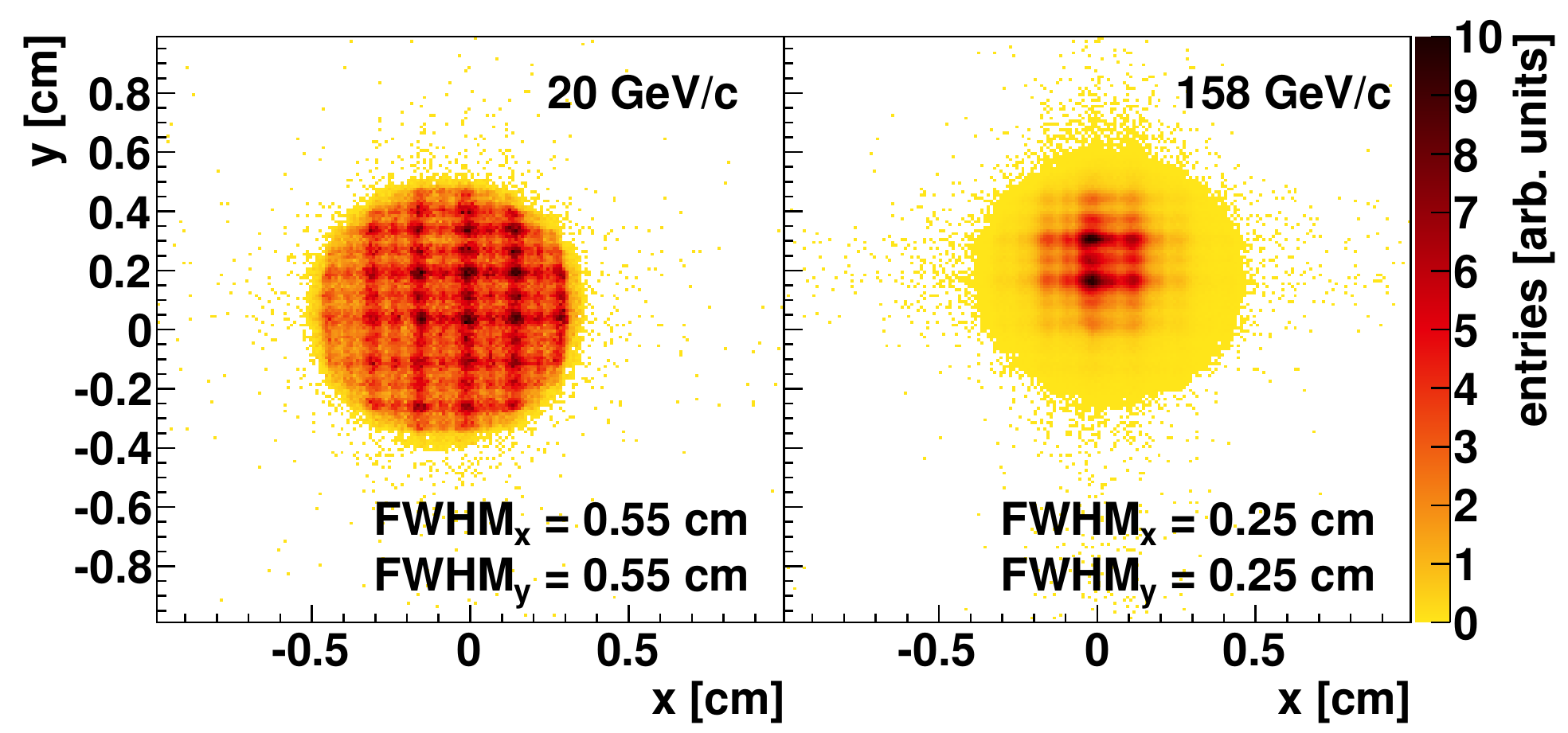}
\includegraphics[width=0.79\linewidth]{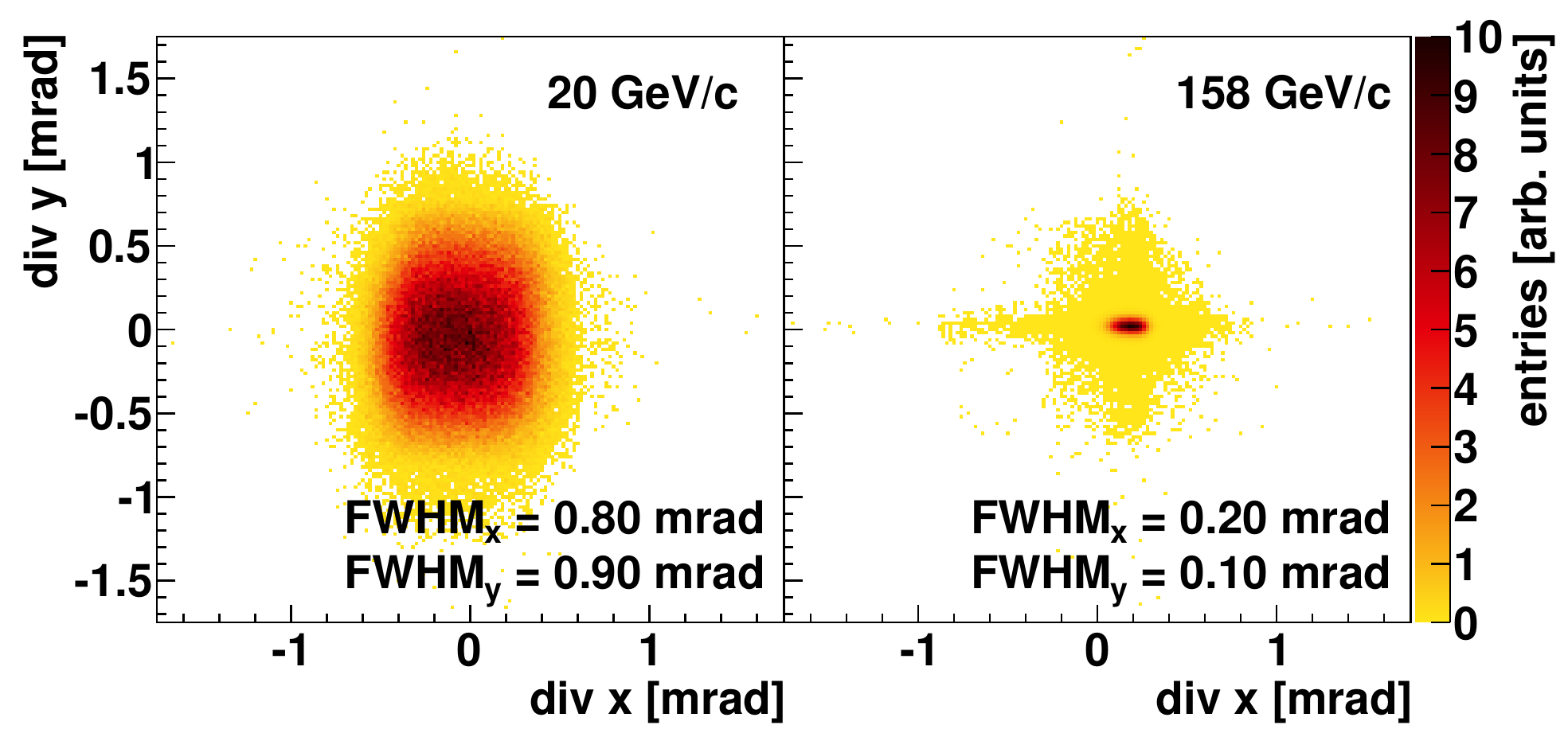}
\caption{
(Colour online)
{\it Top:}
The beam spot as measured by BPD-3 after the $\overline{\textrm{V1}}$ cut
described in the text for 20\GeVc (\emph{left}) and 158\GeVc (\emph{right}) 
beams.
{\it Bottom:}
The beam divergence in $x$ and $y$ for 20\GeVc (\emph{left}) 
and 158\GeVc (\emph{right}) beams.
All distributions were arbitrarily scaled to the full colour scale.
Widths of the distributions are given in the legend.
} 
\label{fig:profile}
\end{figure*}

Two scintillation counters, S1 and S2, provide
beam definition, together with the three veto
counters V0,  V1 and V1$^\mathrm{p}$ with a 1~cm diameter hole,
which are defining the beam before the target.
The S1 counter provides also the timing (start time for all counters).
Beam protons are then selected by the coincidence
$\textrm{S1}
\wedge\textrm{S2}
\wedge\overline{\textrm{V0}}
\wedge\overline{\textrm{V1}}
\wedge\overline{\textrm{V1}^\mathrm{p}}
\wedge\textrm{CEDAR}
\wedge\overline{\textrm{THC}}$.
Trajectories of individual beam particles were measured
in a telescope of beam position detectors placed along the beam line
(BPD-1/2/3 in Fig.~\ref{fig:detector}).
These counters are small ($4.8\times4.8$~cm$^2$)
proportional chambers with cathode strip readout,
providing a resolution of about 100~$\mu$m in two orthogonal
directions, see Ref.~\cite{na61nim} for more details.
The beam profile and divergence obtained from the
BPD measurements are presented in Fig.~\ref{fig:profile}.
Due to properties of the H2 beam line both the beam width and 
divergence at the \NASixtyOne target increase with decreasing beam momentum.

For data taking on p+p interactions 
a liquid hydrogen target  of 20.29~cm length 
(2.8\%~interaction length) and 3~cm diameter
placed 88.4~cm upstream of VTPC-1 was used.
The Liquid Hydrogen Target facility (LHT) 
filled the target cell with para-hydrogen obtained 
in a closed-loop liquefaction system
which was operated at 75~mbar overpressure with respect to the atmosphere. 
At the atmospheric pressure of 965~mbar the liquid hydrogen density
is $\rho_\mathrm{LH} = 0.07$~g/cm$^3$.
The boiling rate in the liquid hydrogen was not monitored during the data taking
and thus the liquid hydrogen density is known only approximately.
It has however no impact on the results presented in this paper as
they are determined from particle yields per selected event and thus
they are independent of the target density.
Data taking with inserted and removed liquid hydrogen in the LHT 
was alternated in order to calculate a data-based correction for 
interactions with the material surrounding the liquid hydrogen.

Interactions in the target are selected by the trigger system 
by an anti-coincidence of the incoming
beam protons with
a small, 2~cm diameter, scintillation counter (S4) placed on the beam
trajectory
between the two vertex magnets (see Fig.~\ref{fig:detector}). This
minimum bias trigger is based on the disappearance of the incident
proton. 
In addition, unbiased proton beam events were recorded with a
frequency typically by a factor of 10 lower than the frequency of 
interaction events.

\section{Data processing, simulation and detector performance}
\label{sec:data}

\begin{figure*}
\centering
  \includegraphics[width=0.79\textwidth]{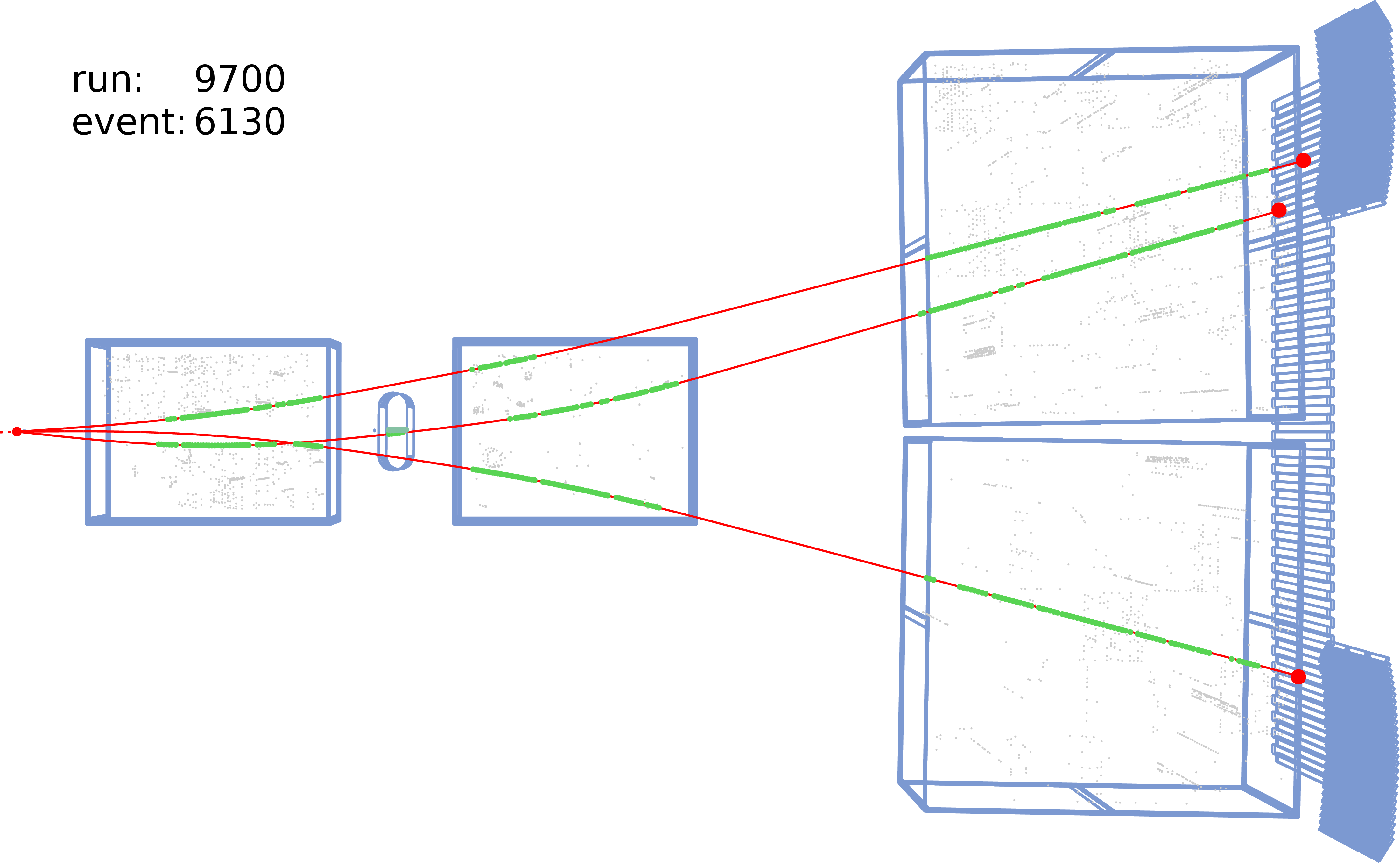}
  \caption{
  (Colour online)
  An example of a p+p interaction at 40\GeVc measured in the \NASixtyOne 
detector. The measured points (green) are used to fit tracks (red lines) to the 
interaction point.
The grey dots show the noise clusters.
Due to the central gap of the VTPCs only a small part of the trajectory of 
the negatively charged particle is seen in VTPC-1.
}
\label{fig:real_data_event}
\end{figure*}

Detector parameters were optimised by a data-based calibration 
procedure which also took into account their time dependences. 
Small adjustments were determined in consecutive steps for:
\begin{enumerate}[(i)]
  \item detector geometry, TPC drift velocities and distortions due to the 
        magnetic field inhomogeneities in the corners of VTPCs,
  \item magnetic field setting,
  \item specific energy loss measurements,
  \item time-of-flight measurements.
\end{enumerate}
Each step involved reconstruction of the data required to optimise a given
set of calibration constants and time dependent corrections 
followed by verification procedures.
Details of the procedure and quality assessment are presented in
Ref.~\cite{Status_Report_2008}.
The resulting performance in the measurements of  quantities relevant for this 
paper is discussed below.

The main steps of the data reconstruction procedure are:
\begin{enumerate}[(i)]
  \item cluster finding in the TPC raw data, calculation of the cluster
  centre-of-gravity and total charge,
  \item reconstruction of local track segments in each TPC separately,
  \item matching of track segments into global tracks,
  \item track fitting through the magnetic field and determination of track
  parameters at the first measured TPC cluster,
  \item determination of the interaction vertex using the beam trajectory ($x$ 
  and $y$ coordinates) fitted in the BPDs and the trajectories of tracks
  reconstructed in the TPCs  ($z$ coordinate),
  \item
  \begin{sloppypar}
  refitting the particle trajectory using the interaction vertex as an
  additional point and determining the particle momentum at the interaction
  vertex,
  \end{sloppypar}
  \item matching of ToF hits with the TPC tracks.
\end{enumerate}
An example of a reconstructed p+p interaction at 40\GeVc 
is shown in Fig.~\ref{fig:real_data_event}.
Long tracks of one negatively charged and 
two positively charged particles are seen.
All particles leave signals in the ToF detectors. 

\begin{sloppypar}
A simulation of the \NASixtyOne detector response is used to correct the 
reconstructed data.
Several MC models were compared with the \NASixtyOne results on p+p, p+C and 
$\pi$+C interactions: FLUKA2008, URQMD1.3.1, VE\-NUS4.12, EPOS1.99, 
GHEISHA2002, QGSJetII-3 and 
Sibyll2.1~\cite{pionpC,M.UngerfortheNA61/SHINE:2013aha,Unger:2010ze,Agnieszka}. 
Based on these comparisons  
and taking into account continuous support and documentation from the 
developers the EPOS model~\cite{EPOS} was selected for the MC simulation.
The simulation consists of the following steps (see Ref.~\cite{Nicolas} for 
more details):
\end{sloppypar}
\begin{enumerate}[(i)]
\begin{sloppypar}
  \item generation of inelastic p+p interactions using the EPOS 
  model~\cite{EPOS},
\end{sloppypar}
  \item propagation of outgoing particles through the detector material
  using the GEANT 3.21 package~\cite{Geant3} which takes into account the
  magnetic field as well as relevant physics processes, such as particle
  interactions and decays,
  \item
  \begin{sloppypar}
  simulation of the detector response using dedicated \NASixtyOne packages
  which introduce distortions corresponding to all corrections applied to
  the real data,
  \item simulation of the interaction trigger selection by checking whether
  a charged particle hits the S4 counter, see Sec.~\ref{sec:beam},
  \end{sloppypar}
  \item storage of the simulated events in a file which has the same format
  as the raw data,
  \item reconstruction of the simulated events with the same reconstruction
  chain as used for the real data and
  \item matching of the reconstructed tracks to the simulated ones 
  based on the cluster positions.
\end{enumerate}

It should be underlined that only inelastic p+p interactions in
the hydrogen in the target cell were simulated and reconstructed.
Thus the Monte Carlo based corrections (see Sec.~\ref{sec:analysis}) 
can be applied only for inelastic events.  
The contribution of elastic events is removed by the event selection cuts
(see Sec.~\ref{sec:event_cuts}), whereas the contribution of off-target
interactions is subtracted based on the data (see Sec.~\ref{sec:subtraction}).

\begin{figure*}
  \includegraphics[width=0.49\textwidth]{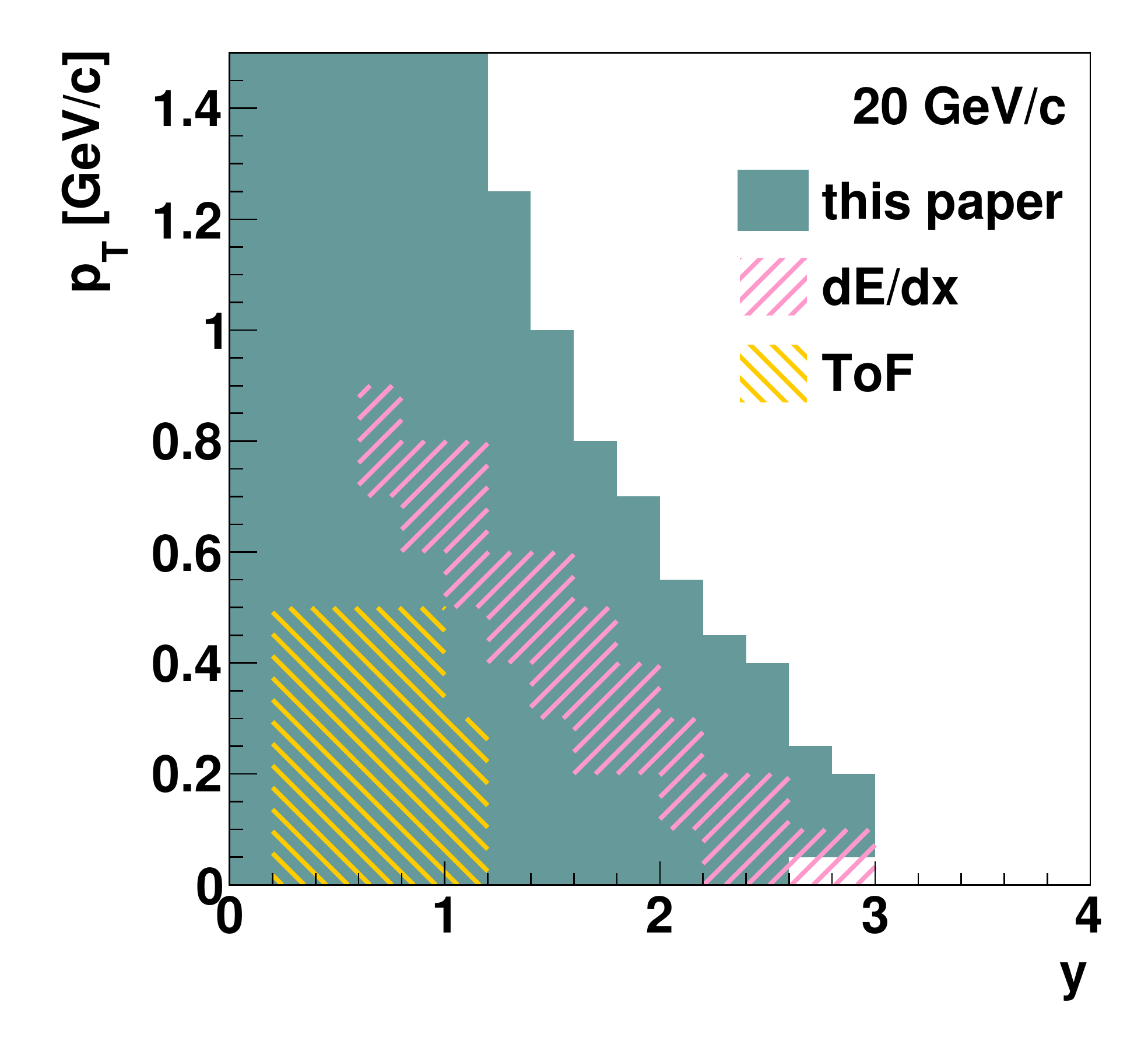}
  \includegraphics[width=0.49\textwidth]{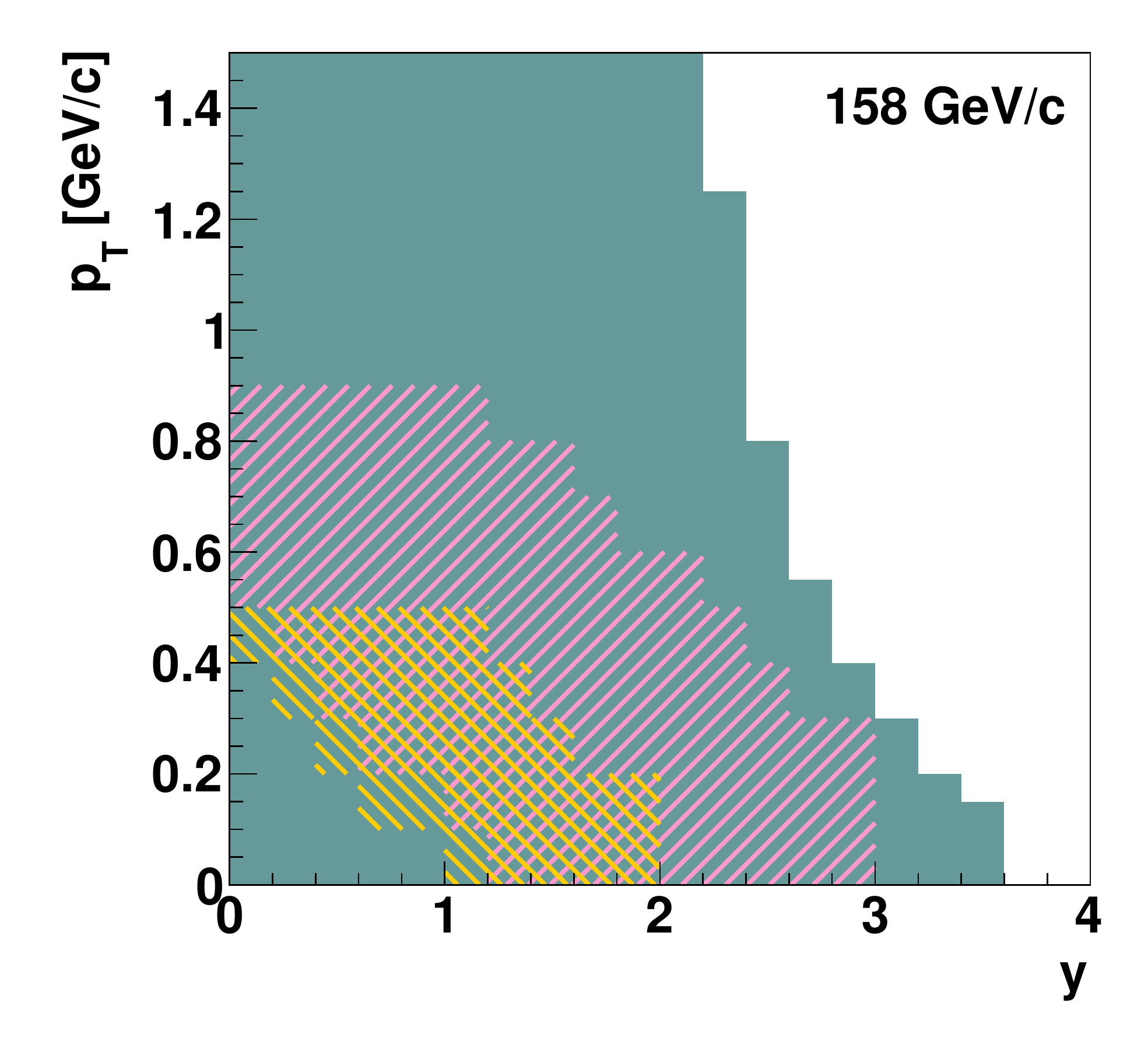}
  \caption{
    Typical acceptance regions for \pim meson spectra in p+p interactions
    at 20\GeVc (\emph{left}) and 158\GeVc (\emph{right})
    for different analysis methods: the method used in this paper
    which does not require an explicit pion identification,
    the method which identifies pions via their energy loss  (\dedx)
    and, in addition, their time-of-flight (\tof). 
  }
  \label{fig:hminus_acceptance}
\end{figure*}

Spectra of \pim mesons presented in this paper were derived
from spectra of all negatively charged hadrons corrected for
a small contamination of mostly \km mesons and anti-protons.
The typical acceptance in rapidity and transverse momentum
is presented in Fig.~\ref{fig:hminus_acceptance} for p+p
interactions at 20 and 158\GeVc.
This figure also shows acceptance regions for 
methods based on explicit pion identification using
\dedx and \tof measurements. They are limited due to
the geometrical acceptance of the ToF detectors, the finite
resolution of the \dedx measurements and limited
data statistics.

\begin{figure}
\centering
\includegraphics[width=0.4\textwidth]{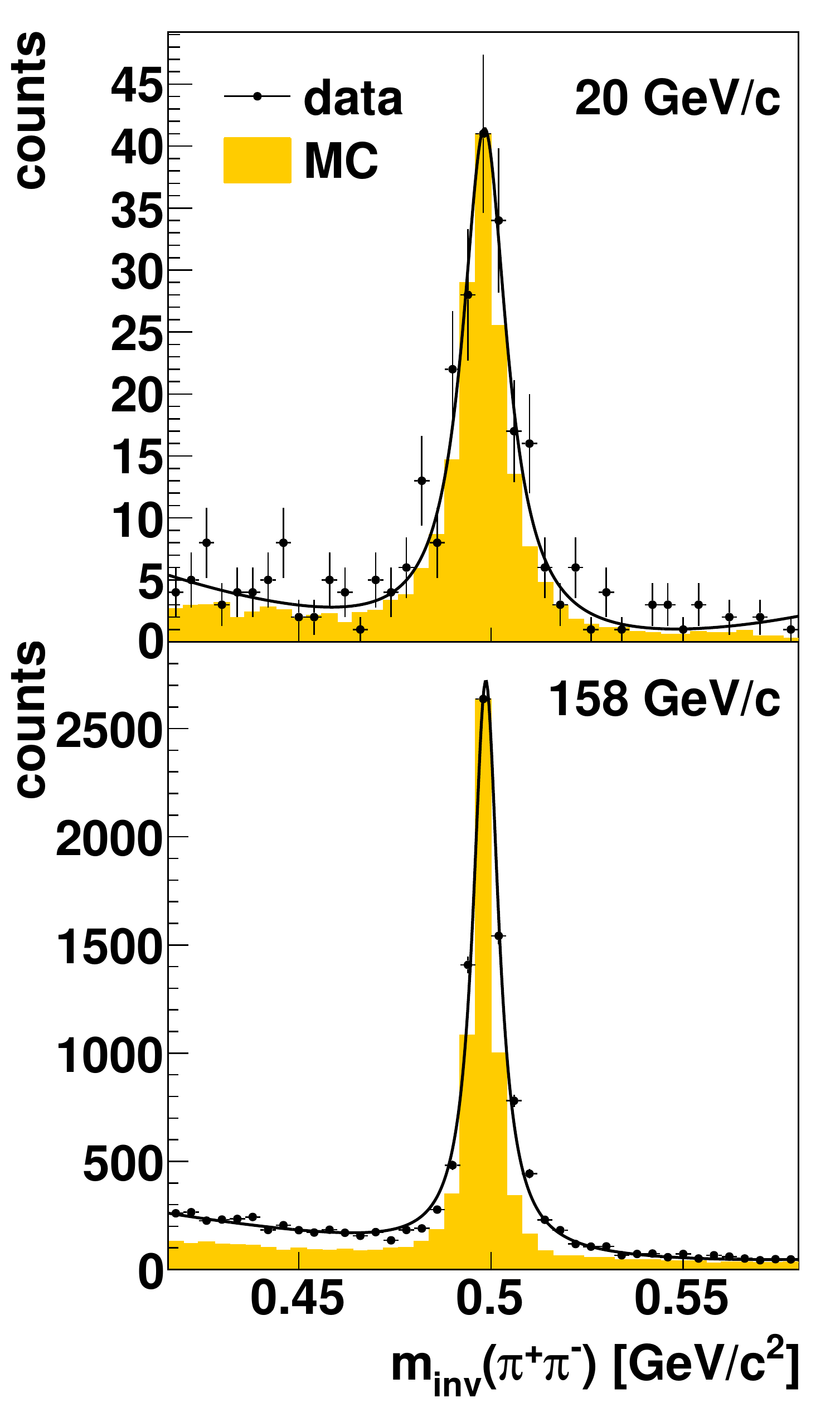}
\caption{
(Colour online)
Invariant mass distribution of reconstructed \kzeros candidates in p+p 
interactions at 20 (\emph{top}) and 158\GeVc (\emph{bottom}) for the 
measured data and EPOS model based Monte Carlo simulations. 
The MC plot was normalised to the peak height of the data.
The \kzeros candidates were selected within $0<y<-1$ and 
$0<\pt<0.5$\GeVc for 20\GeVc and $-1<y<0$ and $0<\pt<0.5$\GeVc for 158\GeVc.
The distribution was fitted with the sum of a Lorentzian function 
(signal) and a second order polynomial (background).
}
\label{fig:V0_study}
\end{figure}

\begin{sloppypar}
The quality of measurements was studied by reconstructing 
masses of \kzeros
particles from their V$^0$ decay topology.
As an example the invariant mass distributions of \kzeros 
candidates found in p+p interactions at 20 and 158\GeVc are 
plotted in Fig.~\ref{fig:V0_study}.
The differences between the measured peak positions 
and the literature value of the \kzeros mass~\cite{PDG:2012}
are smaller than 1\MeVcc. 
The width of the distributions, related to the detector resolution, 
is about 25\% smaller for the Monte Carlo than for the data.
This implies that statistical and/or systematic uncertainties 
of track parameters reconstructed from the data
are somewhat underestimated in the simulation.
Systematic bias due to this imperfectness was estimated by varying 
the selection cuts and was found to be below 2\% (see Sec.~\ref{sec:Syst}).
\end{sloppypar}

\begin{figure*}
\centering
\includegraphics[width=0.79\linewidth]{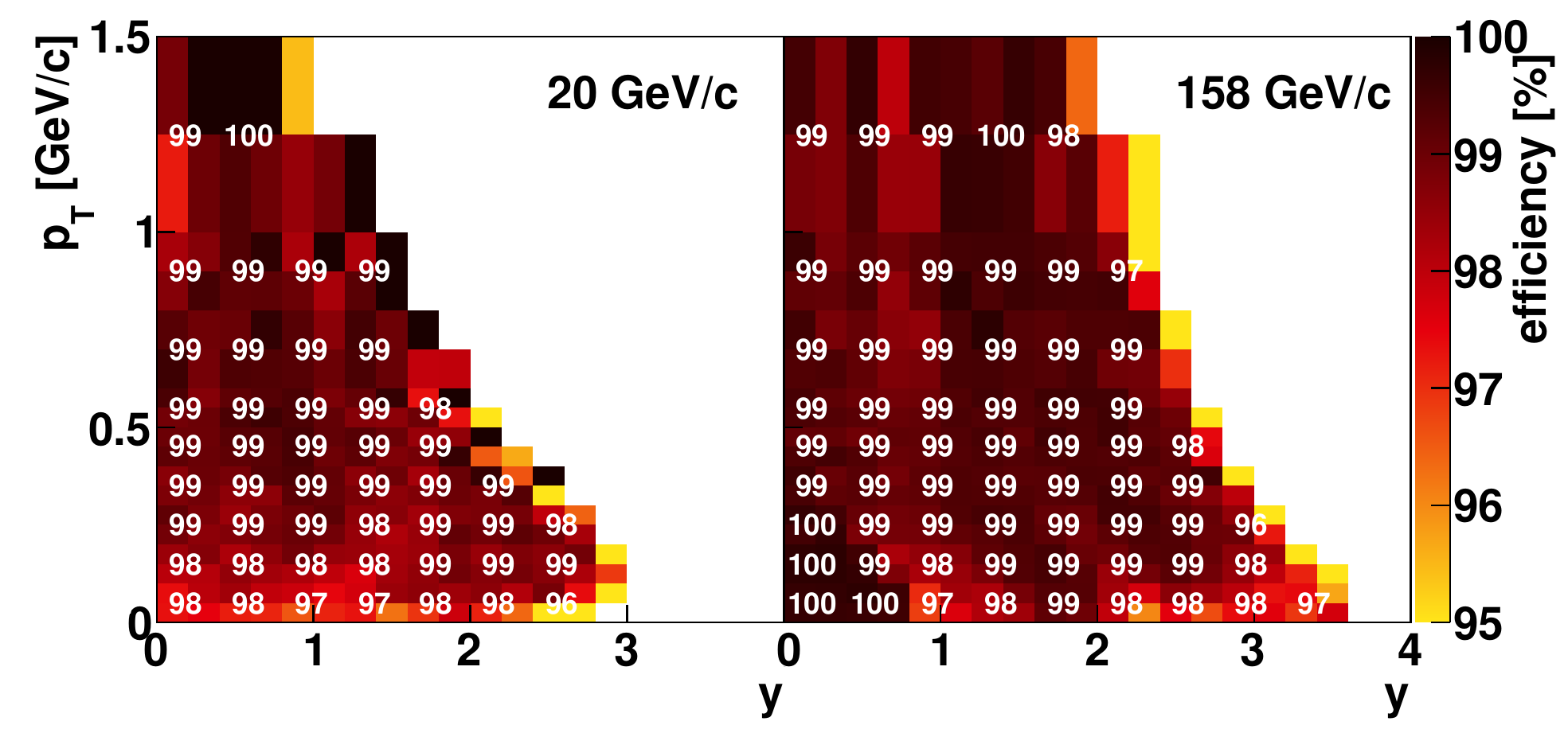}
\caption{
(Colour online)
Reconstruction efficiency of negatively charged pions produced
in p+p interactions at 20 (\emph{left}) and 158\GeVc (\emph{right}) as a 
function of  rapidity and transverse momentum.
It was calculated by dividing the number of tracks passing
the track selection cuts specified in Sec.~\ref{sec:track_cuts}
by the number of the generated tracks. The selection criteria
include the requirement of at least 90\% reconstruction efficiency.
} 
\label{fig:track_eff}
\end{figure*}

\begin{figure*}
\centering
\includegraphics[width=0.79\linewidth]{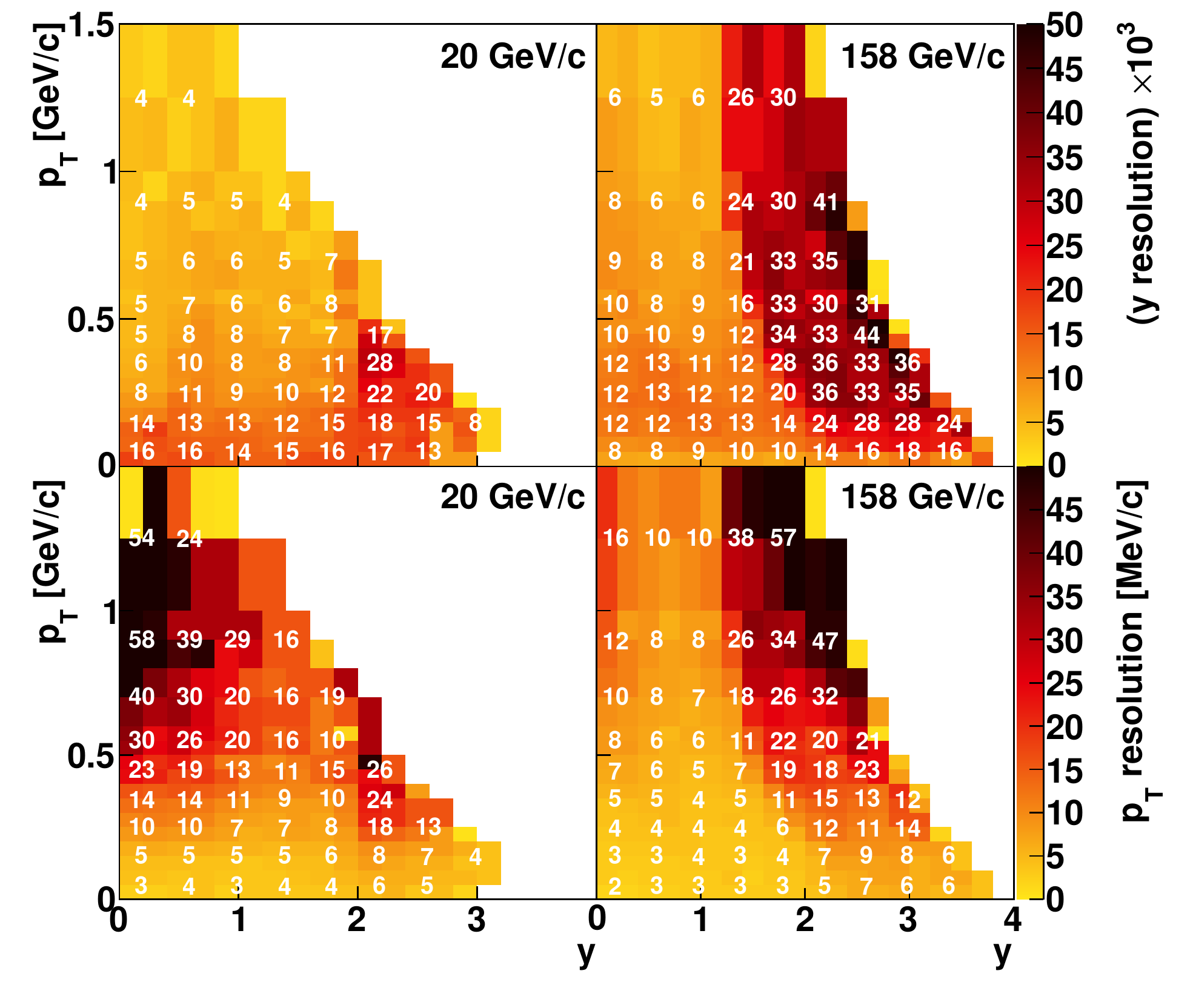}
\caption{
(Colour online)
Resolution of rapidity (\emph{top}, scaled by $10^3$) and transverse 
momentum (\emph{bottom}) measurements for negatively charged pions produced in 
p+p interactions at 20 
(\emph{left}) and 158\GeVc (\emph{right}) as a function of pion rapidity 
and transverse momentum.
The results are obtained using the track selection cuts specified in
Sec.~\ref{sec:track_cuts}.
} 
\label{fig:resolution}
\end{figure*}

The track reconstruction efficiency and the resolution of kinematic quantities
were calculated by matching reconstructed tracks to their generated partners.
In only 0.1--0.2\% of cases a single generated track is matched to more than 
one reconstructed partner, typically due to failure of matching reconstructed 
track segments. This effect is taken into account in the correction described 
in Sec.~\ref{sec:event_track_losses}.
As examples, the reconstruction efficiency as a function of rapidity and
transverse momentum for
negatively charged pions produced in p+p interactions at 20 and 158\GeVc is
shown in Fig.~\ref{fig:track_eff}.
The resolution of  rapidity and transverse momentum measurements
is illustrated in Fig.~\ref{fig:resolution}.
The resolution was calculated as the FWHM
of the distribution of the difference between the 
generated and reconstructed $y$ and \pt.
These results were obtained for negatively charged pions passing the
track selection criteria described in Sec.~\ref{sec:track_cuts}.
Resolution of the transverse momentum is worse at low beam momenta.
This is caused by the lower magnetic field and by the fact that the same 
rapidity region 
in the centre of mass frame corresponds to lower momenta in the laboratory 
frame.

\begin{figure}
\centering
  \includegraphics[width=0.49\textwidth]{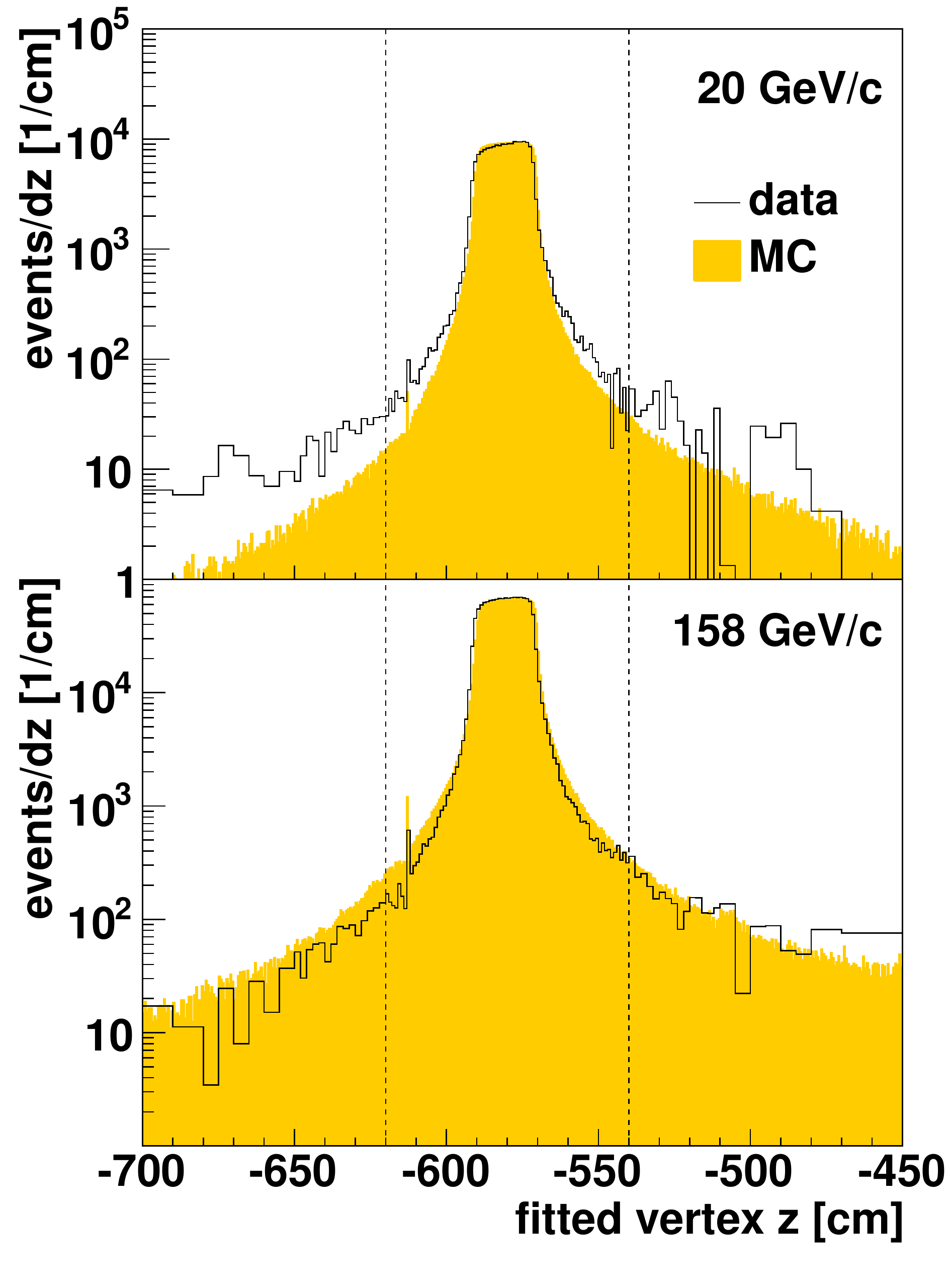}
  \caption{
  (Colour online)
  Distribution of fitted vertex $z$ coordinate for p+p interactions at   
  20 (\emph{top}) and 158\GeVc (\emph{bottom}). The black line shows the data 
  after target removed subtraction 
  (see Sec.~\ref{sec:subtraction}). The filled area shows the distribution for
  the reconstructed  Monte Carlo simulation. This distribution
   was normalised to the total integral of 
  the data plot. The dashed vertical lines show the $z$ vertex selection 
  range.
  }\label{fig:vertex_z_MC}
\end{figure}

\begin{figure*}
\centering
  \includegraphics[width=0.79\linewidth]{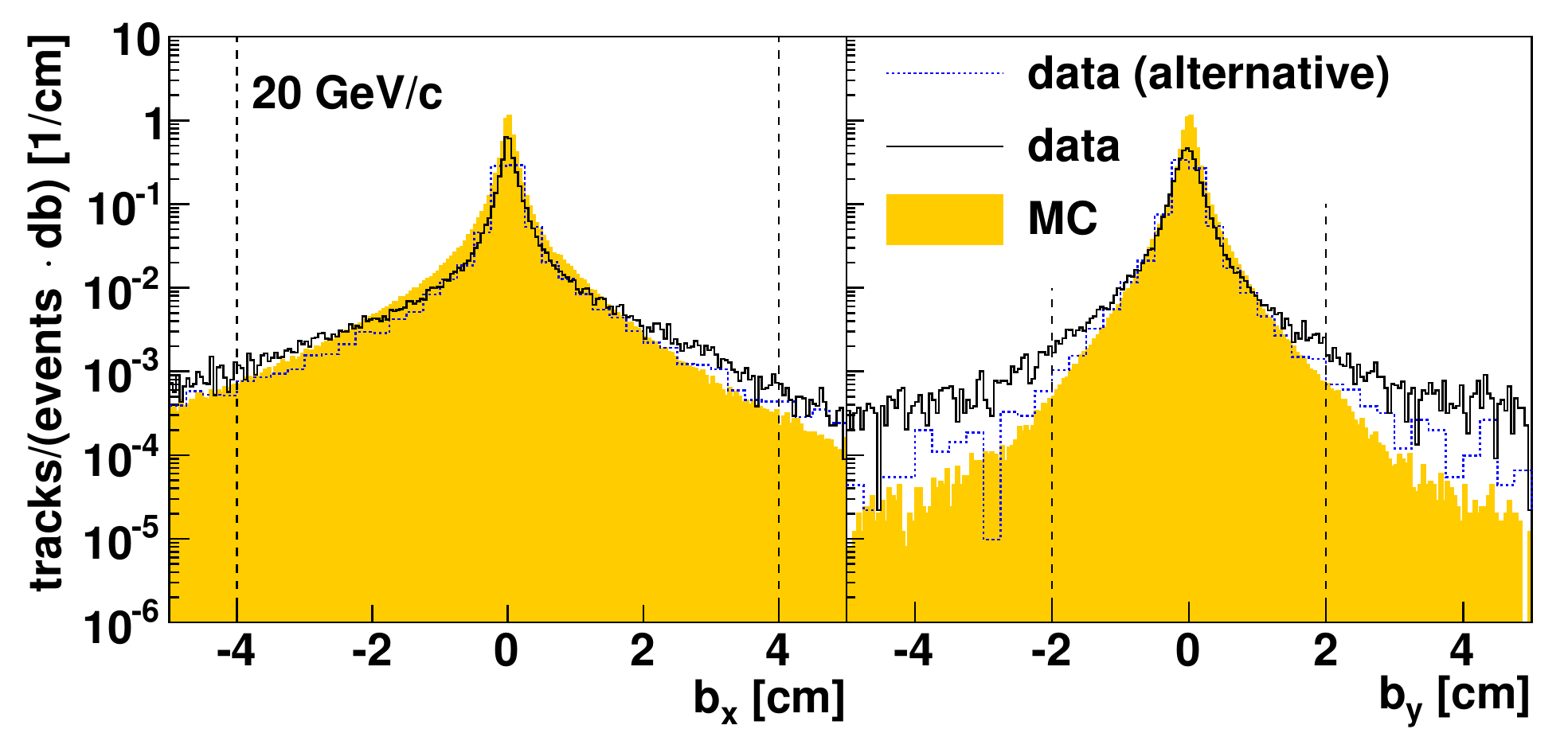}
  \includegraphics[width=0.79\linewidth]{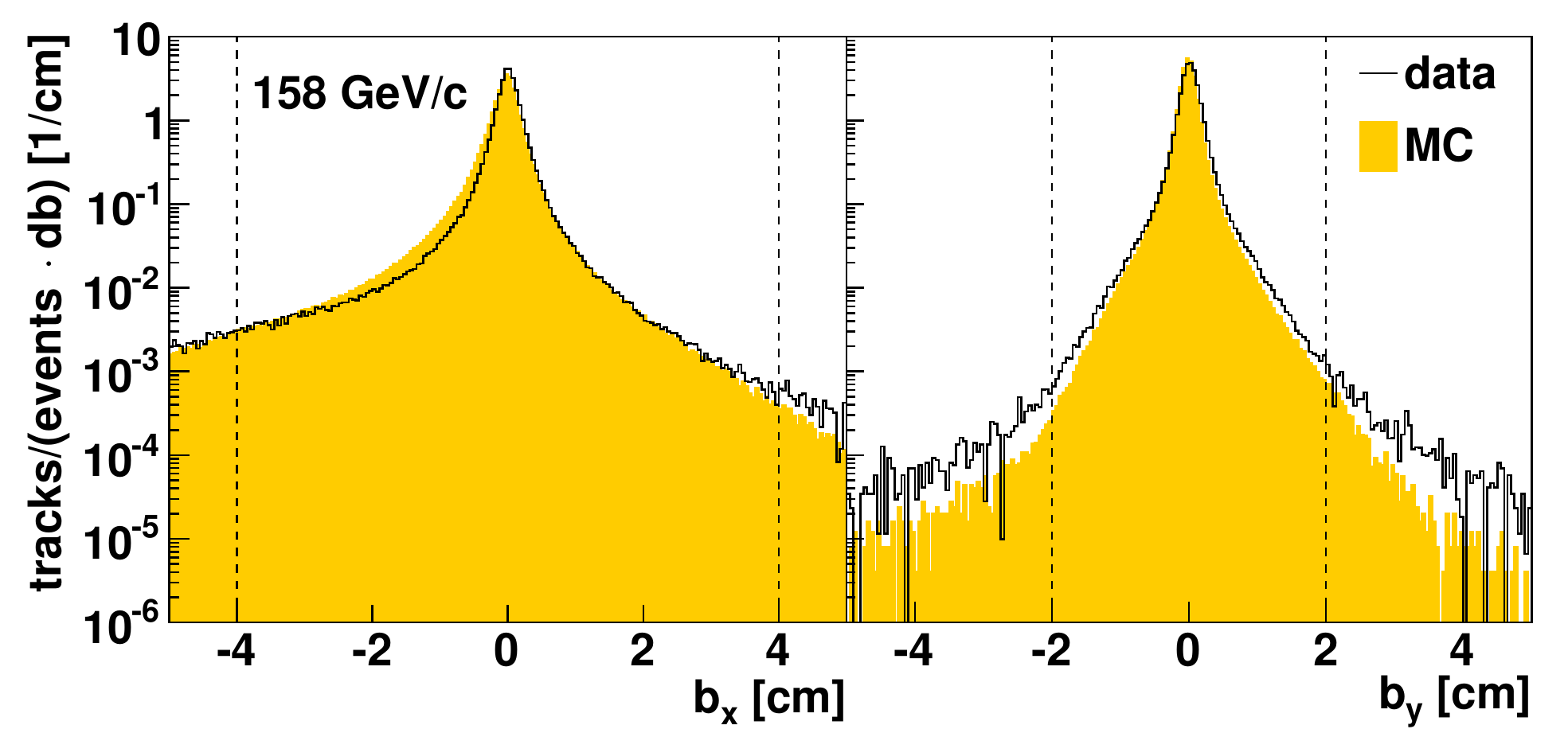}
  \caption{
  (Colour online)
  Distribution of the impact parameter in the $x$ (\emph{left}) and $y$ 
  (\emph{right}) coordinate for p+p interactions at 20 (\emph{top}) and 
  158\GeVc (\emph{bottom}). The black line 
  shows the data after target removed subtraction (see 
  Sec.~\ref{sec:subtraction}). The filled area shows the reconstructed 
  Monte Carlo simulation. The dashed vertical lines show the accepted range 
  (see Sec.~\ref{sec:track_cuts}).
  The dotted blue line in the 20\GeVc plots show the distribution obtained
  using the alternative event selection. Namely only events with no off-time
  beam particles within the time window of  $\pm 6\ \mu$s around the trigger
  particle time were accepted.
  }\label{fig:impact}
\end{figure*}

Figures~\ref{fig:vertex_z_MC} and~\ref{fig:impact} show further
examples of the comparison between data and simulation.
Distributions of the $z$ coordinate of 
the fitted vertex are presented in Fig.~\ref{fig:vertex_z_MC}. 
Distributions of the distance between the track trajectory
extrapolated to the $z$ coordinate of the vertex and
the vertex in the $x$--$y$ plane ($b_x$ and $b_y$ impact parameters) 
are given in Fig.~\ref{fig:impact}.  
Differences visible in the tails of distributions are partially due to  
imperfect simulation of the detector response
and, in case of the impact parameter, partially due to the contribution of 
background tracks from off-time beam particles which are not included in the 
simulation.
The difference is smaller for events selected using more
restrictive cuts on the off-time beam particles.  
A possible small bias due to these effects was estimated by varying the
impact parameter cuts and was found to be below 1\%.

\section{Analysis technique}
\label{sec:analysis}

This section presents the procedures used for data analysis
consisting of the following steps:
\begin{enumerate}[(i)]
  \item applying event and track selection criteria,
  \item determination of spectra of negatively charged hadrons
        using the selected events and tracks,
  \item evaluation of corrections to the spectra based on
        experimental data and simulations,
  \item calculation of the corrected spectra.
\end{enumerate}  

Corrections for the following biases were evaluated and applied:
\begin{enumerate}[(i)]
 \item geometrical acceptance,
 \item contribution of off-target interactions,
 \item contribution of particles other than
       negatively charged pions produced  in inelastic p+p interactions,
 \item losses of inelastic p+p interactions as well
       as of negatively charged pions produced in accepted interactions
       due to the trigger and the
       event and track selection criteria employed in the analysis.
\end{enumerate}

These steps are described  in the successive subsections.

The final results refer to \pim mesons 
produced in inelastic p+p interactions 
by strong interaction processes and in  electromagnetic
decays of produced hadrons. 
Such pions are referred to as \emph{primary} \pim.
The term \emph{primary} will be used in the above meaning also for
other particles.

The analysis was performed independently in ($y$, \pt) and ($y$, \mt) bins.
The bin sizes were selected taking into account the statistical uncertainties 
as well as the resolution of the momentum reconstruction.
Corrections as well as statistical and systematic uncertainties
were calculated for each bin.

\subsection{Event selection criteria}
\label{sec:event_cuts}
This section presents the event selection criteria.
The number of events selected by the
trigger (see Sec.~\ref{sec:beam}) and used in the analysis is shown in 
Table~\ref{tab:statistics}.
The fraction of events selected for the analysis increases with the interaction 
energy, mostly due to lower beam intensity (see Table~\ref{tab:beams}) and 
resulting smaller off-time particle contamination, and smaller fraction of the 
low multiplicity events for which no tracks are found within the acceptance.

\begin{table}
  \caption{
  Number of events recorded with the interaction trigger (all) and
selected for the analysis (selected).
}
\label{tab:statistics}
\centering
  \begin{tabular}{r | r r | r r}
  \hline\hline
$p_\mathrm{beam}$ & \multicolumn{2}{c|}{target inserted}  & 
\multicolumn{2}{c}{target removed}\\
{[\GGeVc]} & \multicolumn{1}{c}{all} & \multicolumn{1}{c|}{selected} & 
\multicolumn{1}{c}{all} & \multicolumn{1}{c}{selected}\\
    \hline 
20 & 1\,324\,k & 233\,k & 123\,k & 4\,k\\
31 & 3\,145\,k & 843\,k & 332\,k & 15\,k\\
40 & 5\,239\,k & 1\,578\,k & 529\,k & 44\,k\\
80 & 4\,038\,k & 1\,543\,k & 429\,k & 54\,k\\
158 & 3\,502\,k & 1\,650\,k & 427\,k & 51\,k\\
\hline\hline
  \end{tabular}
\end{table}

The following event selection criteria were applied to the events 
recorded with the interaction trigger:
\begin{enumerate}[(i)]
  \item no off-time beam particle is detected within $\pm 2\ \mu$s around the 
        trigger particle,
  \item the beam particle trajectory is
        measured in at least one of BPD-1 or BPD-2 and in
        the BPD-3 detector
        positioned just in front of the LHT,
  \item there is at least one track reconstructed in the TPCs and 
        fitted to the interaction vertex,
  \item the vertex $z$ position (fitted using the beam and TPC tracks)
        is not farther away than 40~cm from the centre of the LHT,
  \item events with a single, well measured positively charged track with 
        absolute momentum close to the beam momentum are rejected. The momentum 
        thresholds are listed in Table~\ref{tab:momentum_cut}.
\end{enumerate}

\begin{figure*}
  \centering
  \includegraphics[width=0.79\textwidth]{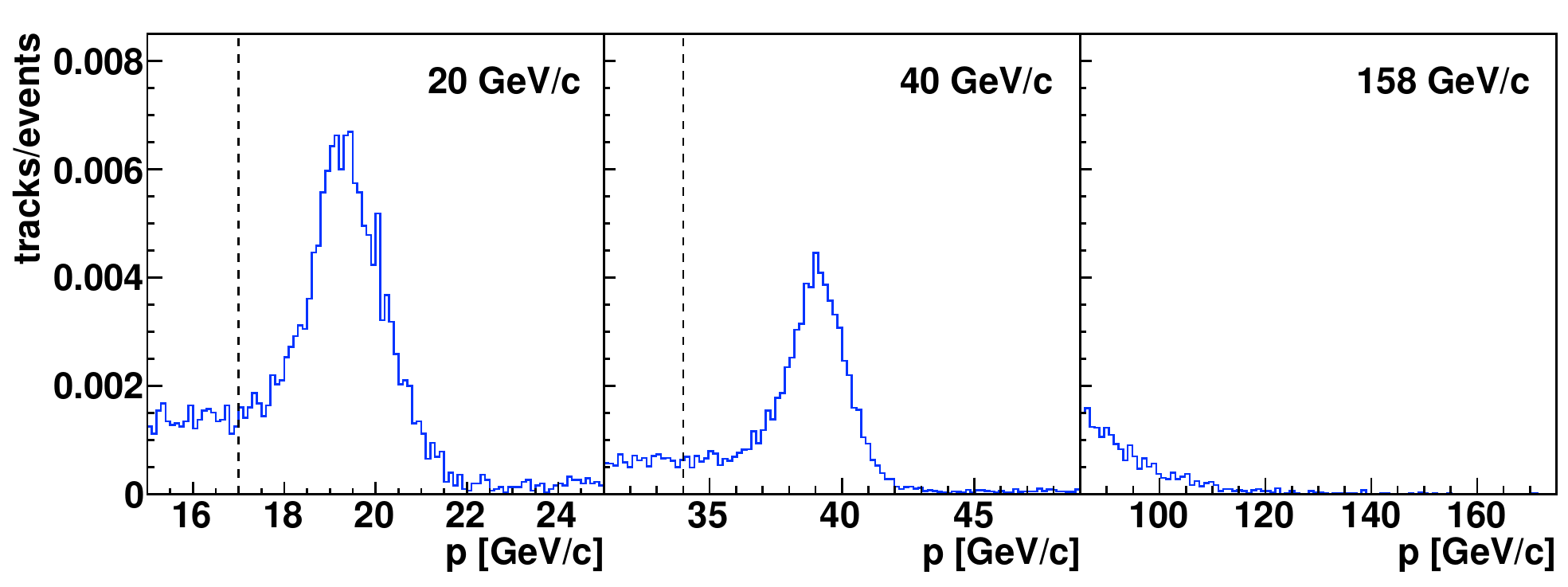}
  \caption{
  Momentum distributions at 20 (\emph{left}), 40 (\emph{middle}) and 
158\GeVc (\emph{right}) of the positively charged tracks in events
passing selection cuts (i)--(iv),
containing a single track, which is positively charged and measured in the 
GAP TPC and MTPC. The distributions were normalised to all events.
The vertical dashed lines at 20 and 40\GeVc show the momentum threshold used to 
remove elastic events (cut (v)).
}\label{fig:momentum_cut}
\end{figure*}

\begin{table}
\caption{Momentum thresholds used to reject elastic interactions 
(cut (v)).}
\label{tab:momentum_cut}
\centering
 \begin{tabular}{r c c c c c}
 \hline\hline
  Beam momentum [\GGeVc]      & 20 & 31 & 40 & 80 & 158\\
  Threshold momentum [\GGeVc] & 17 & 28 & 35 & 74 & --\\
  \hline\hline
 \end{tabular}
\end{table}

The off-line (listed above) and on-line (the interaction trigger condition, 
see Sec.~\ref{sec:beam}) event  cuts 
select a large fraction of well measured (cuts (i) and (ii))
inelastic (cut (iii)) p+p interactions.
The cut (iii) removes part of elastic interactions.
However in some elastic events at beam momenta up to 80\GeVc the beam particle 
is deflected enough to be measured in the detector.
This is demonstrated in the momentum distributions shown in 
Fig.~\ref{fig:momentum_cut}.
Such events are removed by cut (v).

Moreover
cut (iv)
significantly suppresses interactions outside the hydrogen in the 
target cell. 
The corrections for the contribution of interactions outside the hydrogen in 
the target cell and the loss of inelastic events are presented in 
Secs.~\ref{sec:subtraction} and~\ref{sec:event_track_losses}.

\subsection{Track selection criteria}\label{sec:track_cuts}

\begin{figure*}
  \includegraphics[width=0.49\textwidth]{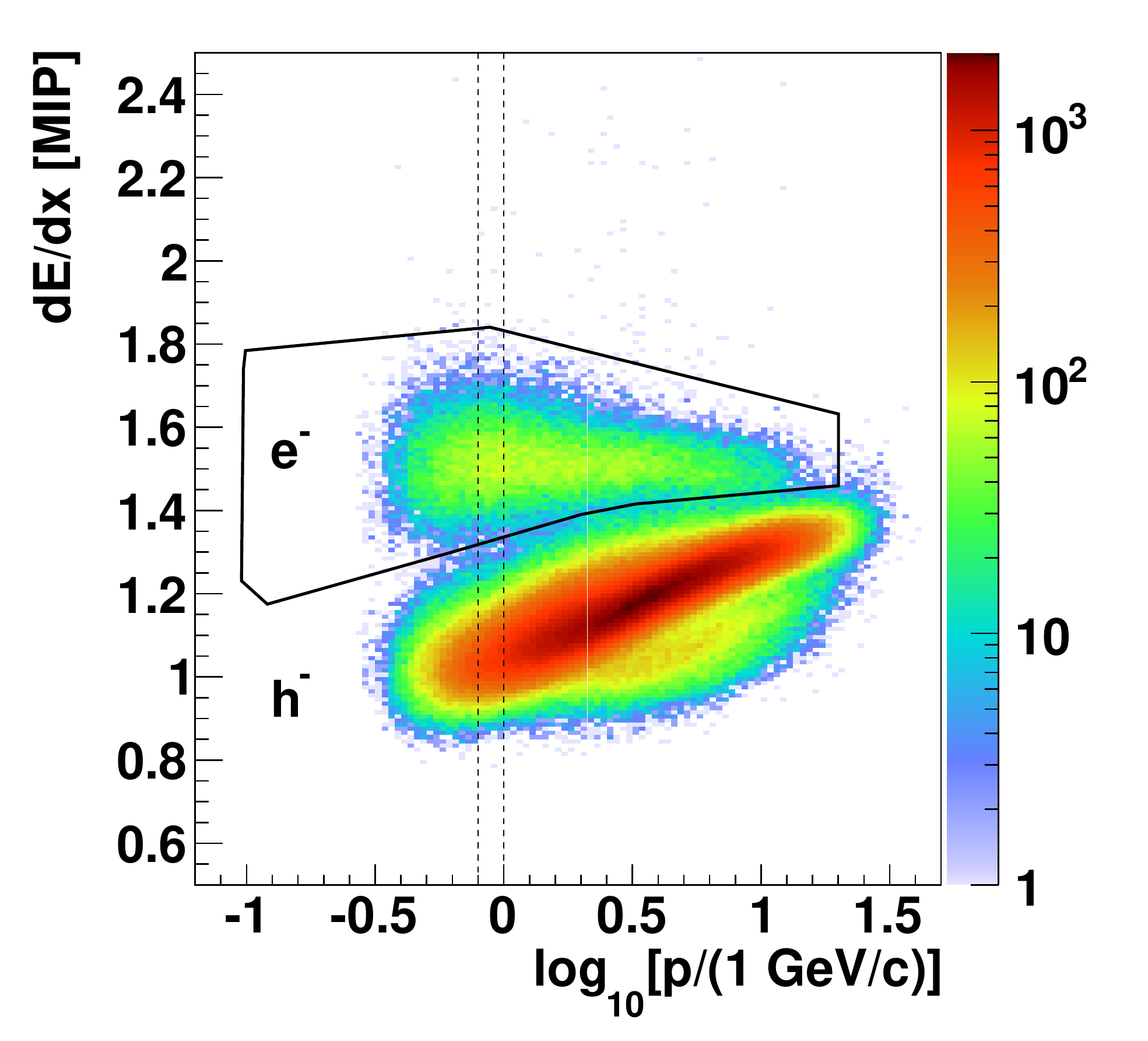}
  \includegraphics[width=0.49\textwidth]{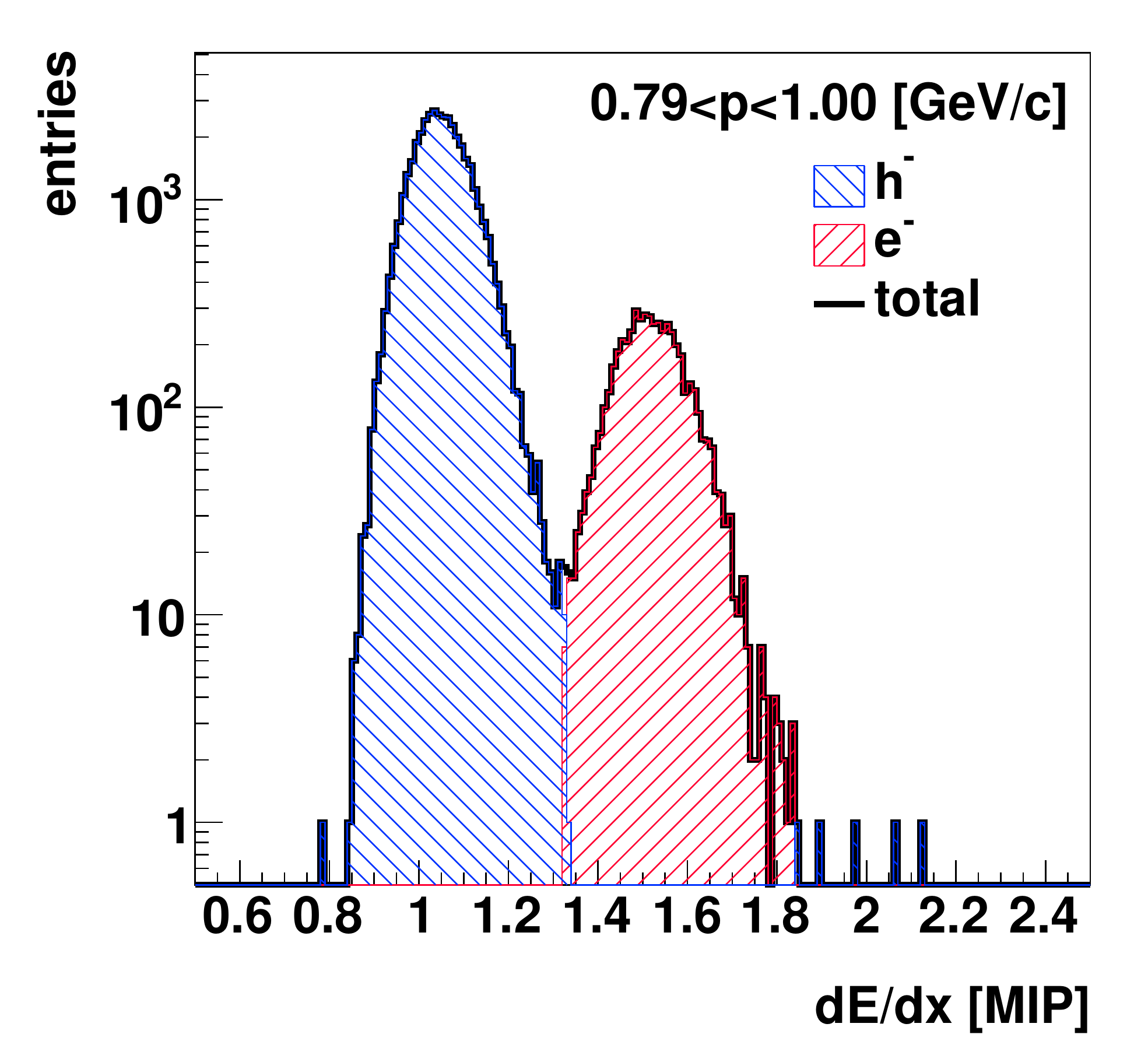}
  \caption{
  \emph{Left:} Distribution of particle energy loss as a function of the 
  logarithm of total momentum, for negatively charged
  particles produced in p+p interactions at 40\GeVc.
  The black contour shows the electron selection region.
  \emph{Right:} \dedx distribution in the momentum range indicated in the 
  figure and marked with vertical dashed lines in the \emph{left} panel.
  Electrons and negatively charged hadrons show separated peaks.
  }
  \label{fig:dedx_cut}
\end{figure*}

In order to select well-measured tracks of primary negatively charged
hadrons  as well as to reduce the contamination of tracks from secondary 
interactions, weak decays and off-time interactions the following track 
selection criteria were applied:
\begin{enumerate}[(i)]
  \item the track momentum fit at the interaction vertex should have converged,
  \item the fitted track charge is negative,
  \item the fitted track momentum component $p_x$ is negative. This selection
        minimises the angle between the track trajectory and the TPC pad
        direction for the chosen magnetic field direction.
        This reduces statistical and systematic uncertainties  of the 
        cluster position, energy deposit and track parameters,
        \label{item:rst_selection}
  \item the total number of reconstructed points on the track should be greater 
        than~30,
  \item
  \begin{sloppypar}
        the sum of the number of reconstructed points in VTPC-1 and VTPC-2 
        should be greater than~15 or the number of reconstructed points 
        in the GAP-TPC should be greater than~4,
  \end{sloppypar}
  \item the distance between the track extrapolated to the interaction plane 
        and the interaction point (impact parameter) should be smaller 
        than 4~cm in the horizontal (bending) plane and 2~cm in the 
        vertical (drift) plane,
  \item the track should be measured in a high ($\geq90\%$) TPC acceptance 
        region (see Sec.~\ref{sec:acc}),
  \item tracks with \dedx{} and total momentum values characteristic for 
        electrons are rejected.
        The electron contribution to particles with momenta above 20\GeVc is 
        corrected using the simulation.
        The electron selection procedure is visualised in 
        Fig.~\ref{fig:dedx_cut}.
\end{enumerate}

The spectra of negatively charged particles after track and event selections
were obtained in 2-dimensional bins of ($y$, \pt) and ($y$, \mt). 
The spectra were evaluated in the centre-of-mass frame after rotation of the
$z$ axis into the proton beam direction measured event-by-event by the BPDs.

\subsection{Correction for detector acceptance}\label{sec:acc}
\begin{figure*}
  \centering
  \includegraphics[width=0.49\textwidth]{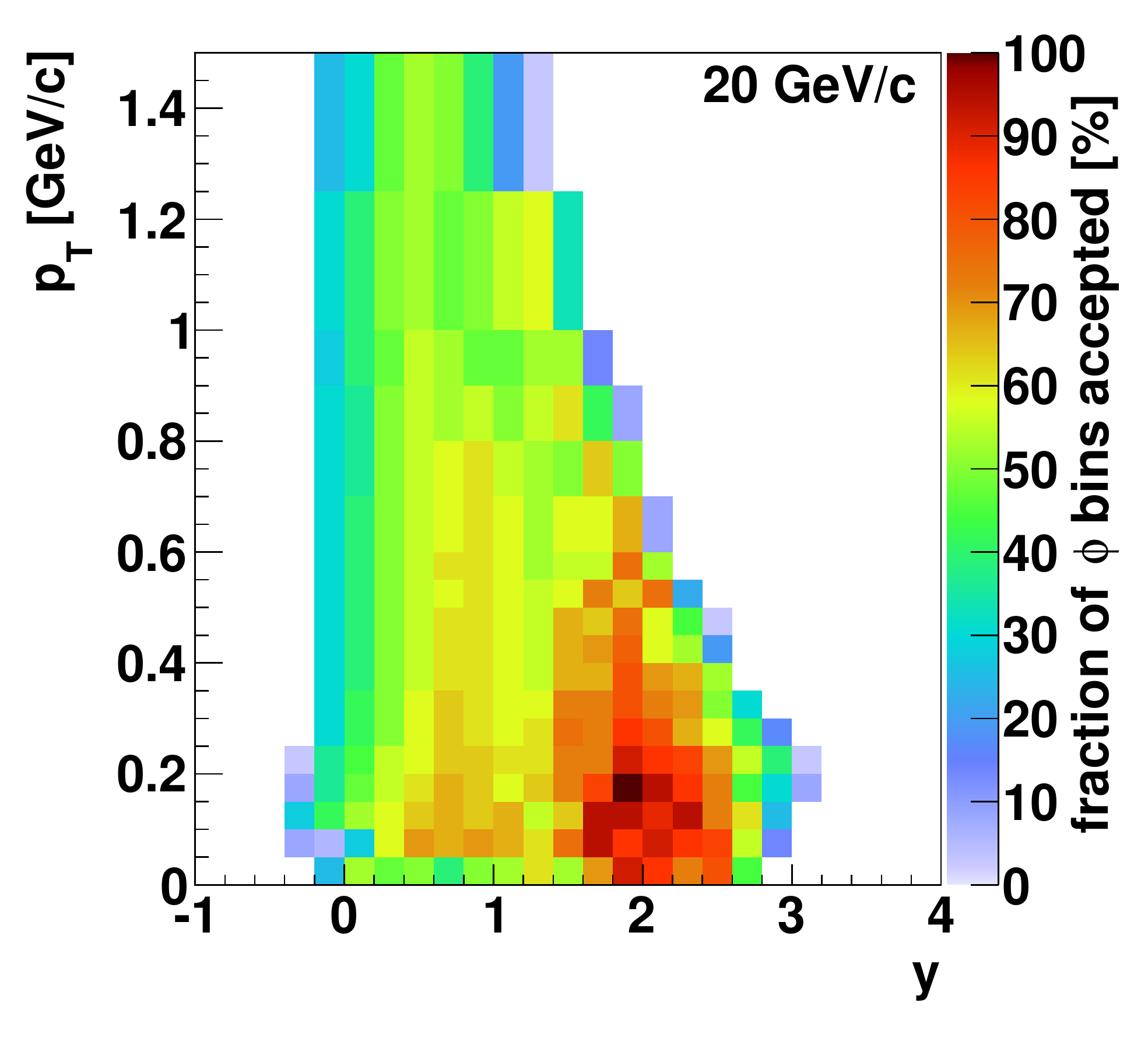}
  \includegraphics[width=0.49\textwidth]{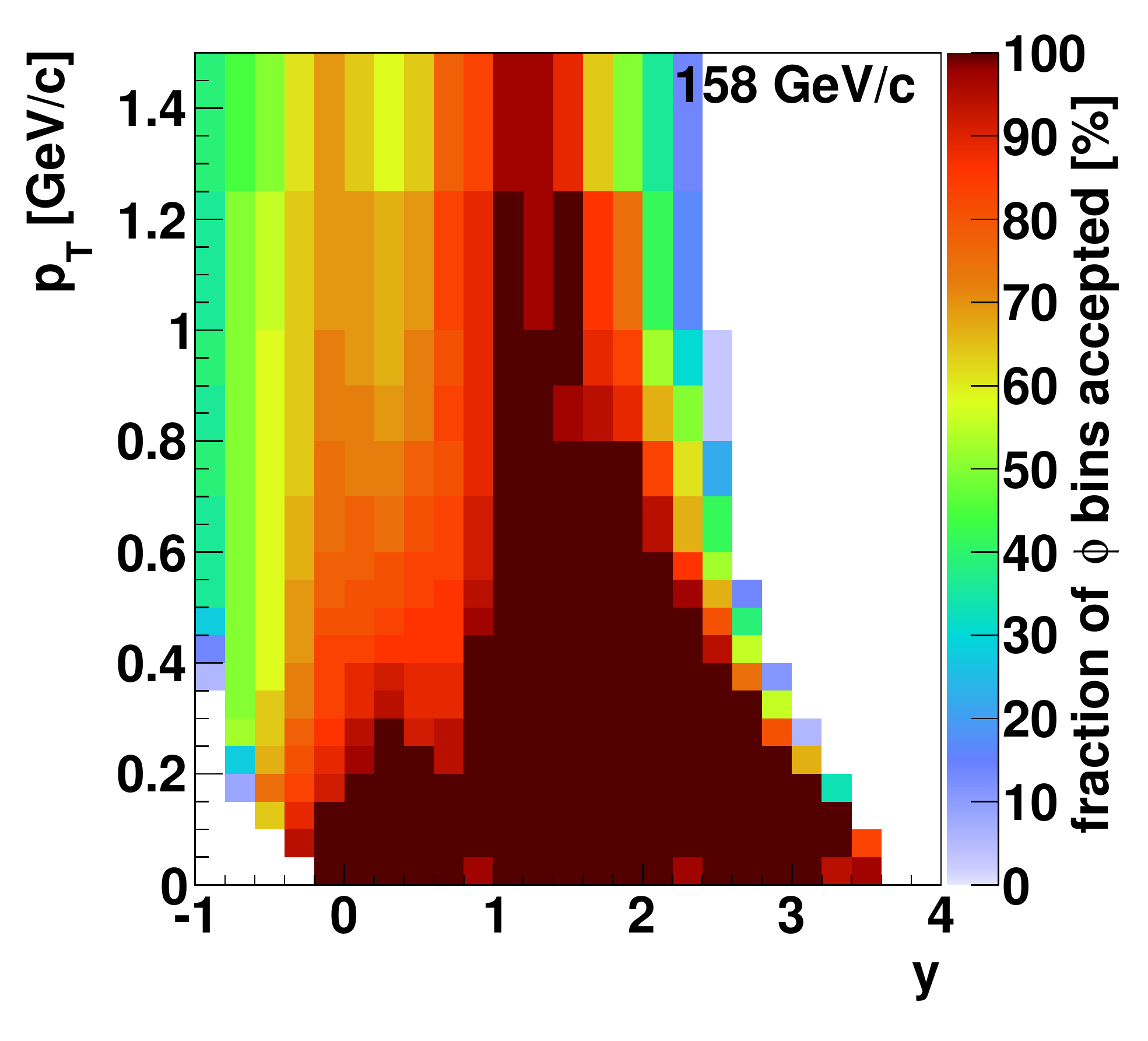}
  \caption{
  Detector acceptance at 20 (\emph{left}) and 158\GeVc (\emph{right}).
  The acceptance is calculated as fraction of ($y$, \pt, $\phi$) bins accepted 
  for given $y$ and \pt
  for tracks with $p_x<0$ selected for this analysis.
  }
  \label{fig:acceptance}
\end{figure*}

The detection and reconstruction inefficiencies are corrected using the 
simulation described in Sec.~\ref{sec:data}.
However, in order to limit the impact of possible inaccuracies of this 
simulation, only regions are accepted where the reconstruction efficiency
(defined as the ratio of the number of reconstructed and matched MC \pim 
tracks passing the track selection criteria to the number of generated \pim) equals 
at least 90\%. These regions were identified using a separate, statistically 
independent simulation in 
three-dimensional bins of $y$, \pt or \mt and the azimuthal angle $\phi$ 
($5^\circ$ bin width).
The resulting acceptance maps are shown in Fig.~\ref{fig:acceptance}.
The acceptance calculated in the $y<0$ region, not used for the final results, 
is shown also for comparison.
We chose an upper limit of  1.5\GeVc for the transverse momentum spectra, 
because beyond the admixture of background tracks reaches a level which cannot 
be handled by the correction procedures used in this paper.
Future publications will be devoted to the high \pt region.

\begin{sloppypar}
Since neither target nor beam are polarized, we can assume a uniform 
distribution of particles in $\phi$.
The data falling into the accepted bins is summed over $\phi$ 
bins and the ($y$, \pt/\mt) bin content is multiplied by a correction 
factor to compensate for the rejected $\phi$ ranges.
The acceptance correction also compensates for the $p_x < 0$ selection (see 
Sec.~\ref{sec:track_cuts}, point~(\ref{item:rst_selection})).
\end{sloppypar}

Even a small deviation of the beam direction from the nominal axis ($z$)
results in a 
non-negligible bias in the reconstructed transverse momentum. The beam 
direction 
is 
measured in the BPDs, and the particle momenta are recalculated to the frame 
connected with the beam direction. However, the detector acceptance depends 
on the momentum in the detector frame. Therefore the acceptance selection is 
done in the detector frame, and the acceptance correction is applied as a 
weight 
to each track. 
The weights are used to obtain particle spectra corrected for the
detector acceptance in the frame connected with the beam direction.

\subsection{Correction for off-target interactions}
\label{sec:subtraction}
The spectra were derived for events with liquid hydrogen in 
($_\mathrm{I}$) and removed ($_\mathrm{R}$) from the LHT. 
The latter data set represents interactions outside the liquid hydrogen 
(interactions with materials downstream and
upstream of the liquid hydrogen). The differential inclusive yield of 
negatively charged particles per event in interactions of beam protons
with the liquid hydrogen inside the LHT ($\n_\mathrm{T}[\hm]$) is calculated as:
\begin{linenomath} 
\begin{equation}\label{eq:subtraction}
\n_\mathrm{T}[\hm] =
\frac{1}{1-\epsilon} \cdot 
\left(
  \n_\mathrm{I}[\hm] - \epsilon \cdot \n_\mathrm{R}[\hm]
\right)\ ,
\end{equation}
\end{linenomath}
where:
\begin{enumerate}[(i)]
  \item $\n_\mathrm{I}[\hm]$ and $\n_\mathrm{R}[\hm]$ is the number of tracks 
in a given bin per 
  event selected for the analysis (see Sec.~\ref{sec:track_cuts}) for the data 
  with the liquid hydrogen inserted and removed, respectively,
  \item $\epsilon$ is the ratio of the interaction probabilities for the 
  removed and inserted target operation. 
\end{enumerate}

\begin{figure*}
\centering 
  \includegraphics[width=0.79\linewidth]{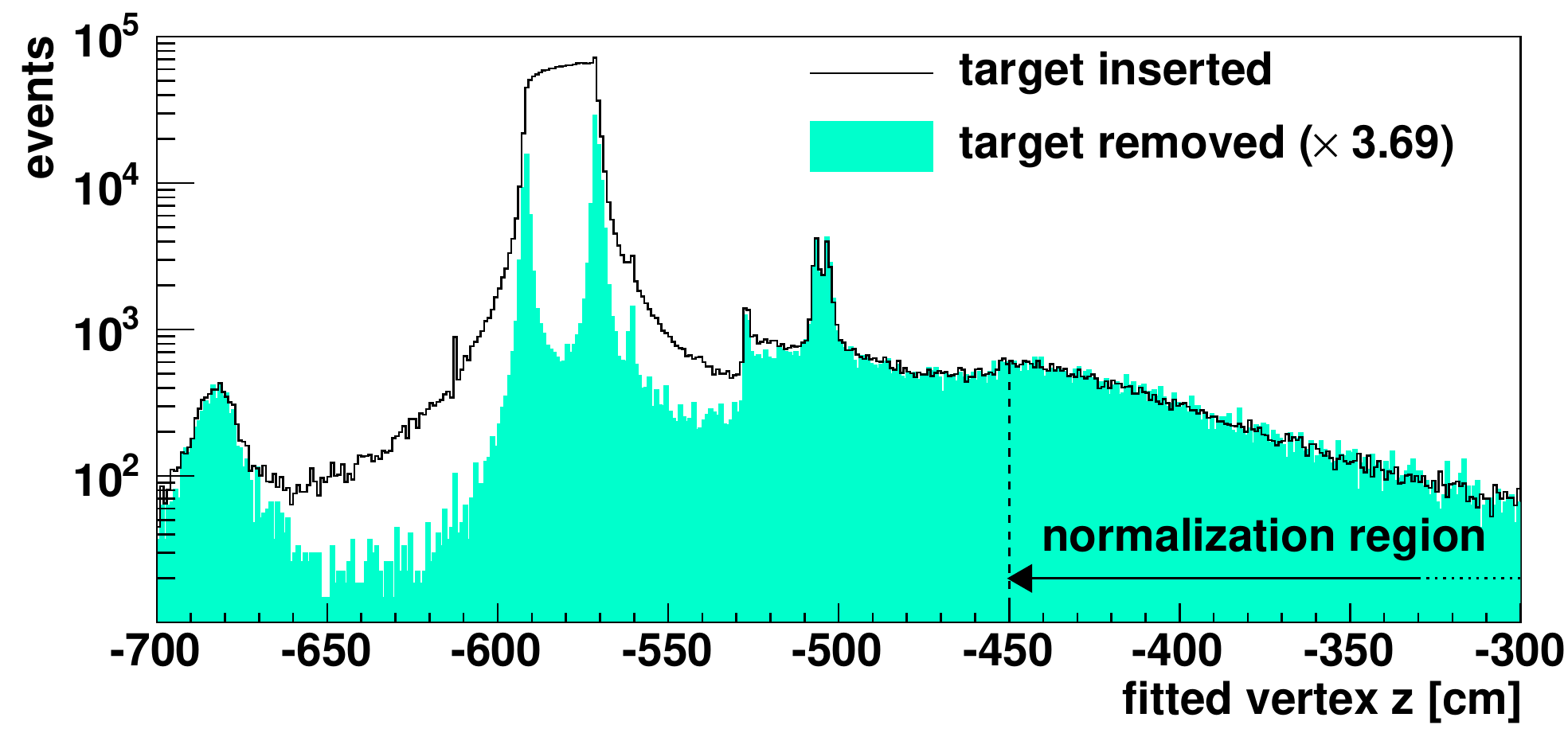}
  \caption{
  (Colour online)
  Distribution of fitted vertex $z$ coordinate for data on 
  p+p interactions at  40\GeVc. The  
distribution for the data recorded with the removed liquid hydrogen
was multiplied by a factor of $N_\mathrm{I}[z>-450\mathrm{\ 
cm}]/N_\mathrm{R}[z>-450\mathrm{\ cm}]$.
  }\label{fig:vertex_z}
\end{figure*}

$\epsilon$ was derived based on the distribution of the fitted $z$  coordinate 
of the interaction vertex.
All vertices far away from the target originate from interactions 
with the beam-line and detector materials. 
Neglecting the beam attenuation in the target one gets:
\begin{linenomath} 
\begin{equation}\label{eq:epsilon}
  \epsilon = 
\frac{N_\mathrm{R}}{N_\mathrm{I}}
\cdot\frac{N_\mathrm{I}[z>-450\mathrm{\ cm}]}{N_\mathrm{R}[z>-450\mathrm{\ 
cm}]}\ ,
\end{equation}
\end{linenomath}
where $N[z>-450\mathrm{\ cm}]$ is the number of events with fitted vertex 
$z>-450$~cm. 
Examples of distributions of $z$ of the fitted vertex for events recorded with 
the
liquid hydrogen inserted and removed are shown in Fig.~\ref{fig:vertex_z}.
Values of $\epsilon$ are listed in Table~\ref{tab:epsilon}.

\begin{table}
 \caption{
  The ratio of the interaction probabilities, $\epsilon$, for the
  removed and inserted target operation for data taking on p+p
  interactions at 20, 31, 40, 80 and 158\GeVc.
 }\label{tab:epsilon}
 \centering
  \begin{tabular}{c r}
 \hline\hline
$p_\mathrm{beam}$ [\GGeVc] & \multicolumn{1}{c}{$\epsilon$ [\%]} \\\hline
 20 & $8.0  \pm 0.3$ \\
 31 & $7.1  \pm 0.1$ \\
 40 & $10.4 \pm 0.1$ \\
 80 & $12.7 \pm 0.1$ \\
158 & $12.6 \pm 0.1$\\
  \hline\hline
 \end{tabular}
\end{table}

The correction for the off-target interactions changes the yields obtained from 
the target inserted data by less than $\pm 5\%$, except in the regions where 
the statistical uncertainty is high.

\subsection{The correction for contamination of primary \pim mesons}
\label{sec:hminus}

More than 90\% of primary negatively charged particles produced in p+p 
interactions in the SPS energy range are \pim 
mesons~\cite{na49_pp_pions,na49_pp_protons,na49_pp_kaons}.
Thus \pim meson spectra can be obtained by subtracting the estimated non-pion
contribution from the spectra of negatively charged particles and 
additional particle identification is not required.

The simulation described in Sec.~\ref{sec:data} was used to 
calculate corrections for the admixture of particles other than primary \pim 
mesons to the reconstructed negatively charged particles.
The dominating contributions are primary $\km$ and \pbar, and secondary \pim 
from weak decays of $\Lam$ and \kzeros (feed-down) and from secondary 
interactions, incorrectly fitted to the primary vertex.

\begin{sloppypar}
The EPOS spectra
were adjusted based on the existing data~\cite{Agnieszka,Antoni}.
Preliminary \NASixtyOne results were used to scale double differential spectra 
of \km, and \pbar~\cite{Szymon}.
EPOS spectra of \pim were replaced by the preliminary \NASixtyOne 
results~\cite{Antoni_preliminary} normalised to the multiplicity from the world
data compilation~\cite{pp_compil}.
Spectra of \Lam and \kzeros were scaled by a constant factor derived at each 
energy using the world data compilation~\cite{kaon_lambda} of total 
multiplicities.
The impact of the adjustments on the final spectra ranges from $-2\%$ to $+5\%$ 
in most regions, except of the low \pt region at the low beam momenta, where 
it reaches $+20\%$.
\end{sloppypar}

\begin{sloppypar}
As it was found in~\cite{Agnieszka} 
the yields of $\km$ and \pbar are strongly correlated with the \pim yield.
Thus the correction for the contribution of primary hadrons 
is performed via  the multiplicative factor $c_\mathrm{K}$.
On the contrary the  contribution due to weak decays and secondary
interactions is mostly located in the low 
\pt region, and it is weakly correlated with the primary pion yield in
this region.
Thus this feed-down contribution is corrected for using the additive
correction $c_\mathrm{V}$.
The total correction is applied in as:
\begin{linenomath} 
\begin{equation}
\n_\mathrm{prim}[\pim] =
\left( \n_\mathrm{T}[\hm] - c_\mathrm{V} \right)\cdot c_\mathrm{K}\ ,
\end{equation}
\end{linenomath}
where
\begin{linenomath} 
\begin{equation}
  c_\mathrm{V} = \left(
  \n[\pim_\Lam]+\n[\pim_{\kzeros}]+\n[\mathrm{other}]
  \right)^\mathrm{MC}_\mathrm{sel}\ ,
\end{equation}
\end{linenomath}
\begin{linenomath} 
\begin{equation}
  c_\mathrm{K} =
  \left(
  \frac{\n[\pim]}
  {\n[\km]+\n[\pbar]+\n[\pim]}
  \right)^\mathrm{MC}_\mathrm{sel}
\ .
\end{equation}
\end{linenomath}
The 
spectrum of a particle $x$ is denoted as $\n[x]$ whereas 
$\n[\mathrm{other}]$ stands for all primary and secondary particles other than 
$\km$, \pbar, \pim and feed-down from $\Lam$ and \kzeros. 
The spectrum $\n[\mathrm{other}]$ of all other particles originates mostly
from secondary interactions with $>$90\% occurring in the hydrogen target. 
This contribution  was taken  
from the simulations without an additional  adjustment.
The superscript $^{\mathrm{MC}}$ marks adjusted EPOS spectra.
The subscript $_{\mathrm{sel}}$ indicates that
the event and track selection criteria were applied and then the
correction for the detector acceptance was performed;
the reconstructed tracks were identified by matching.
\end{sloppypar}

\subsection{Correction for event as well as track losses and migration}
\label{sec:event_track_losses}

The multiplicative correction $c_\mathrm{loss}$ for 
losses of inelastic events as well as losses and bin-to-bin
migration of  primary \pim mesons emitted within
the acceptance 
is calculated using the Monte Carlo simulation as:
\begin{linenomath} 
\begin{equation}
  c_\mathrm{loss} =
  \n[\pim]_\mathrm{gen}^\mathrm{MC}\ /\
  \n[\pim]_\mathrm{sel}^\mathrm{MC}\ ,
\end{equation}
\end{linenomath}
where 
the subscript $_{\mathrm{gen}}$ indicates the
generated spectrum of  primary \pim mesons 
binned according to the generated 
momentum vector.
Then the final, corrected \pim meson spectrum in inelastic p+p interactions 
is calculated as
\begin{linenomath} 
\begin{equation}
 \n[\pim] =
 c_\mathrm{loss}\cdot \n_\mathrm{prim}[\pim]\ .
\end{equation}
\end{linenomath}

The dominating effects contributing to the $c_\mathrm{loss}$  correction are
\begin{itemize}
 \item losses of inelastic events due to the trigger and off-line
       event selection,
 \item the pion migration between analysis bins,
 \item the pion reconstruction inefficiency.
\end{itemize}

\begin{figure*}
\centering
  \includegraphics[width=0.44\textwidth]{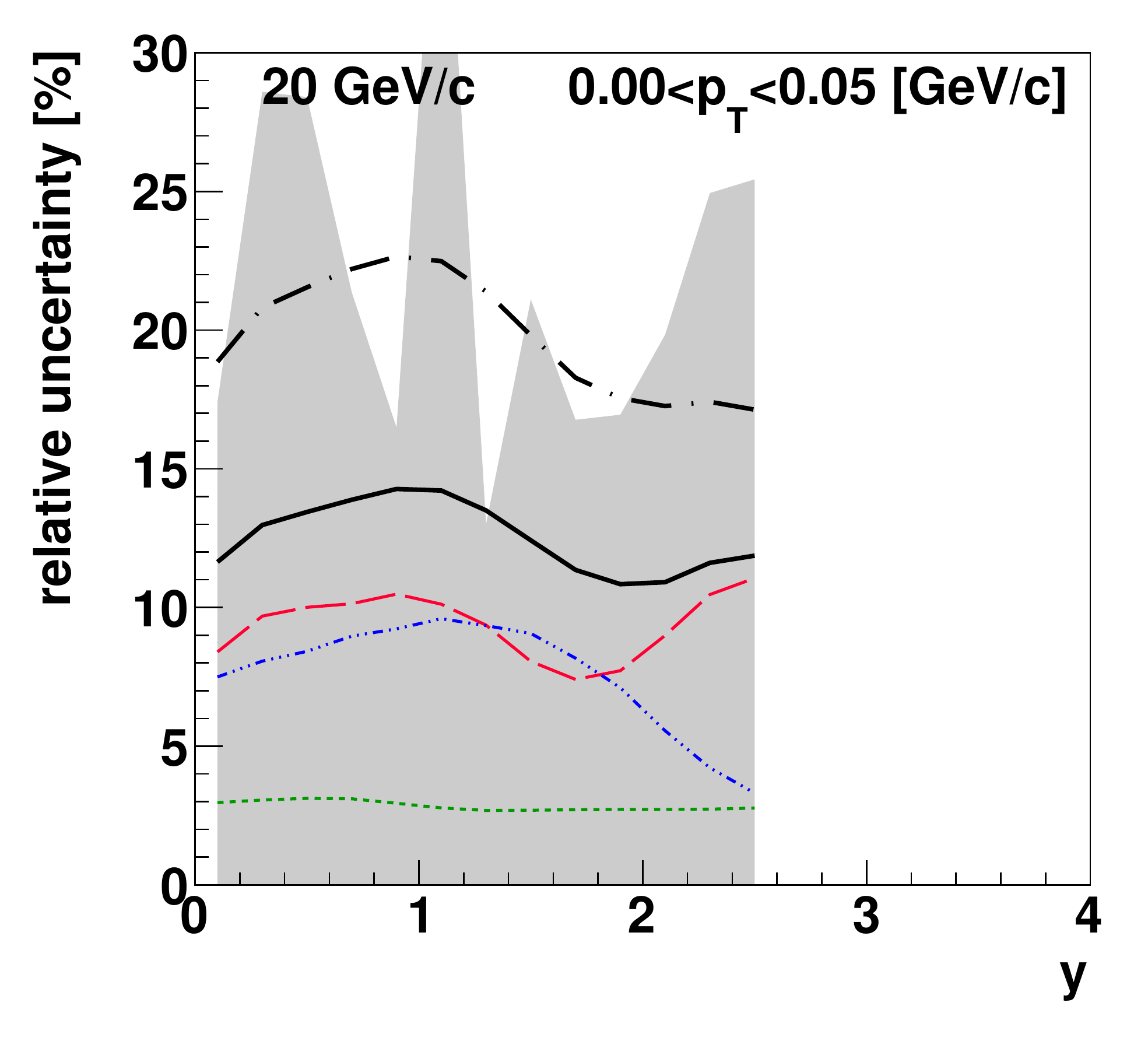}
  \includegraphics[width=0.44\textwidth]{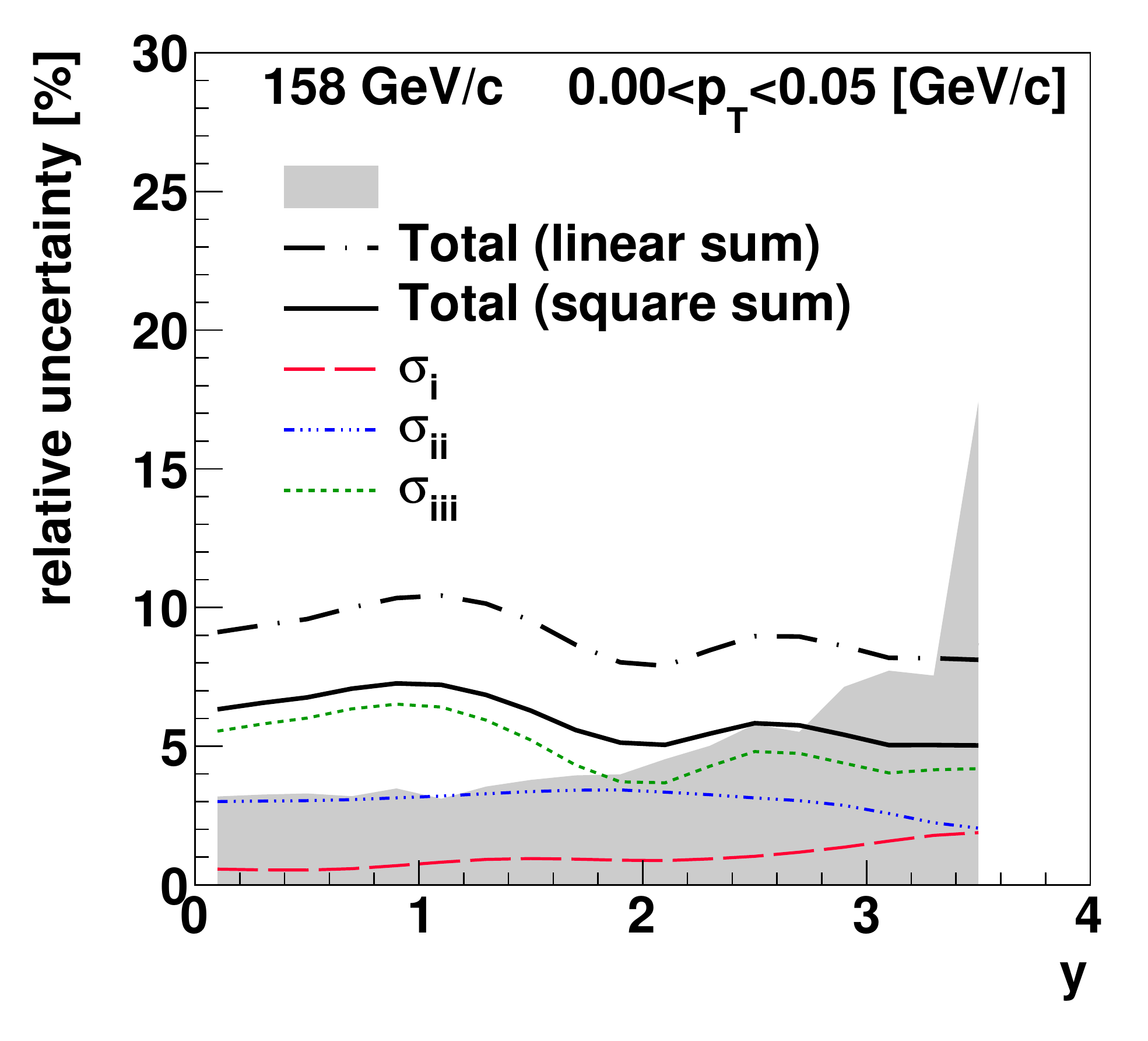}\\
  \includegraphics[width=0.44\textwidth]{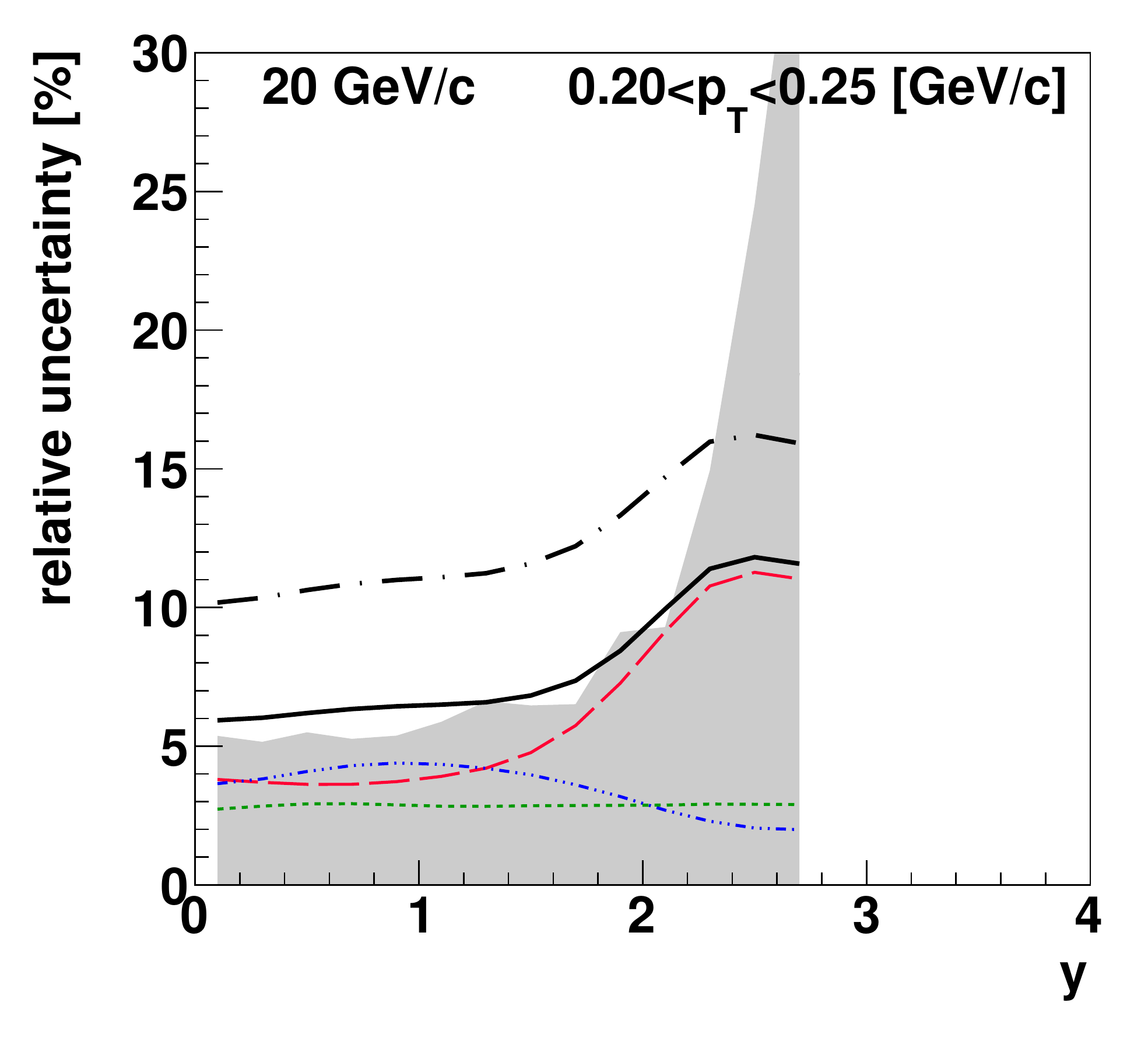}
  \includegraphics[width=0.44\textwidth]{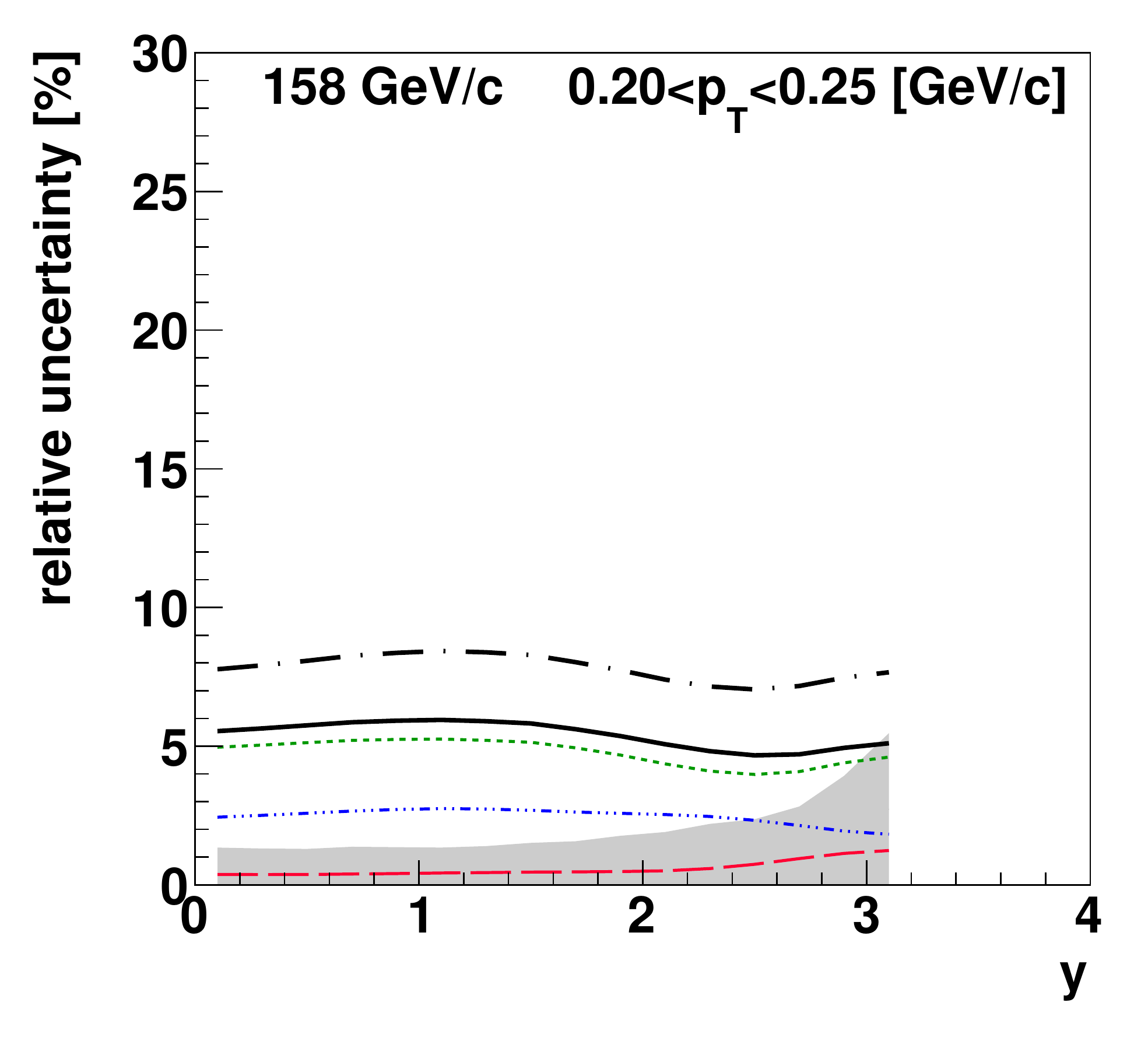}\\
  \includegraphics[width=0.44\textwidth]{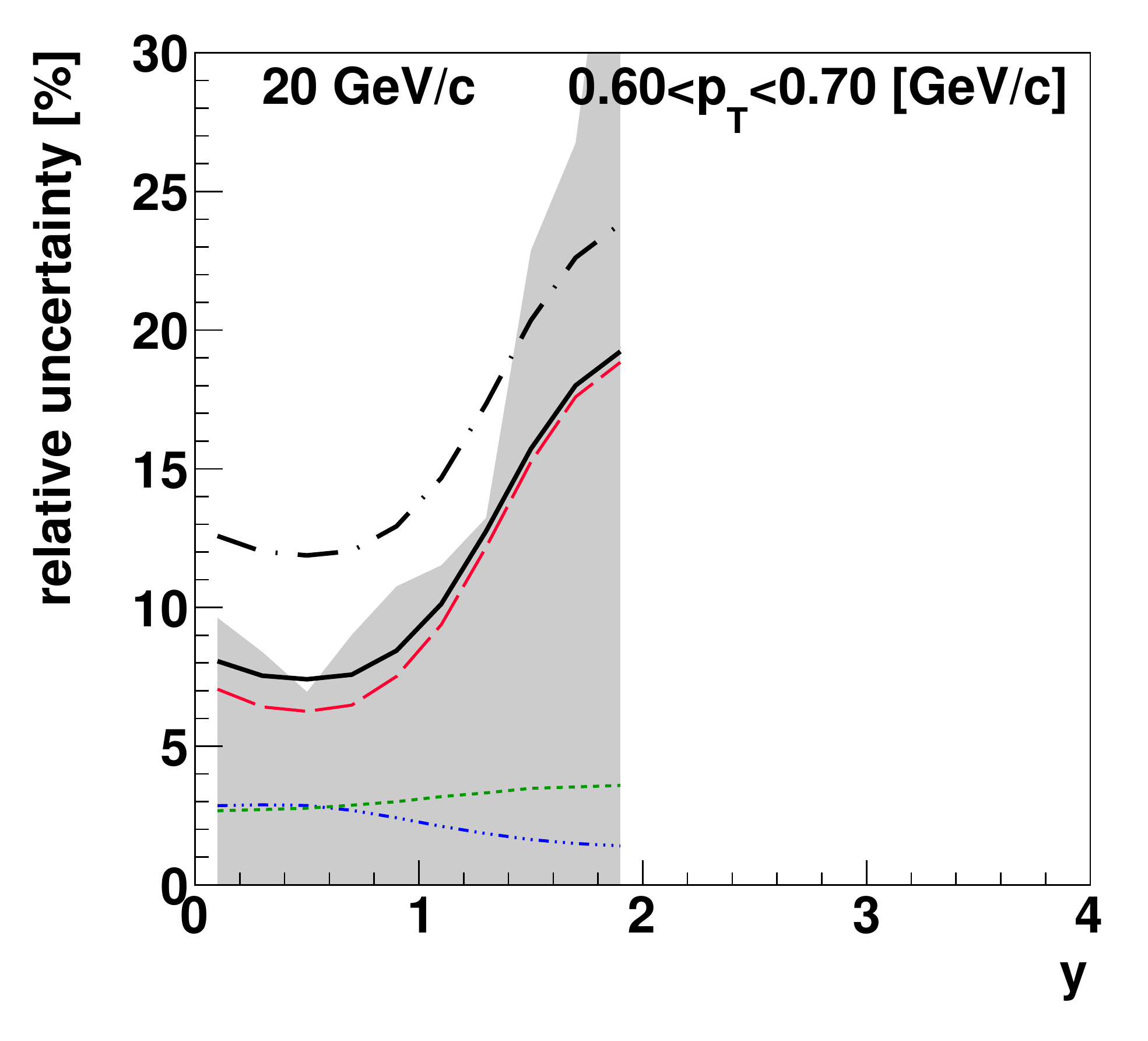}
  \includegraphics[width=0.44\textwidth]{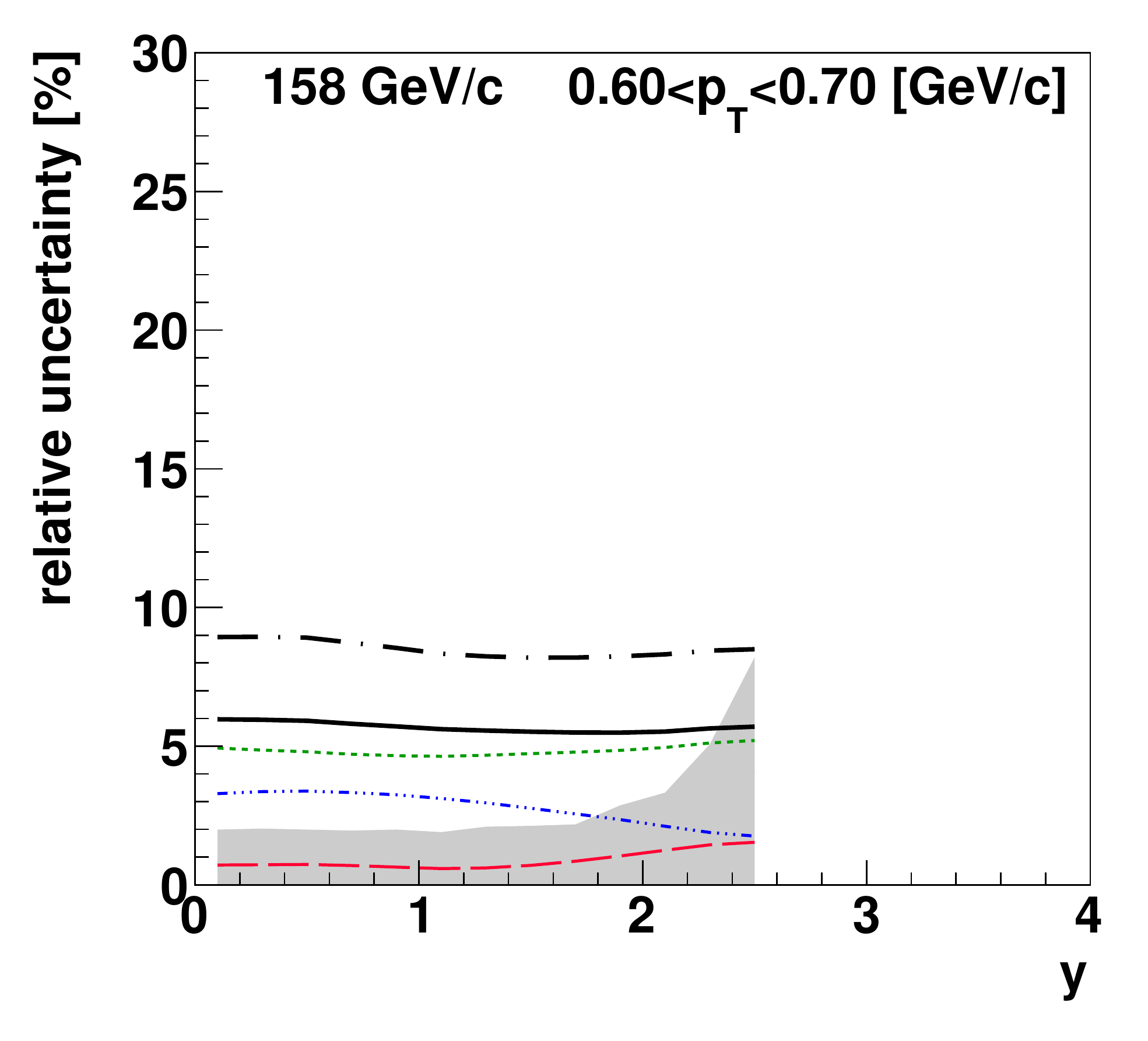}
  \caption{
  (Colour online)
  Statistical and systematic uncertainties in selected bins of \pt for 
20 (\emph{left}) and 158\GeVc (\emph{right}) p+p data. The shaded band shows 
the 
statistical uncertainty. The coloured thin lines show the contributions 
to the systematic uncertainty listed 
in Sec.~\ref{sec:Syst}. The thick black lines show the total systematic 
uncertainty, which was calculated by adding the contributions in quadrature 
(continuous line) or linearly (dashed/dotted line, shown for 
comparison).
}
  \label{fig:uncertainty}
\end{figure*}
\subsection{Statistical and systematic uncertainties}
\label{sec:errors}
\subsubsection{Statistical uncertainties}
\label{sec:errors_stat}
Statistical errors receive contributions from the finite
statistics of both the data as well as the simulated events
used to obtain the correction factors.
The dominating contribution is the uncertainty of the data which is 
calculated assuming a Poisson probability distribution for the number of entries
in a bin. The Monte Carlo statistics was higher than the data 
statistics. Also the uncertainties of the Monte Carlo corrections are
significantly smaller than the uncertainties of the number of entries in bins. 

\subsubsection{Systematic uncertainties}
\label{sec:Syst}
Systematic errors presented in this paper were calculated taking into
account contributions from the following effects.
\begin{enumerate}[(i)]
 \item \label{item:syst_cuts1} Possible biases due to event and track cuts 
which are not corrected for. These are:
 \begin{itemize}
  \item a possible bias due to the \dedx{} cut applied to remove electrons,
  \item a possible bias related to the removal of events with off-time beam 
        particles close in time to the trigger particle.
 \end{itemize}
 The magnitude $\sigma_\mathrm{i}$ of possible biases was estimated by varying 
 values of the corresponding cuts.
 The \dedx{} cut  was changed by $\pm 0.01$~\dedx units (where 1 
 corresponds to a minimum ionising particle, and 0.04 is a typical width of a 
 single particle distribution), and the 
 off-time interactions cut was varied from a $\pm 1~\mu$s to a $\pm 3~\mu$s 
 time window.
 The assigned systematic uncertainty was calculated as the maximum of the 
 absolute differences between the results obtained for both cut values.
 This contribution is drawn with a long-dashed red line 
 (\textcolor{pink}{--~--})  in Fig.~\ref{fig:uncertainty}.

 \item
 \label{item:syst_hminus} Uncertainty of the correction
       for contamination of the primary \pim mesons.
  The systematic uncertainty $\sigma_\mathrm{ii}$ of this correction was 
  assumed as 20\% 
  (for 40, 80 and 158\GeVc) and 40\% (for 20 and 31\GeVc) of the absolute 
  value of
  the correction. At the low beam momenta there was less data available to 
  adjust 
  the simulated spectra, which was the reason to increase the uncertainty.
 This contribution is drawn with a dashed-dotted blue line 
 (\textcolor{blue}{-$\cdot\!\cdot\!\cdot$-}) in 
  Fig.~\ref{fig:uncertainty}.
  The absolute correction is small thus the related systematic uncertainty 
  is small also.

 \item \label{item:syst_events} Uncertainty of the correction for 
the event losses.
  The uncertainty was estimated using 20\% of the correction magnitude and a 
comparison with the correction calculated using the VENUS~\cite{Venus} model:
\begin{linenomath} 
  \begin{equation}
   \sigma_\mathrm{iii} = 
   0.2\cdot\left|1 - c_\mathrm{loss}^\mathrm{EPOS}\right| +
   \left|c_\mathrm{loss}^\mathrm{EPOS}-c_\mathrm{loss}^\mathrm{VENUS}\right|\ .
  \end{equation}
\end{linenomath}
  This contribution is drawn with a short-dashed green line 
  (\textcolor{darkgreen}{-~-~-})  in 
  Fig.~\ref{fig:uncertainty}.
  
  \item \label{item:syst_cuts2} Uncertainty  related to the track selection 
method. It was estimated by varying the track selection cuts: removing the 
impact parameter cut and decreasing the minimum number of required points to 25 
(total) and 10 (in VTPCs)
and by checking symmetries with respect to $y=0$ and $\pt=0$.
The potential bias is below 2\% and the corresponding contribution was 
neglected.
\end{enumerate}

\begin{sloppypar}
The total systematic uncertainty was calculated by adding in quadrature
the contributions
$\sigma_\mathrm{sys} = 
\sqrt{\sigma_\mathrm{i}^2+\sigma_\mathrm{ii}^2+\sigma_\mathrm{iii}^2}$.
This uncertainty is listed in the tables including 
numerical values and it is visualised by a shaded band around
the data points in plots presenting the results.
Statistical and systematic uncertainties in selected example regions are shown 
in Fig.~\ref{fig:uncertainty}.
Systematic biases in different bins are correlated,
whereas statistical fluctuations are almost independent.
\end{sloppypar}

\begin{figure}
\centering
\includegraphics[width=0.49\textwidth]{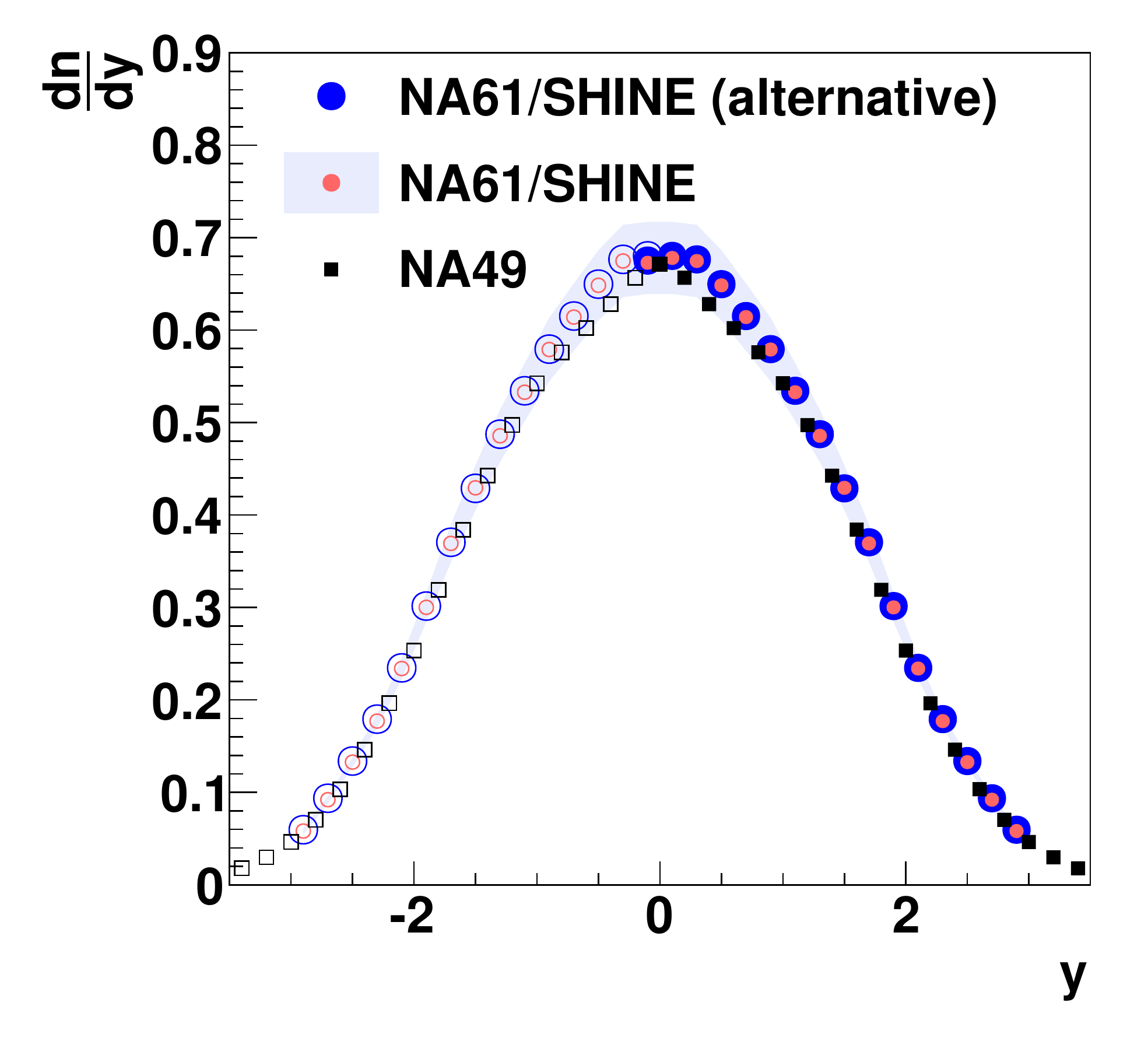}
  \caption{
  (Colour online)
  Rapidity distribution of \pim mesons produced in inelastic p+p interactions
  at 158\GeVc. 
  The big blue points show the results obtained 
  with an alternative method: without vertex fit requirement and rejection of 
  events with a single very high momentum positively charged track.
  The results of \NASixtyOne (this paper, red dots) are compared
  with the NA49 measurements~\cite{na49_pp_pions}~(black squares).
  The open symbols show points reflected with respect to
  mid-rapidity. A single \NASixtyOne point measured at 
  $y<0$ is also shown for comparison.
  The shaded band shows the \NASixtyOne systematic uncertainty.
}
  \label{fig:rapidity_na49}
\end{figure}

\begin{sloppypar}
Figure~\ref{fig:rapidity_na49} presents a comparison of the rapidity
spectrum of \pim mesons produced in inelastic p+p interactions at 158\GeVc
(for details see Sec.~\ref{sec:results}) from the present analysis with
the corresponding spectrum measured by NA49~\cite{na49_pp_pions}.
Statistical and systematic uncertainties of the NA49 spectrum are
not explicitly given but the published information implies that the systematic
uncertainty dominates and amounts to several~\%.
The results agree within the systematic uncertainties of the \NASixtyOne
spectra.
\end{sloppypar}

The analysis method of p+p interactions at 158\GeVc performed by 
NA49~\cite{na49_pp_pions} 
differed from the one used in this paper.
In particular, pions were identified by \dedx measurement and
the NA49 event selection criteria did not include
the selection according to the fitted $z$ coordinate of the interaction
vertex and the rejection of elastic interactions.
Namely,  all events passing the trigger selection and off-line quality cuts
were used for the analysis. 
For comparison, this event selection procedure was applied to the
\NASixtyOne data. As a result 20\% more 
events were accepted. Approximately half of them were unwanted
elastic and off-target interactions and half were wanted inelastic
interactions. Then the corrections corresponding to the changed selection
criteria were applied
(the contribution of elastic events was subtracted using 
the estimate from Ref.~\cite{na49_pp_pions}). The fully corrected rapidity
spectrum obtained using this alternative analysis is also
shown in Fig.~\ref{fig:rapidity_na49}. The differences between the 
results for the standard and alternative  methods are below 0.5\% at $y<2$ 
and below 2\% at higher $y$.

\begin{figure}
\centering
\includegraphics[width=0.49\textwidth]{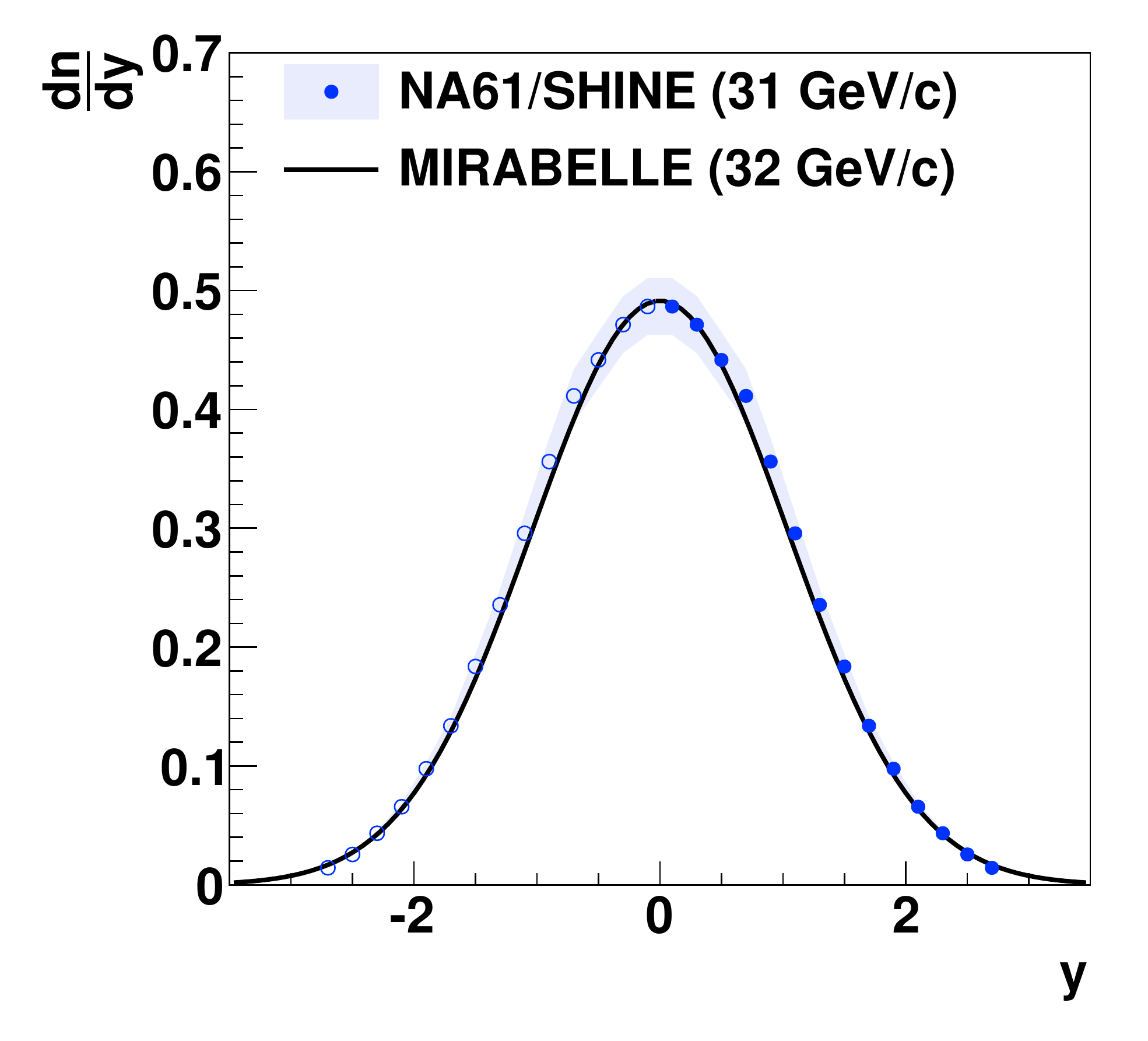}
  \caption{
  (Colour online)
  Rapidity distribution of \pim mesons produced in inelastic p+p 
  interactions.
  The \NASixtyOne results at 31\GeVc (blue points) are compared with 
  the MIRABELLE measurement (parametrised by the black line) at 32\GeVc.
  The shaded band shows the \NASixtyOne systematic uncertainty.
}\label{fig:rapidity_mirabelle}
\end{figure}

\begin{sloppypar}
Figure~\ref{fig:rapidity_mirabelle} shows a comparison of the rapidity 
distribution at 31\GeVc with the MIRABELLE results at 
32\GeVc~\cite{MIRABELLE32}. 
A parametrisation of the distribution and the total \pim multiplicity are 
provided. As the parametrisation appears to be incorrectly normalised, we 
normalised it to the total multiplicity.
The results agree within the \NASixtyOne systematic uncertainties.
\end{sloppypar}

\begin{sloppypar}
The spectra measured in p+p interactions should obey
reflection symmetry with respect 
to mid-rapidity. 
As the \NASixtyOne acceptance extends somewhat below mid\hyp{}rapidity
a check of the reflection symmetry can be performed and used
to validate the measurements.
It was verified that the yields measured for $y < 0$ agree with those measured 
for $y > 0$ in the reflected acceptance within 1.5\%. 
A similar agreement was also found at lower beam momenta.
The measurements above
mid-rapidity are taken as the final results. Nevertheless, 
for comparison the points at $y<0$ were added in 
Figs.~\ref{fig:rapidity_na49} and~\ref{fig:rapidity}
in the regions where the \pt acceptance extends to zero.
\end{sloppypar}

\begin{figure*}
\centering
  \includegraphics[width=0.465\textwidth]{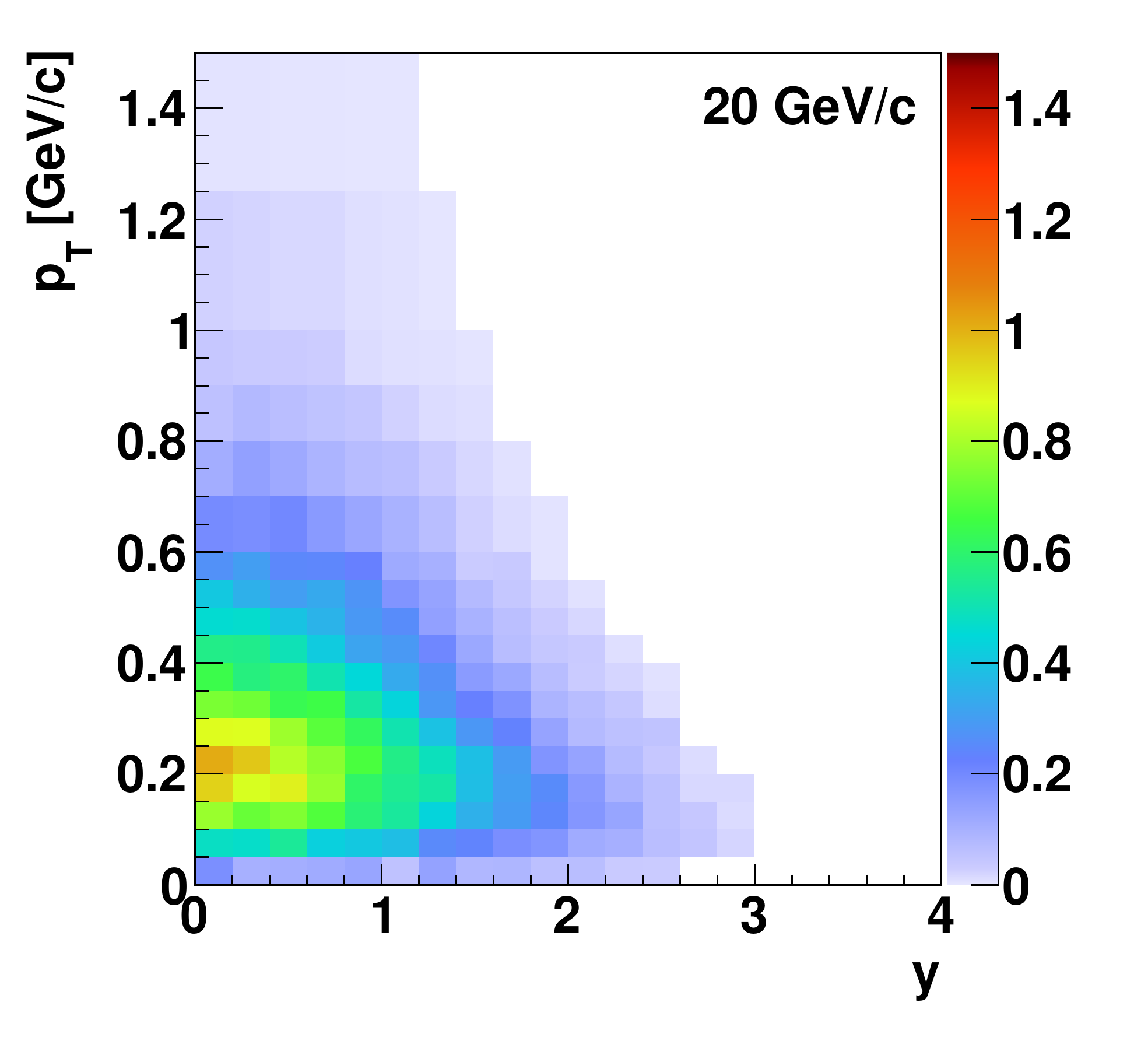}
  \includegraphics[width=0.465\textwidth]{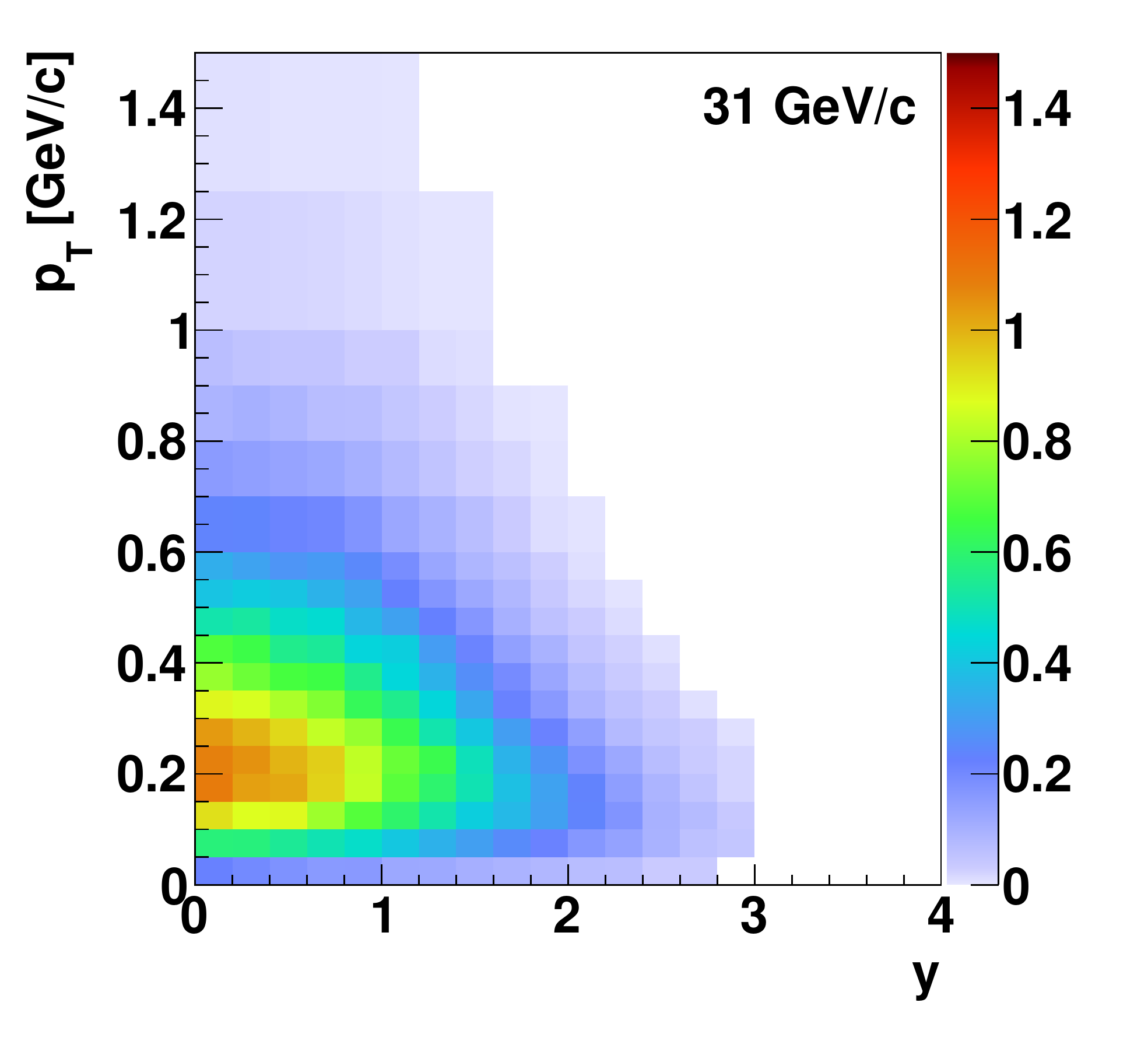}\\
  \includegraphics[width=0.465\textwidth]{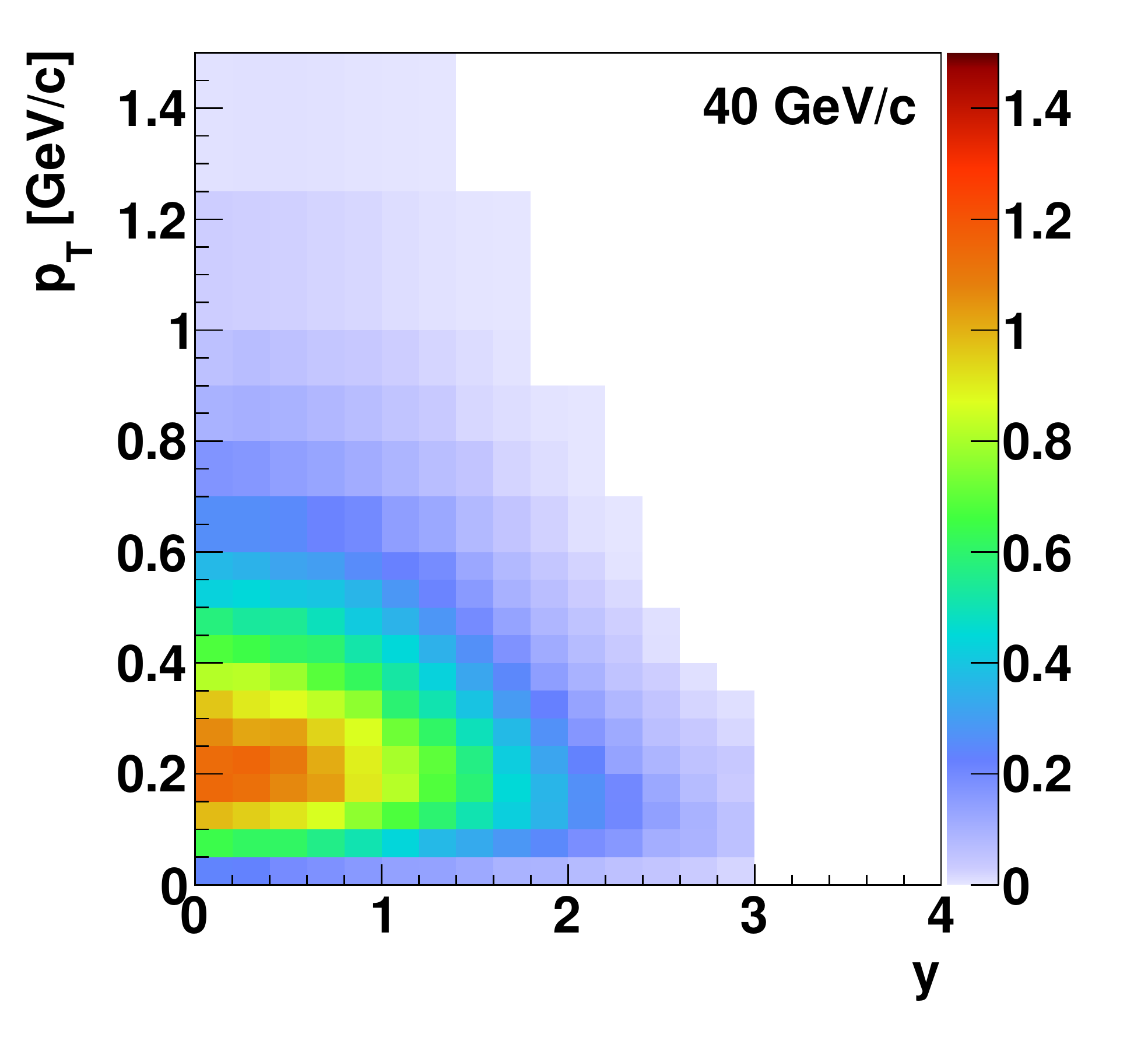}
  \includegraphics[width=0.465\textwidth]{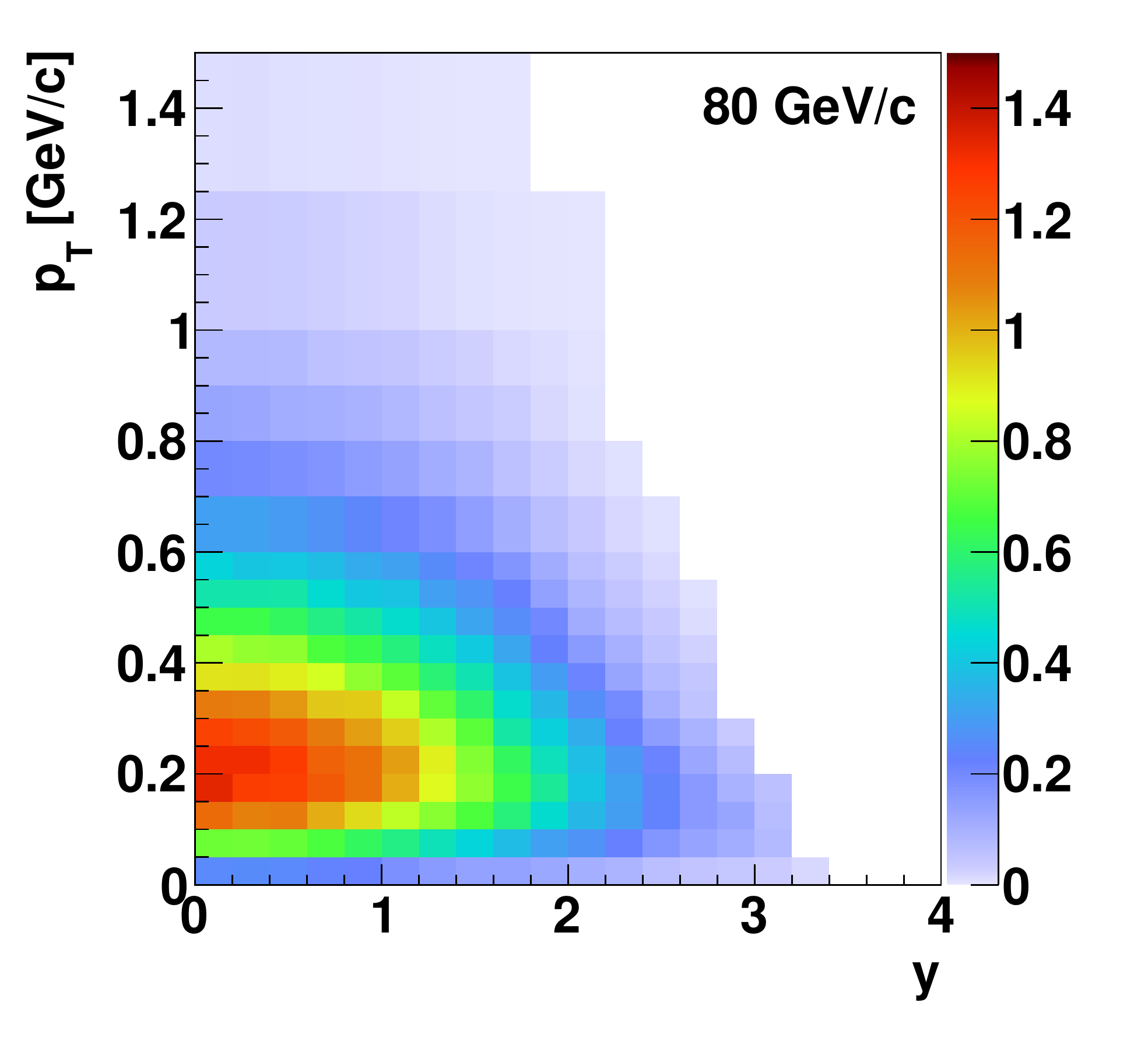}\\
  \includegraphics[width=0.465\textwidth]{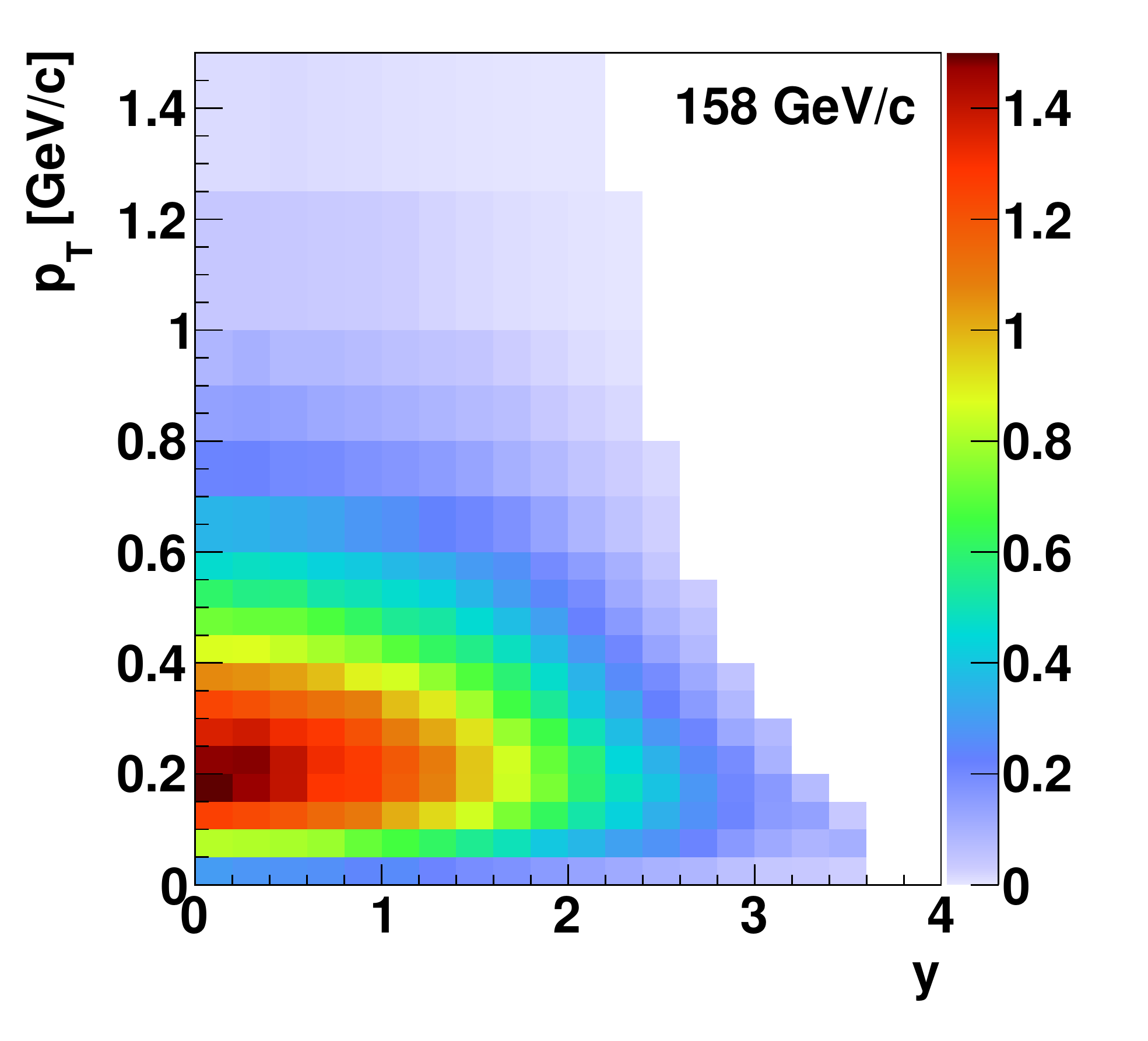}
  \phantom{\includegraphics[width=0.465\textwidth]{plots/y_pt_158.pdf}}
  \caption{
  (Colour online)
  Double differential spectra $\dd^2n/(\dd y\,\dd\pt)$~[(\GGeVc)$^{-1}$]
  of \pim mesons produced in inelastic p+p interactions at 20, 31, 40, 80 
and 
158\GeVc.}
  \label{fig:2Dspectra}
\end{figure*}

\begin{figure*}
\centering
\includegraphics[width=0.465\textwidth]{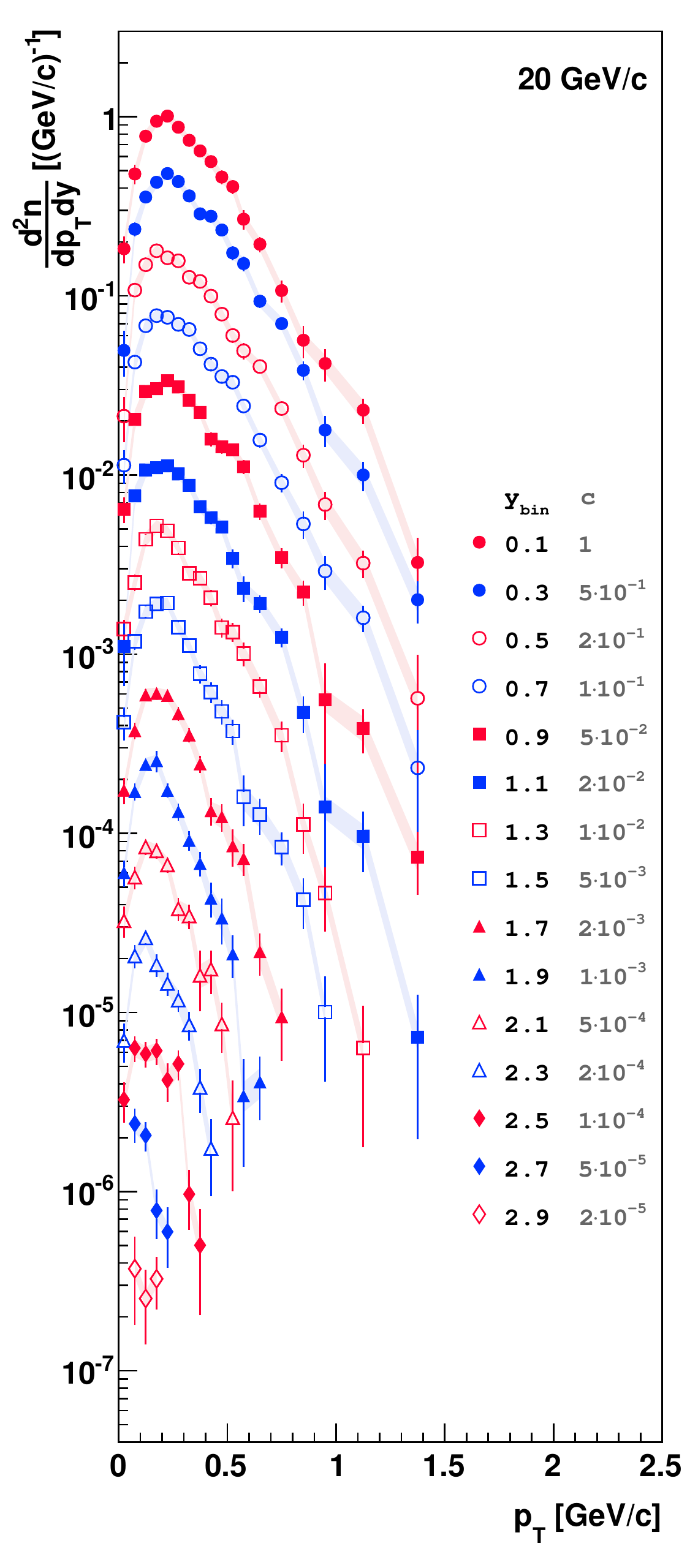}
\includegraphics[width=0.465\textwidth]{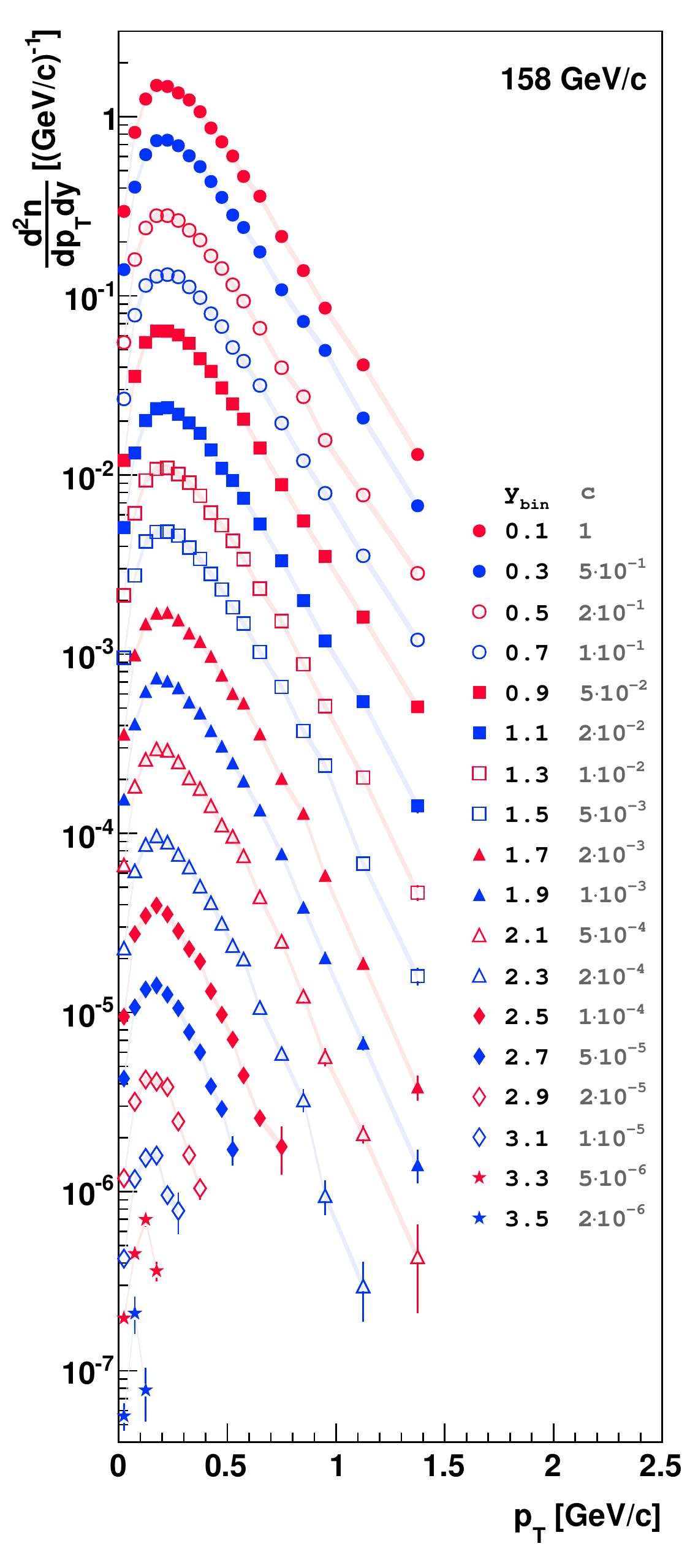}
  \caption{
  (Colour online)
  Transverse momentum spectra of \pim mesons produced in inelastic p+p 
  interactions at 20 (\emph{left}) and 158\GeVc (\emph{right}) in various 
  rapidity ranges.
  The legend provides the centres of the rapidity bins, $y_\mathrm{bin}$ and 
  the scaling factor $c$ used to separate the spectra visually.
  }
  \label{fig:spectra_pt_slices}
\end{figure*}

\section{Results}
\label{sec:results}

\begin{sloppypar}
This section presents
results on inclusive \pim meson spectra
in inelastic p+p interactions at beam momenta of 20, 31, 40, 80 and 158\GeVc.
The spectra refer to pions produced by strong interaction processes 
and in  electromagnetic decays of produced hadrons. 
\end{sloppypar}

Numerical results corresponding to the plotted spectra 
as well as their statistical and systematic uncertainties
are given in Ref.~\cite{na61_edms}.

\subsection{Double differential spectra}

The double differential inclusive spectra of \pim mesons in  
rapidity and transverse momentum produced in inelastic p+p interactions at 20, 
31, 40, 80 and 158\GeVc are shown in Fig.~\ref{fig:2Dspectra}.
The transverse momentum distributions at 20 and 158\GeVc are plotted in
Fig.~\ref{fig:spectra_pt_slices}.
Here $\frac{\dd^2n}{\dd y\,\dd\pt}$ or $\frac{\dd^2n}{\dd y\,\dd\mt}$ were 
calculated
by dividing the fully corrected bin contents $\n[\pim]$ (see 
Sec.~\ref{sec:analysis}) by the bin size. 
The spectra in ($y$, \mt) are not shown here but they are given in
the compilation of the numerical values~\cite{na61_edms}.

\subsection{Transverse mass spectra}

\begin{figure*}
  \includegraphics[width=0.49\textwidth]{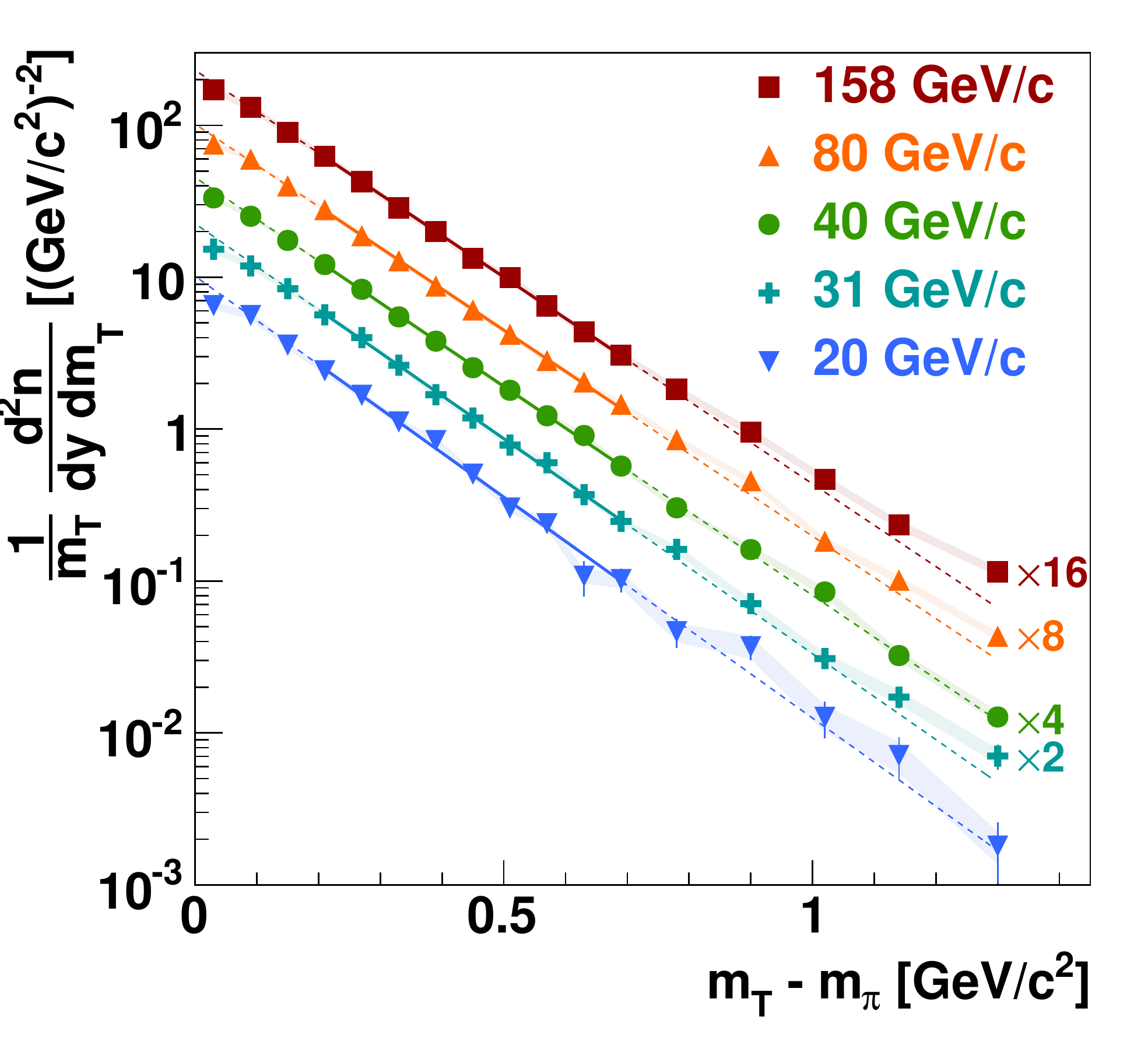}
  \includegraphics[width=0.49\textwidth]{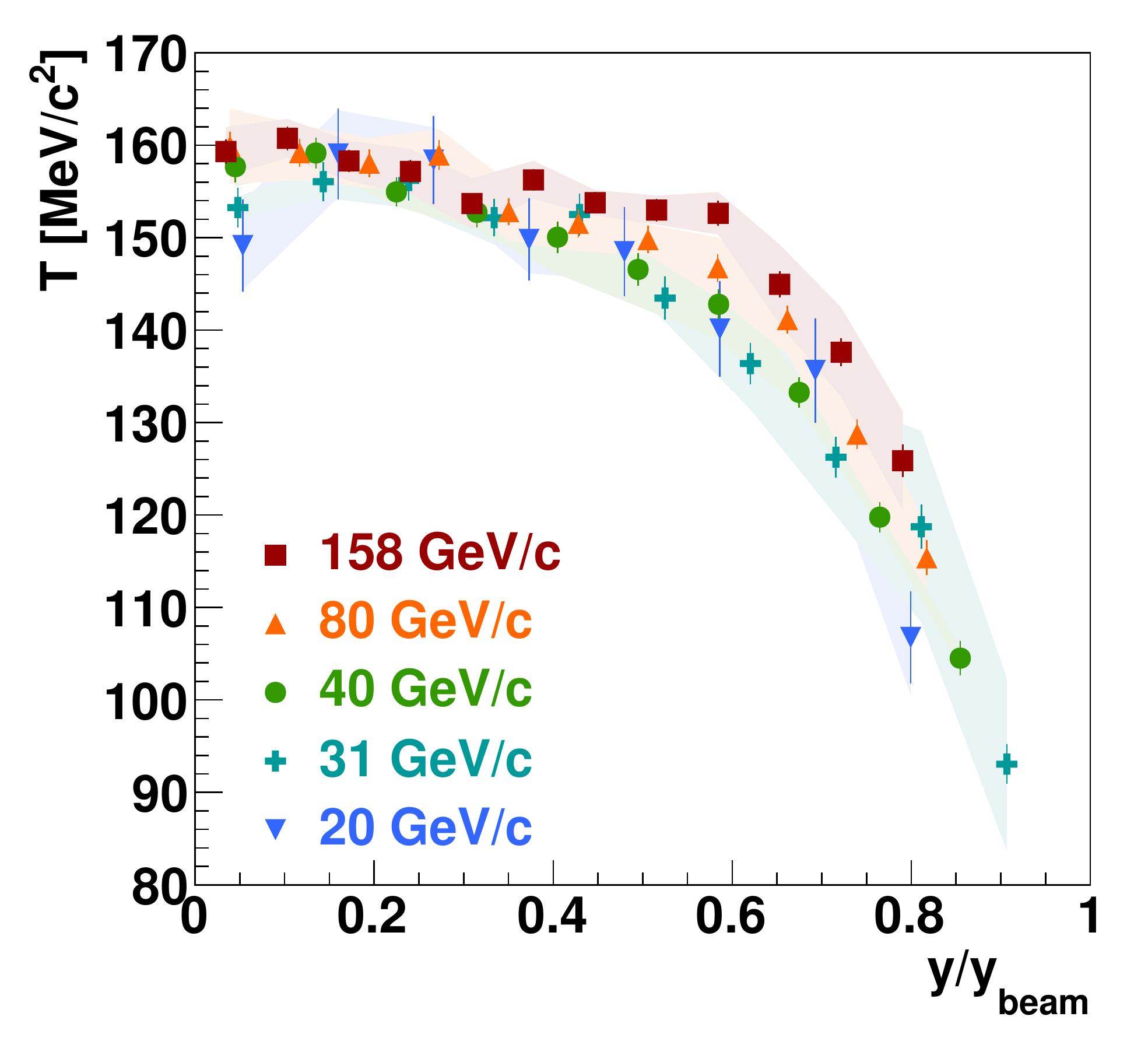}
  \caption{
  (Colour online)
  \emph{Left:} Transverse mass spectra at mid-rapidity ($0 < y < 0.2$).
  The fitted exponential function Eq.~\eqref{eq:mt} is indicated by solid lines
  in the fit range $0.2 < \mt - m_\pi < 0.7$\GeVcc and dashed lines
  outside the fit range. The data points for different beam momenta
  were scaled for better readability.
  \emph{Right:} The inverse slope parameter $T$ of the transverse mass spectra
  as a function of rapidity divided by the projectile rapidity. The fit range 
is 
  $0.2 < \mt - m_\pi < 0.7$\GeVcc.
  The results refer to \pim mesons produced in inelastic p+p interactions 
  at 20, 31, 40, 80 and 158\GeVc.
  }\label{fig:mt}
\end{figure*}

The transverse mass spectra at mid-rapidity 
($0 < y < 0.2$) are shown in Fig.~\ref{fig:mt} 
(\emph{left}).
A function
\begin{linenomath} 
\begin{equation}
\label{eq:mt}
  \frac{\dd n}{\dd\mt} = A\cdot \mt\cdot\exp\left(-\frac{\mt}{T}\right)
\end{equation}
\end{linenomath}
was fitted in the range $0.2 < \mt - m_\pi < 0.7$\GeVcc and
is indicated by lines in Fig.~\ref{fig:mt} (\emph{left}).
The fitted parameters were the normalisation $A$ and the inverse slope $T$.
They minimise the $\chi^2$ function which was calculated using statistical 
errors only.
In the $\chi^2$ calculation a measured bin content ($\dd n/\dd\mt$) was
compared with the integral of the fitted function in a 
bin divided by the bin width.

Similar fits were 
performed to spectra in  other rapidity bins containing data in the fit range. 
The rapidity dependence of 
the fitted inverse slope parameter $T$ is presented in Fig.~\ref{fig:mt} 
(\emph{right}).
The $T$ parameter decreases significantly when going from mid-rapidity
to the projectile rapidity\linebreak
($y_\mathrm{beam}$ = 1.877, 2.094, 2.223,  2.569 and 2.909
at 20, 31, 40, 80 and 158\GeVc, respectively).

\subsection{Rapidity spectra}
\label{rapidity_spectra}

\begin{figure*}
  \includegraphics[width=0.49\textwidth]{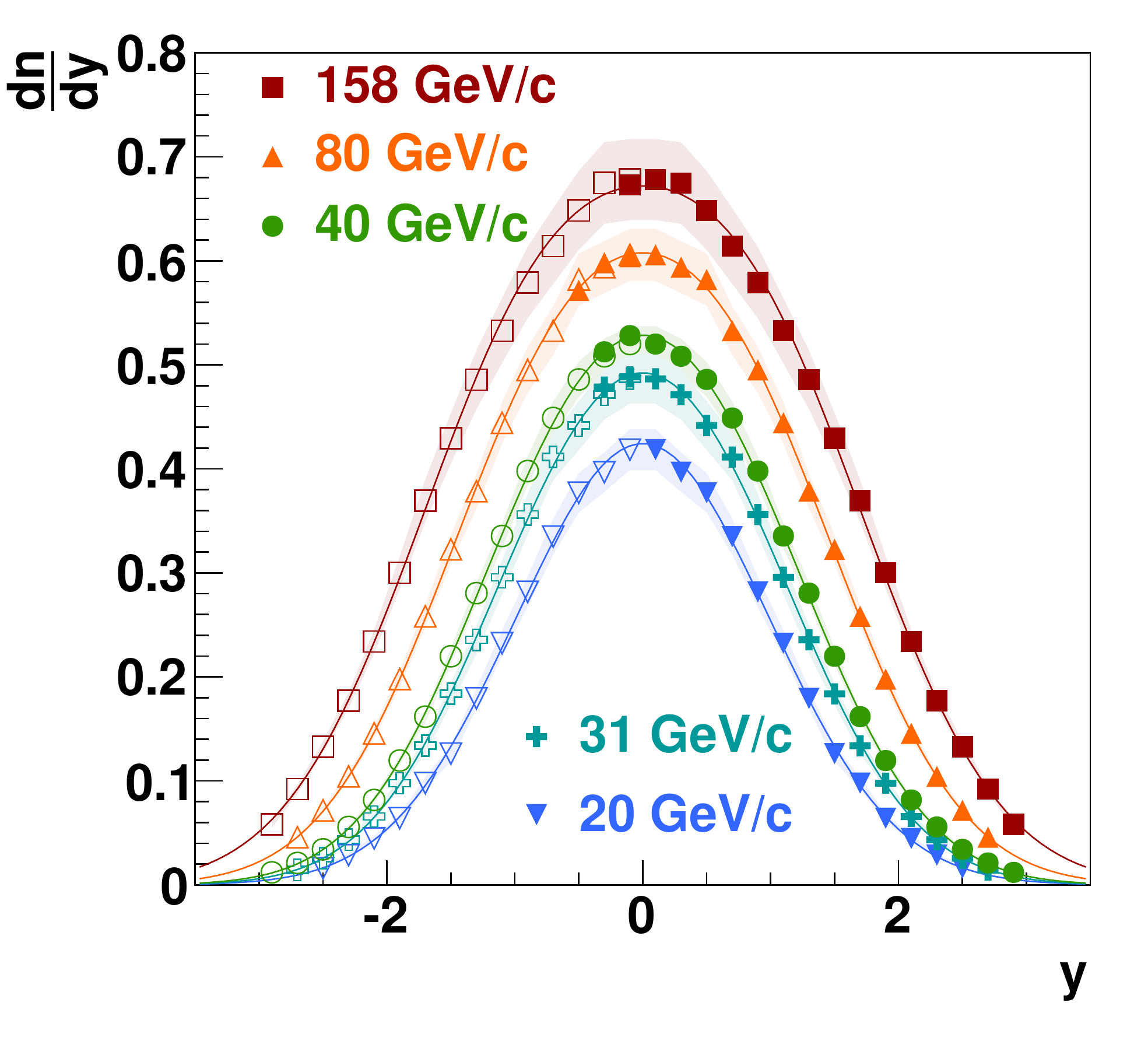}
  \includegraphics[width=0.49\textwidth]{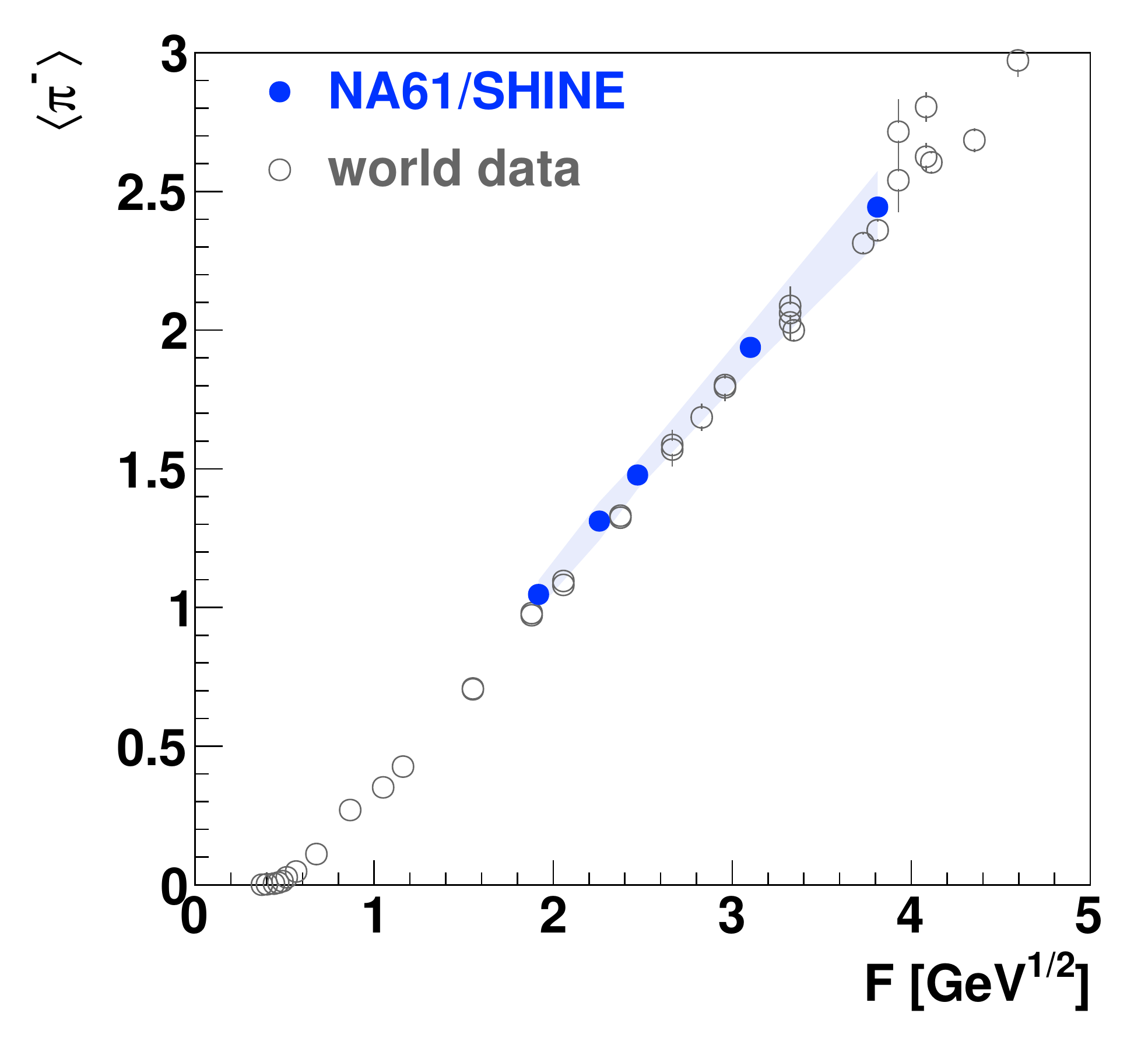}
  \caption{
  (Colour online)
  \emph{Left:}
  Rapidity spectra obtained from sums of the measured and extrapolated \mt 
  spectra. 
  Closed symbols indicate measured points, open points are reflected
  with respect to mid-rapidity.
  The measured points at $y<0$ are shown for systematic comparison only.
  The plotted statistical errors are smaller than the symbol size.
  The systematic uncertainties are indicated by the coloured bands.
  The lines indicate fits of the sum of two symmetrically 
  displaced Gaussian functions (see Eq.~\eqref{eq:rapidity}) to the spectra. 
  The results refer to \pim mesons produced in inelastic p+p interactions
  at 20, 31, 40, 80 and 158\GeVc.
  \emph{Right:} Dependence of the mean total multiplicity of \pim mesons
  produced in inelastic p+p interactions on Fermi's energy measure $F$ (see 
  Eq.~\eqref{eq:fermi}).
  The results of \NASixtyOne are indicated by filled circles and the compilation
  of the world data~\cite{pp_compil,na49_pp_pions} by open circles.
  The plotted statistical errors are smaller than the symbol size.
  The systematic uncertainties are indicated by the coloured band.
  }\label{fig:rapidity}
\end{figure*}

The rapidity spectra are shown in Fig.~\ref{fig:rapidity} (\emph{left}). 
They were obtained by summing the measured \mt spectra and using the
exponential function Eq.~\eqref{eq:mt}.
The function was fitted in the range ending at the maximum measured \mt, and 
starting 0.9\GeVcc below (note this is a different fit from the one shown in 
Fig.~\ref{fig:mt}).
The correction is typically below 0.2\% and becomes significant (several \%) 
only at $y > 2.4$. Half of the correction is added in quadrature to the 
systematic uncertainty in order to take into account a potential imperfectness 
of the exponential extrapolation.
The pion yield increases with increasing collision energy at all 
measured rapidities.

The rapidity spectra are parametrised by the sum of two  
Gaussian functions symmetrically displaced with respect to mid-rapidity:
\begin{linenomath} 
\begin{equation}\label{eq:rapidity}
\begin{split}
  \frac{\dd n}{\dd y} = &
  \frac{\avg{\pim}(y_0,\sigma_0)}{2\sigma_0\sqrt{2\pi}}\cdot\\
  & \cdot \left[
  \exp\left(-\frac{(y-y_0)^2}{2\sigma_0^2}\right)
  +\exp\left(-\frac{(y+y_0)^2}{2\sigma_0^2}\right)
  \right]\ ,
\end{split}
\end{equation}
\end{linenomath}
where $y_0$ and $\sigma_0$ are fit parameters, and the total multiplicity
$\avg{\pim}(y_0,\sigma_0)$ is calculated from the requirement  
that the integral over the 
measured spectrum equals the integral of the fitted function 
Eq.~\eqref{eq:rapidity} in 
the range covered by the measurements.
The $\chi^2$ function was minimised in a similar way as in case of the \mt 
spectra, namely using the integral of the function in a given bin.
The numerical values of the fitted parameters as well as
the r.m.s. width  \mbox{$\sigma = \sqrt{y_0^2 + \sigma_0^2}$}  are given
in Table~\ref{tab:fit_parameters}.

\subsection{Mean multiplicities}
\label{mean_multiplicities}

Mean multiplicities of  \pim mesons, $\avg{\pim}$,
produced in inelastic p+p interactions
at 20, 31, 40, 80 and 158\GeVc were calculated as the
integral of the fitted function Eq.~\eqref{eq:rapidity}.
The extrapolation into the unmeasured region at large $y$ 
contributes about 1\%. Half of it is added in quadrature to the 
systematic uncertainty.

The dependence of the produced average $\avg{\pim}$ multiplicity 
per inelastic p+p collision on the Fermi's energy measure~\cite{Fe:50},
\begin{linenomath} 
\begin{equation}
\label{eq:fermi}
   F \equiv \left[ \frac{(\sqrt{s_\mathrm{NN}} - 2 m_\mathrm{N})^3}
   {\sqrt{s_\mathrm{NN}}} \right]^{1/4} \,
\end{equation}
\end{linenomath}
is plotted in Fig.~\ref{fig:rapidity} (\emph{right}).
The results of \NASixtyOne are in agreement  with a compilation
of the world data~\cite{pp_compil,na49_pp_pions}.

\begin{table*}
  \caption{
  Numerical values of the parameters fitted to rapidity (see 
  Eq.~\eqref{eq:rapidity}) 
  and transverse mass (see Eq.~\eqref{eq:mt}) spectra of \pim mesons produced
  in inelastic p+p interactions at 20, 31, 40, 80 and 158\GeVc. 
  In case of the rapidity fit parameters $\avg{\pim}$, $\sigma$, 
  $\sigma_0$ and $y_0$, the systematic uncertainty dominates. The uncertainties
  written in the table are the quadrature sum of the statistical and systematic 
  uncertainties. All uncertainties are given numerically in~\cite{na61_edms}.  
  For $T$ and $\langle \mt\rangle$ the statistical uncertainty is written first 
  and the systematic one second.
}
\label{tab:fit_parameters}
\centering
\begin{tabular}{c | c c c c | c c}
\hline\hline
$p_\mathrm{beam}$ & $\langle\pi^-\rangle$ & $\sigma$ & $\sigma_0$ & $y_0$ & 
T($y=0$) & $\langle m_T\rangle(y=0) - m_\pi$\\
{[}\GGeVc] &  &  &  &  & [\MMeVcc] & [\MMeVcc]\\\hline
 20 & $1.047 \pm 0.051$ & $0.981 \pm 0.017$ & $0.921 \pm 0.118$ & $0.337 \pm 
0.406$ & $149.1 \pm 5.0 \pm 4.8$ & $237.8 \pm 6.4 \pm 2.3$\\
 31 & $1.312 \pm 0.069$ & $1.031 \pm 0.016$ & $0.875 \pm 0.050$ & $0.545 \pm 
0.055$ & $153.3 \pm 2.2 \pm 1.2$ & $246.1 \pm 2.7 \pm 0.9$\\
 40 & $1.478 \pm 0.051$ & $1.069 \pm 0.014$ & $0.882 \pm 0.045$ & $0.604 \pm 
0.044$ & $157.7 \pm 1.7 \pm 2.1$ & $247.3 \pm 2.0 \pm 0.9$\\
 80 & $1.938 \pm 0.080$ & $1.189 \pm 0.026$ & $0.937 \pm 0.019$ & $0.733 \pm 
0.010$ & $159.9 \pm 1.5 \pm 4.1$ & $253.5 \pm 1.9 \pm 1.1$\\
158 & $2.444 \pm 0.130$ & $1.325 \pm 0.042$ & $1.007 \pm 0.051$ & $0.860 \pm 
0.021$ & $159.3 \pm 1.3 \pm 2.6$ & $253.6 \pm 1.6 \pm 1.4$\\
\hline\hline
\end{tabular}
\end{table*}


\begin{figure}
\centering
  \includegraphics[width=0.49\textwidth]{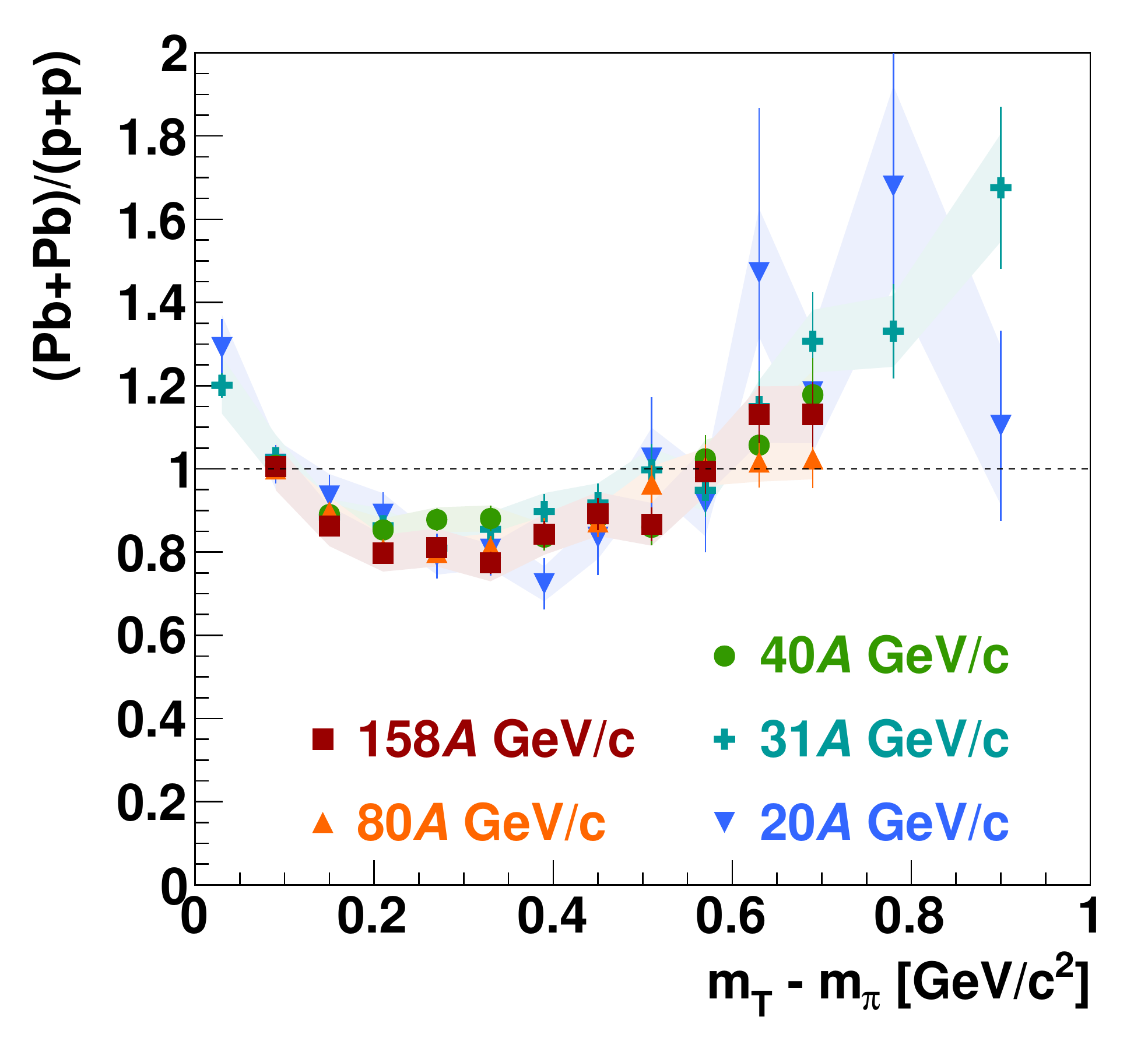}
  \caption{
  (Colour online)
  The ratio of the normalised transverse mass spectra of \pim mesons at 
  mid-rapidity produced in central Pb+Pb collisions 
  and inelastic p+p interactions at the same collision energy per nucleon.
  The coloured bands represent the systematic uncertainty of the p+p data.
  }\label{fig:mt_comp}
\end{figure}

\section{Comparison with central Pb+Pb collisions}
\label{sec:PbPb}

In this section the \NASixtyOne results on inelastic p+p interactions are
compared with the corresponding data on central Pb+Pb collisions
published previously by NA49~\cite{onseta,onsetb}.
Pion production properties which are different and similar
in p+p interactions and central Pb+Pb collisions are identified.
For completeness selected plots include the compilation of the world data on
inelastic p+p interactions~\cite{pp_compil,na49_pp_pions}, as well as
results on central  Au+Au  collisions 
from AGS~\cite{AGS1,AGS2} and RHIC~\cite{RHIC1,RHIC2,RHIC3,RHIC4,RHIC5}, as 
processed in~Ref.~\cite{onsetb}.

Figure~\ref{fig:mt_comp} shows 
the ratio of transverse mass spectra of \pim mesons produced
at mid-rapidity ($0 < y < 0.2$)
in central Pb+Pb collisions and p+p interactions at 
the same collision energy per nucleon.
The spectra were normalised to unity before dividing.
First, one observes that the ratio is not constant implying that
the spectral shapes are different in p+p interactions and central Pb+Pb 
collisions.
Second, it is seen that the ratio depends weakly, if at all,
on collision energy.
The ratio is higher than unity in the left ($\mt - m_{\pi} < 0.1$\GeVcc) and
right ($\mt - m_{\pi} > 0.5$\GeVcc) parts of the \mt range.
It is below unity in the central region $0.1 < \mt - m_{\pi} < 0.5$\GeVcc.

\begin{figure*}
  \includegraphics[width=0.49\textwidth]{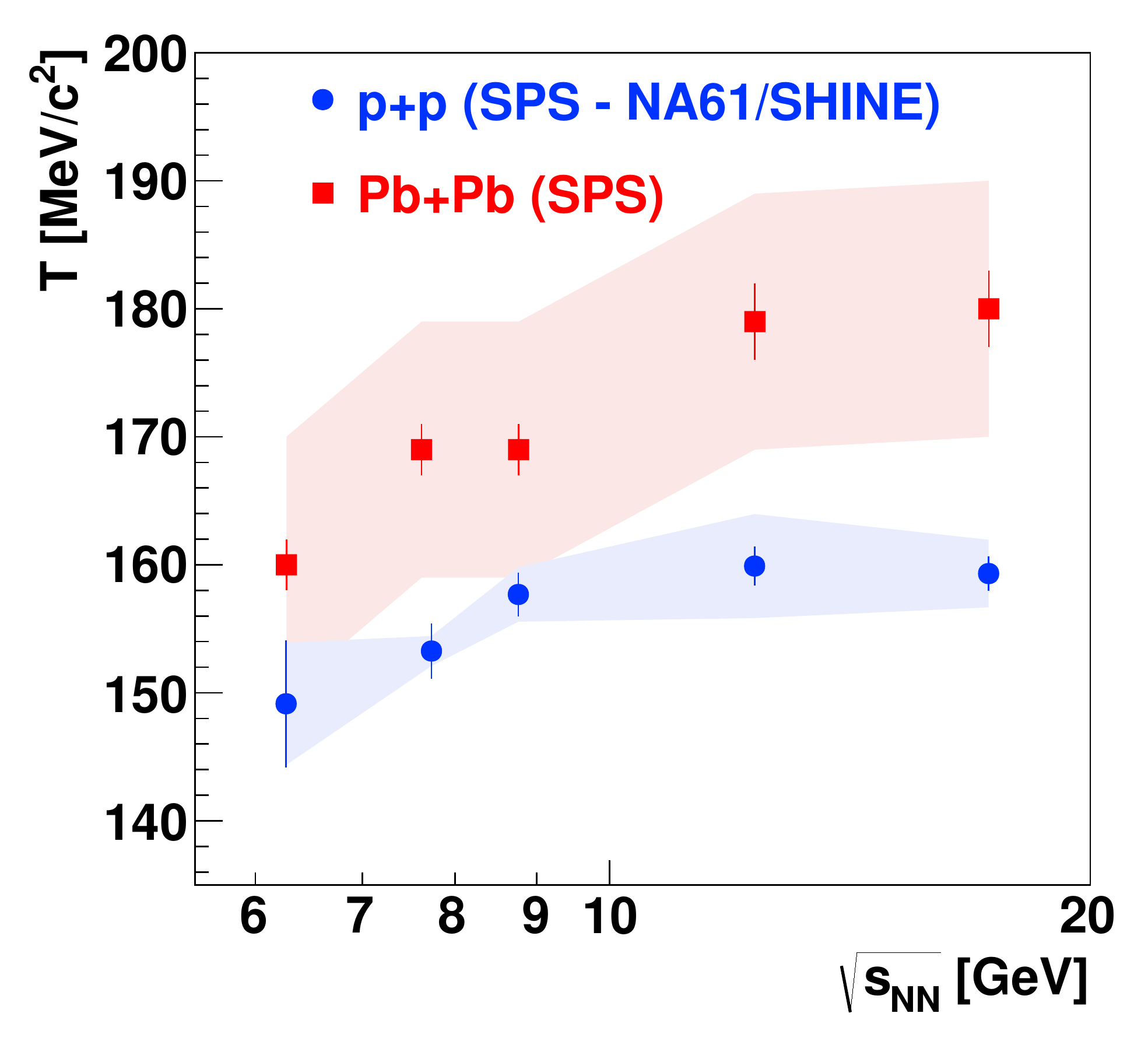}
  \includegraphics[width=0.49\textwidth]{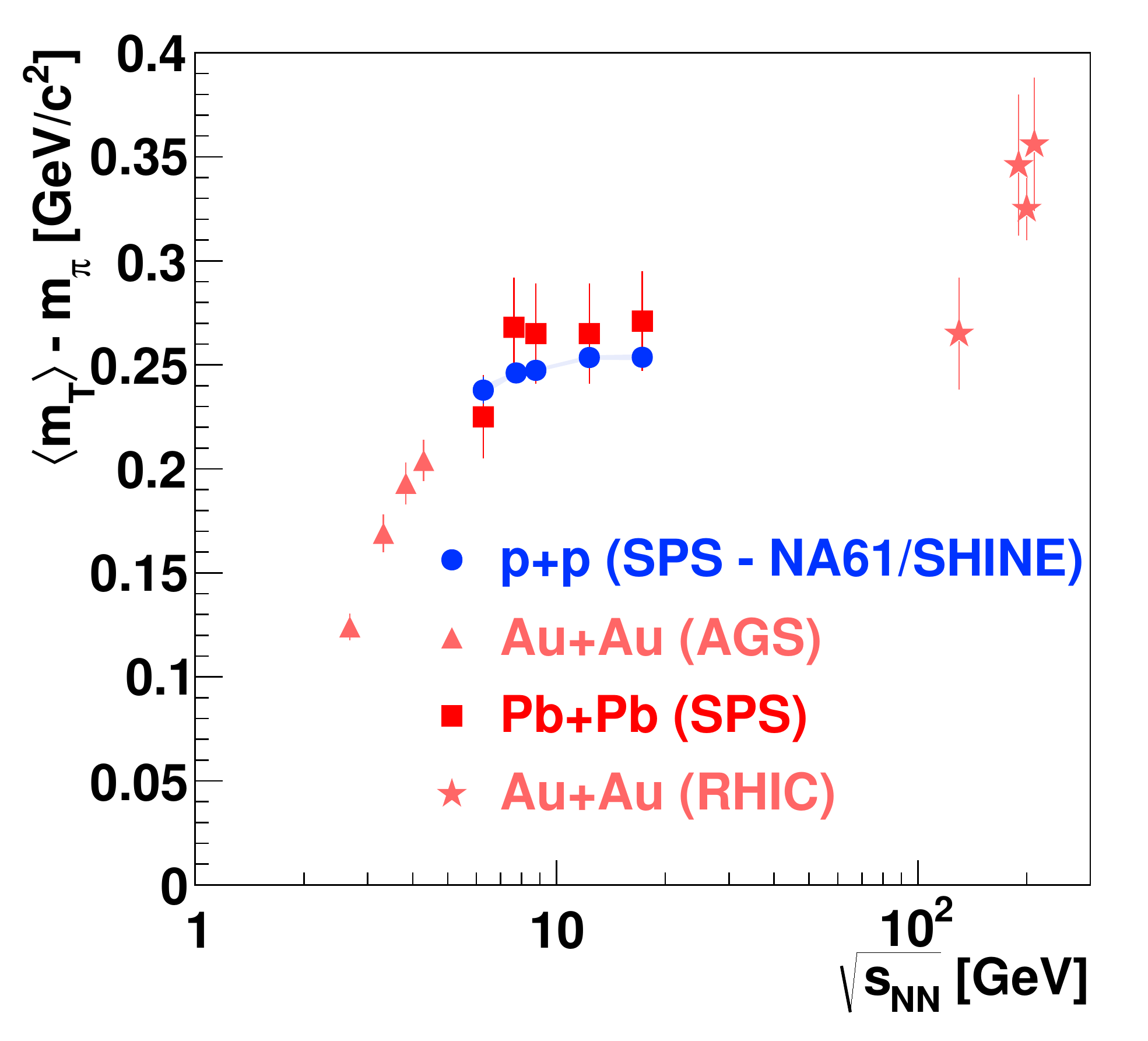}
  \caption{
  (Colour online)
  \emph{Left:} Inverse slope parameter $T$ of the transverse mass spectra at 
  mid-rapidity ($0 < y < 0.2$) plotted against the collision energy 
  per nucleon.
  The parameter $T$ was fitted in the range $0.2 < \mt - m_\pi < 
  0.7$\GeVcc.
  The systematic uncertainty for the two lowest energy points for Pb+Pb, not 
  given in \cite{onsetb} is assumed to be the same as for the higher 
  energies~\cite{onseta}.
  \emph{Right:} Mean transverse mass $\avg{\mt}$ at mid-rapidity 
  ($0 < y < 0.2$) versus the collision energy.
  The results on inelastic p+p interactions are compared with the corresponding
  data on central Pb+Pb (Au+Au) collisions.
  }\label{fig:slope_comp}
\end{figure*}

The inverse slope parameter $T$ of transverse mass spectra 
fitted in the range $0.2 < \mt - m_\pi < 0.7$\GeVcc is plotted
versus the collision energy in Fig.~\ref{fig:slope_comp} (\emph{left}).
The $T$ parameter is larger by about 10--20\MeVcc in central Pb+Pb 
collisions than in p+p interactions.
 
The transverse mass spectra measured by \NASixtyOne and NA49
allow a reliable calculation of mean transverse mass.
A small correction to the measured value for the high \mt
region not covered by the measurements was applied based on
the exponential extrapolation of the tail of the distributions.
Half of the correction was added to the systematic uncertainty on $\avg{\mt}$.
In spite of the different shapes of the \mt spectra
the mean transverse mass calculated for p+p interactions
and central Pb+Pb collisions is similar, see 
Fig.~\ref{fig:slope_comp} (\emph{right}).
This is because the differences shift the mean \mt in opposite
directions for different regions of \mt and as a result leave it almost 
unchanged. 
Thus the mean transverse mass appears  to be insensitive
to the apparent changes of the pion production properties
observed between p+p interactions and central Pb+Pb collisions.

\begin{figure*}
  \includegraphics[width=0.49\textwidth]{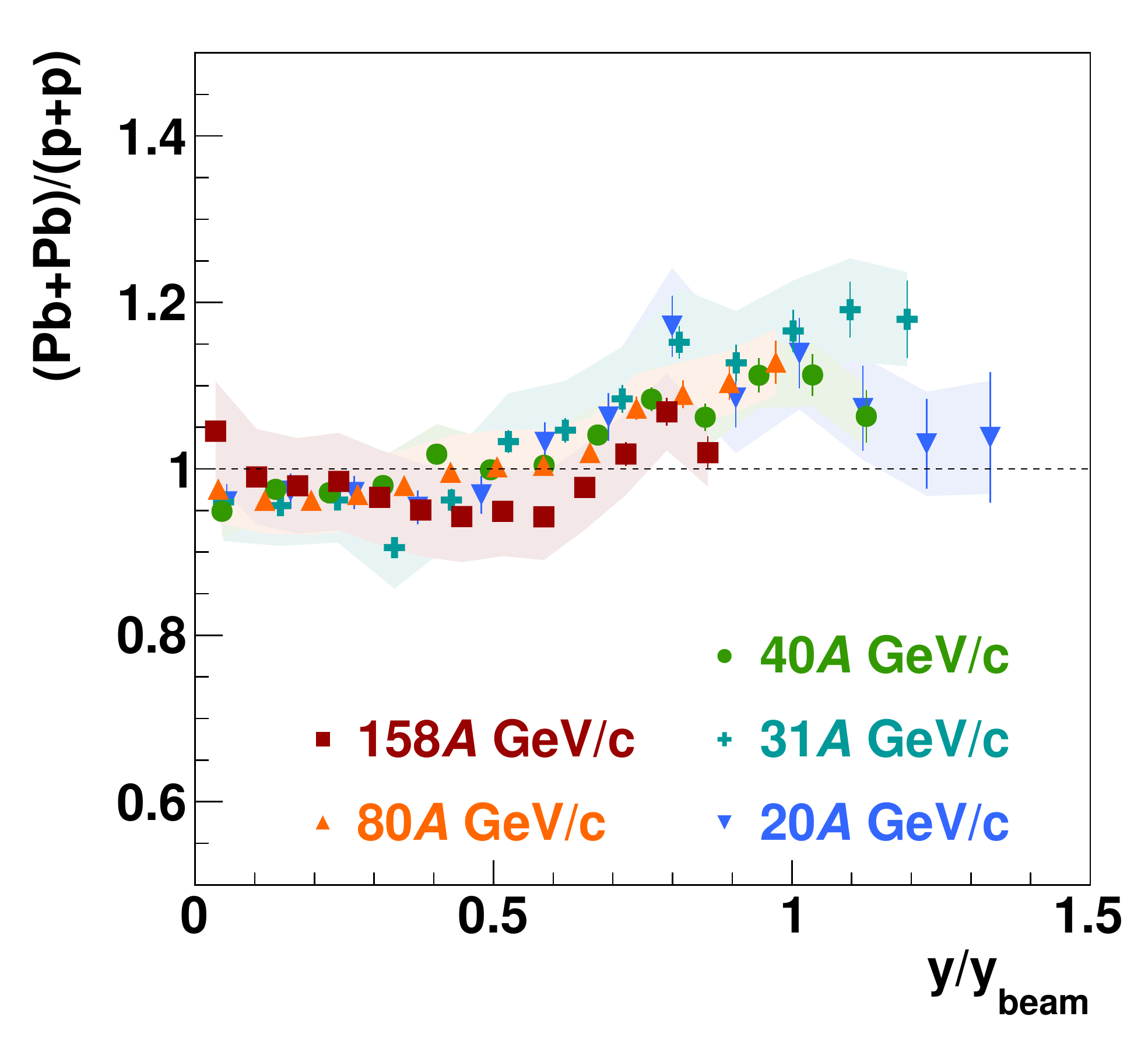}
  \includegraphics[width=0.49\textwidth]{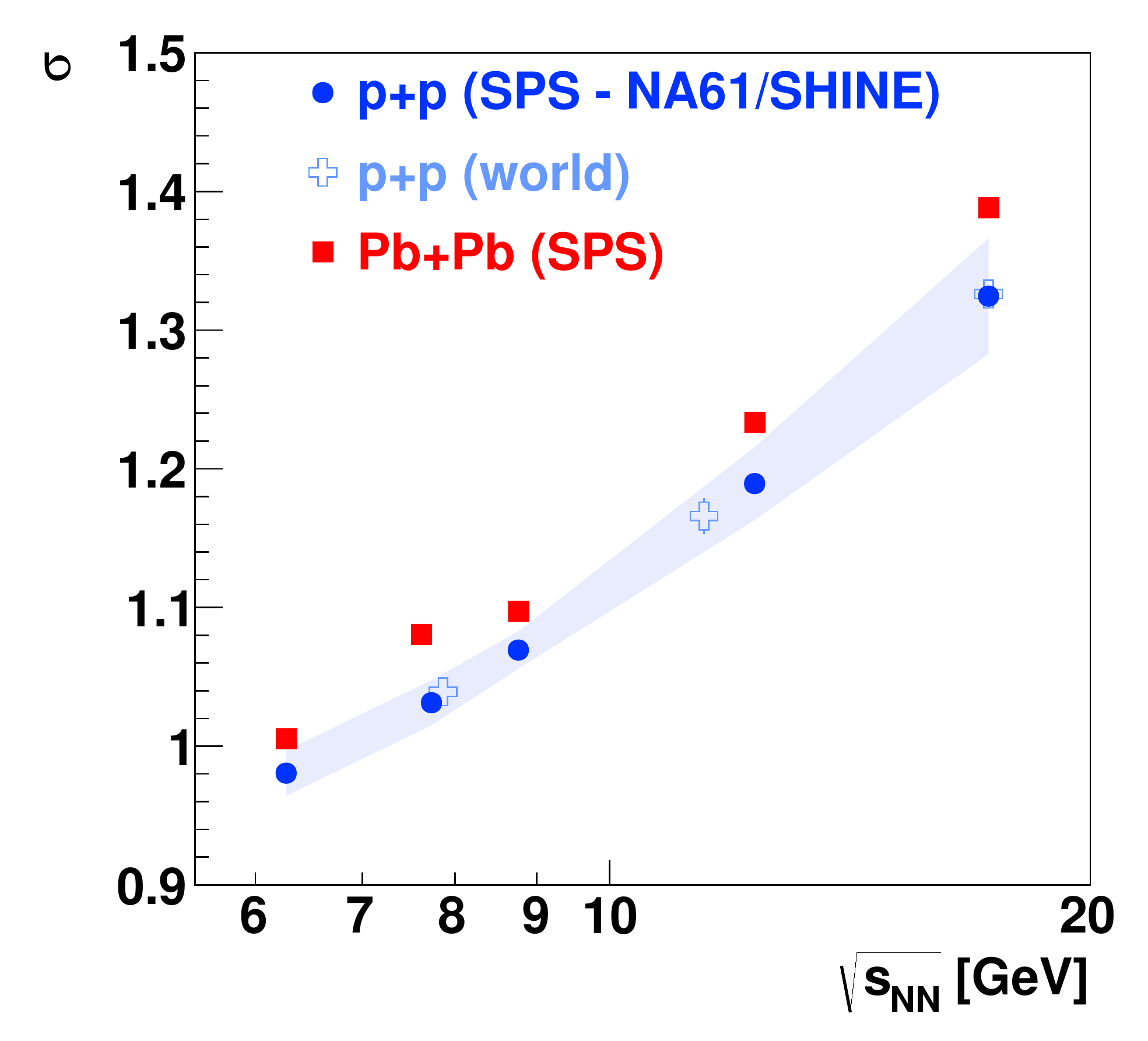}
  \caption{
  (Colour online)
  \emph{Left:} 
  The ratio of normalised rapidity spectra of \pim mesons produced
  in central Pb+Pb collisions 
  and inelastic p+p interactions at the same collision energy per nucleon
  plotted versus the rapidity scaled by the beam rapidity.
  The coloured bands represent the \NASixtyOne systematic uncertainty.
  \emph{Right:}
  Energy dependence of the width of the rapidity distribution
  of \pim mesons produced in p+p interactions and central Pb+Pb   
  collisions. The systematic uncertainty for the Pb+Pb points is not given.
}
\label{fig:rapidity_comp}
\end{figure*}

Figure~\ref{fig:rapidity_comp} (\emph{left}) presents the ratio of the 
normalised \pim rapidity spectra produced in central Pb+Pb and inelastic p+p 
interactions at the same collision energy per nucleon. The spectra are plotted 
versus versus the rapidity scaled by the beam rapidity.
Only weak, if any, energy dependence of the ratio is observed.
Moreover, the ratio is close to unity in the central rapidity
region ($y/y_\mathrm{beam} < 0.6$), whereas it is higher closer
to beam rapidity ($y/y_\mathrm{beam} > 0.6$).

Consequently the r.m.s. width $\sigma$
of rapidity distributions of \pim mesons produced 
in p+p interactions is smaller
than the width in central Pb+Pb collisions.
This is seen in
Fig.~\ref{fig:rapidity_comp} (\emph{right}) where the energy
dependence of $\sigma$ is plotted.
Additionally, p+p data from~\cite{MIRABELLE32,MIRABELLE69,na49_pp_pions} are 
shown; they agree with the \NASixtyOne results.

\begin{figure}
\centering
  \includegraphics[width=0.49\textwidth]{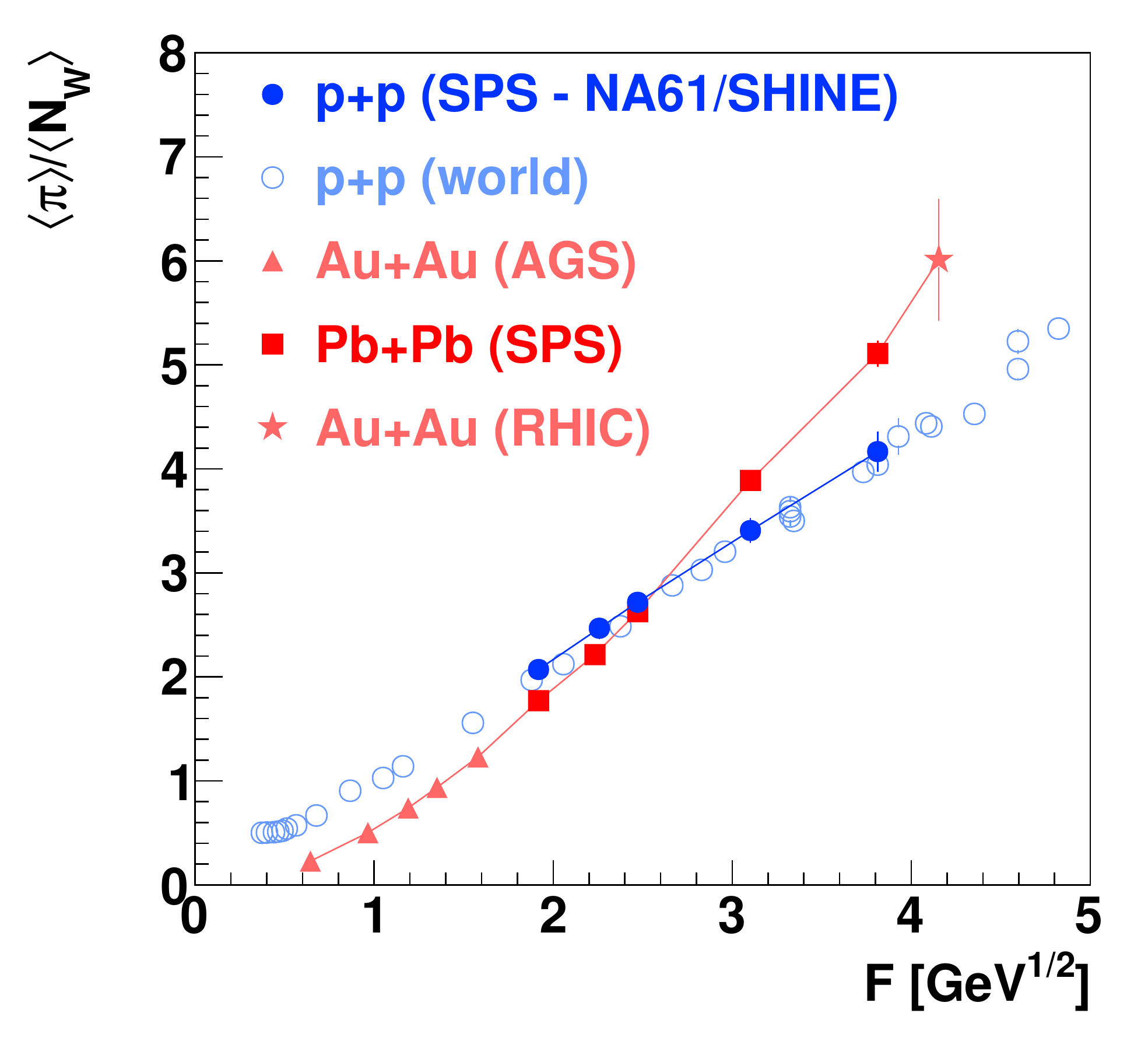}
  \caption{
  (Colour online)
  Mean multiplicity of all pions per wounded nucleon produced in
  inelastic p+p interactions and central Pb+Pb (Au+Au) collisions.
  The vertical lines show the total uncertainty.
}
\label{fig:mean_multiplicity_comp}
\end{figure}

Note, that when interpreting differences between results
obtained for inelastic p+p interactions and central Pb+Pb
collisions the isospin effects should be taken into account.
This concerns both the spectra as well as the total 
multiplicities~\cite{na49_pp_pions}.

In order to reduce their influence the mean multiplicity
of pions is obtained from a sum of mean multiplicities
of negatively and positively charged pions 
using the phenomenological formula~\cite{Sasha}:
\begin{linenomath} 
\begin{equation}
  \avg{\pi} = \frac{3}{2}\left(\avg{\pip} + \avg{\pim}\right)\ .
\end{equation}
\end{linenomath}
The results divided by the mean number of wounded nucleons
($N_\mathrm{W} = 2$ for p+p) are shown in 
Fig.~\ref{fig:mean_multiplicity_comp}  as a function of 
the Fermi energy measure $F$. 
The value of $\avg{\pip}$ for the \NASixtyOne 
results on inelastic p+p interactions 
was estimated from the measured $\avg{\pim}$ multiplicity
assuming $\avg{\pip} = \avg{\pim} + 2/3$.
This assumption is based on the compilation of the world data 
presented in Ref.~\cite{Sasha} and the model presented therein.
At beam momenta lower than 40\AGeVc the $\avg{\pi}/\avg{N_\mathrm{W}}$
ratio is higher in p+p interactions than in central Pb+Pb
collisions. The opposite relation holds for beam
momenta higher than 40\AGeVc. 
The energy dependence for inelastic p+p interactions crosses the one for central
Pb+Pb (Au+Au) collisions at about 40\AGeVc.

\section{Summary}
\label{sec:summary}
\begin{sloppypar}
We presented experimental results on inclusive spectra and mean multiplicities 
of negatively charged pions produced in inelastic p+p interactions
at  20, 31, 40, 80 and 158\GeVc.
Two dimensional spectra in transverse momentum and rapidity
and parameters characterizing them were given.
The results agree with existing sparse measurements, extend their 
range, accuracy and depth of detail.
\end{sloppypar}

The results on inelastic p+p interactions were compared with the corresponding
data on central Pb+Pb collisions obtained by NA49.
The spectra in p+p interactions are narrower both in rapidity and in 
transverse mass, which might be attributed to isospin effects.
The mean pion multiplicity per wounded nucleon in p+p 
interactions increases more slowly with energy in the SPS range and crosses the 
corresponding dependence measured in the Pb+Pb collisions at about 40\AGeVc.

\section{Acknowledgements}

\begin{sloppypar}
This work was supported by
the Hungarian Scientific Research Fund (grants OTKA 68506 and 71989),
the Polish Ministry of Science and Higher Education (grants
667\slash N-CERN\slash 2010\slash 0, NN 202 48 4339 and  NN 202 23 1837),
the National Science Center of Poland (grant UMO-2012\slash 04\slash M\slash 
ST2\slash 00816),
the Foundation for Polish Science -- MPD program, co-financed by the
European Union within the European Regional Development Fund, 
the Federal Agency of Education of the Ministry of Education and Science
of the Russian Federation (grant RNP 2.2.2.2.1547), the Russian Academy of
Science and
the Russian Foundation for Basic Research (grants 08-02-00018, 09-02-00664,
and 12-02-91503-CERN),
the Ministry of Education, Culture, Sports, Science and Technology,
Japan, Grant-in-Aid for Scientific Research (grants 18071005, 19034011,
19740162, 20740160 and 20039012),
the German Research Foundation (grants GA 1480\slash 2-1, GA 1480\slash 2-2),
Bulgarian National Scientific Fondation (grant DDVU 02\slash 19\slash  2010),
Ministry of Education and Science of the Republic of Serbia (grant OI171002),
Swiss Nationalfonds Foundation (grant 200020-117913\slash 1)
and ETH Research Grant TH-01 07-3.

Finally, it is a pleasure to thank
the European Organization for Nuclear Research
for a strong support and hospitality
and, in particular, the operating crews of the CERN SPS accelerator
and beam lines who made the measurements possible.
\end{sloppypar}

\clearpage

\end{document}